\documentclass[aps,twocolumn,nofootinbib,longbibliography,notitlepage,floatfix]{revtex4-2}
\usepackage[T1]{fontenc}
\usepackage{graphicx}
\usepackage{xcolor}
\usepackage{tikz}
\usetikzlibrary{arrows.meta}
\usetikzlibrary{positioning}
\usetikzlibrary{svg.path}
\usepackage{amsmath,mathtools}
\usepackage{amssymb}
\usepackage{amstext,cuted}
\usepackage{amsbsy}
\usepackage{amsthm}
\usepackage{physics}
\usepackage{mathrsfs}
\usepackage{esint}
\usepackage{aligned-overset}
\usepackage{dsfont}
\usepackage{tabularx}
\usepackage{longtable,booktabs}
\usepackage{dcolumn}
\usepackage{multirow}
\usepackage{makecell}
\usepackage{outlines}
\usepackage{subfigure}
\usepackage{url}
\usepackage{bm}
\usepackage{float}
\usepackage{extarrows}
\usepackage[super]{nth}
\usepackage{scalerel}
\usepackage{lineno}

\usepackage{xcolor}

\definecolor{darkblue}{RGB}{0,0,150}
\definecolor{darkred}{RGB}{150,0,0}

\usepackage[
  colorlinks=true,
  linkcolor=darkblue,
  citecolor=darkblue,
  urlcolor=darkred,
  linktocpage=true
]{hyperref}
\usepackage{cleveref}

\tikzset{
    wcirc/.style={draw, circle, minimum size=1ex, inner sep=0pt},
    bcirc/.style={fill, circle, minimum size=1ex, inner sep=0pt},
	wcircL/.style={draw, circle, minimum size=1.5ex, inner sep=0pt},
    bcircL/.style={fill, circle, minimum size=1.5ex, inner sep=0pt}
}

\def\i{\mathrm{i}}

\definecolor{orcidlogocol}{HTML}{A6CE39}
\tikzset{
	orcidlogo/.pic={
		\fill[orcidlogocol] svg{M256,128c0,70.7-57.3,128-128,128C57.3,256,0,198.7,0,128C0,57.3,57.3,0,128,0C198.7,0,256,57.3,256,128z};
		\fill[white] svg{M86.3,186.2H70.9V79.1h15.4v48.4V186.2z}
		svg{M108.9,79.1h41.6c39.6,0,57,28.3,57,53.6c0,27.5-21.5,53.6-56.8,53.6h-41.8V79.1z M124.3,172.4h24.5c34.9,0,42.9-26.5,42.9-39.7c0-21.5-13.7-39.7-43.7-39.7h-23.7V172.4z}
		svg{M88.7,56.8c0,5.5-4.5,10.1-10.1,10.1c-5.6,0-10.1-4.6-10.1-10.1c0-5.6,4.5-10.1,10.1-10.1C84.2,46.7,88.7,51.3,88.7,56.8z};
	}
}

\allowdisplaybreaks[4]
\newcommand\orcidicon[1]{\href{https://orcid.org/#1}{\mbox{\scalerel*{
				\begin{tikzpicture}[yscale=-1,transform shape]
					\pic{orcidlogo};
				\end{tikzpicture}
			}{|}}}}
\tikzset{bold/.style={color=blue, line width=2pt}}
\tikzset{redop/.style={circle,fill=red}}
\tikzset{blueop/.style={circle,fill=blue}}
\graphicspath{{pic/}}

\makeatletter
\newcommand{\rmnum}[1]{\romannumeral #1}
\newcommand{\Rmnum}[1]{\expandafter\@slowromancap\romannumeral #1@}

\makeatother

\begin{document}


\title{Universal Design and Physical Applications of Non-Uniform Cellular Automata on Translationally Invariant Lattices}

\author{Xiang-You Huang}\thanks{These authors contributed equally.}

\author{Jie-Yu Zhang}\thanks{These authors contributed equally.}
 
   \author{Peng Ye\orcidicon{0000-0002-6251-677X}}
\email{yepeng5@mail.sysu.edu.cn}
\affiliation{Guangdong Provincial Key Laboratory of Magnetoelectric Physics and Devices, State Key Laboratory of Optoelectronic Materials and Technologies,
	and School of Physics, Sun Yat-sen University, Guangzhou, 510275, China}

\date{\today}

\begin{abstract}
Motivated by recent theoretical and experimental advances, hyperbolic lattices have emerged as a paradigmatic setting in which geometry becomes an active organizing principle of quantum systems.
Their negative curvature, exponential volume growth, and non-Abelian translation symmetry make them fundamentally distinct from Euclidean lattices and give rise to rich geometry-dependent physics, but also hinder the direct application of well-established analytical and computational approaches originally developed for physical systems defined on Euclidean lattices.
To establish a unified framework for geometry-dependent physics on Euclidean and hyperbolic lattices, we develop \textit{higher-order non-uniform cellular automata} (NUCA) as a local-to-global construction for translationally invariant regular lattices.
This construction derives geometry-dependent update rules through a lattice-deforming procedure that embeds hyperbolic lattices into a Euclidean square lattice, thereby encoding hyperbolic geometry while preserving physical locality.
It thus provides a systematic route toward quantum and classical physics on hyperbolic lattices.
We demonstrate the framework in three applications ranging from quantum many-body physics to non-equilibrium statistical physics.
First, on the hyperbolic $\{5,4\}$ lattice, a linear NUCA generates exactly solvable subsystem symmetry-protected topological (SSPT) models and spontaneous subsystem symmetry-breaking models, in which regular and irregular subsystem symmetries are organized by the underlying negatively curved geometry.
We diagnose the resulting SSPT order using NUCA-generated multi-point strange correlators and derive a sufficient condition for NUCA-generated Hamiltonians to be invariant under non-Abelian translations.
Second, as a quantum generalization, we construct non-uniform Clifford quantum cellular automata (CQCA) for the hyperbolic cluster state, extending the correspondence between CQCA dynamics and subsystem symmetries from Euclidean translation invariance to the hyperbolic setting with non-Abelian translations.
Third, we formulate a probabilistic NUCA for directed percolation (DP) on the hyperbolic lattice, whose stochastic dynamics incorporate the treelike growth induced by negative curvature, from which we estimate bond and site DP thresholds and the phase diagram.
Taken together, these results establish both deterministic and probabilistic NUCA constructions designed here as a local-to-global principle through which translationally invariant curved lattice geometry imposes concrete constraints and enriches the physics of subsystem-symmetric topological matter, quantum cellular automata, and directed percolation.
These constructions open a new avenue toward exotic quantum matter, quantum information, and non-equilibrium statistical physics beyond the conventional Euclidean framework.
\end{abstract}

\maketitle

\tableofcontents

\section{Introduction}
\label{sec_introduction}
Lattice geometry is a fundamental organizing principle of physical systems, shaping phenomena ranging from single-particle band structures to strongly correlated phases of matter.
Beyond specifying microscopic connectivity, geometry imposes local constraints that can reorganize the symmetry structure, excitation content, and long-distance behavior of a many-body system.
A particularly sharp manifestation of this principle is provided by subsystem symmetry.
Unlike ordinary global symmetries, subsystem symmetries act on rigid, non-deformable, and subextensive regions whose geometry is constrained by the underlying lattice~\cite{you2018sspt,Devakul2018classification}.
As a result, they are intrinsically sensitive to lattice geometry and provide a natural route to geometry-dependent quantum phases.
Such symmetries are now known to play a central role in subsystem symmetry-protected topological (SSPT) order~\cite{you2018sspt,Devakul2018classification,Devakul2019fractal,Devakul2019fractalclass,Devakul2018fractaluniversal,Stephen2019subsystem,Daniel2020computational,Raussendorf_2001_MBQC,zhou_2022_strangecor,zhang2024hoca}, fracton topological order~\cite{fu2015fracton,fu2016fracton}, and symmetry-enriched topological order~\cite{Barkeshli2019set,Stephen2020set,Stephen2022fractionalization,zhang2025set}.
For instance, cluster states protected by subsystem symmetries can serve as universal resource states for measurement-based quantum computation~\cite{Raussendorf_2001_MBQC,Devakul2018fractaluniversal,Stephen2019subsystem,Daniel2020computational}, as experimentally demonstrated in various platforms~\cite{Asavanant2019cluster,Larsen2019cluster,Lanyon2013clusters}.
Geometry also plays an essential role in non-equilibrium statistical physics.
A central example is directed percolation (DP), where local propagation rules determined by the lattice geometry control the global phase transition.
DP represents the universality class identified by the Janssen-Grassberger conjecture~\cite{Janssen1981,Grassberger1982,Hinrichsen2000,DK1984}, whose critical behavior has been experimentally observed in various turbulent systems~\cite{Takeuchi_2007_DP,Takeuchi_2009_DP,Lemoult2016,Lemoult2024}.
More recently, DP has also appeared in broad modern physics, ranging from measurement-induced phase transitions to discrete time crystals~\cite{Pizzi2021,MIPT_DP_2021}.
Therefore, investigating these geometry-dependent systems across diverse lattice geometries is essential for understanding how geometry organizes both equilibrium phases of matter and non-equilibrium dynamics.

Among various lattice geometries, translationally invariant lattices play a central role in condensed-matter physics because they provide the basic setting for defining and classifying phases of matter.
A hyperbolic lattice is a discretization of the hyperbolic plane with constant negative curvature, which is closely related to anti-de Sitter space~\cite{Brower_ads2,Brower_ads3}.
Compared with Euclidean lattices, hyperbolic lattices possess intrinsically non-Euclidean geometric structures, most notably non-Abelian translation symmetry, which generalizes translation invariance beyond the conventional Abelian picture~\cite{Crystallography2022,Lux_2023_PBC,Maciejko2020HBT,Maciejko2022AHBT,Lenggenhager2023supercell}.
In recent years, hyperbolic lattices have attracted broad interest in condensed-matter physics, driven in part by experimental advances in simulating quantum systems on such geometries~\cite{Kollr2019,Lenggenhager_2022_circuit,Huang_2024_photonic}.
On the theoretical side, hyperbolic geometry has been shown to support phenomena beyond the scope of Euclidean lattice models, including non-interacting fermions~\cite{Maciejko2020HBT,Maciejko2022AHBT,Lenggenhager2023supercell,Vidal2023dos,chen_2023_h_haldane,Tummuru_2024_hyperbolicsemimetal,A_Chen_2024_localization,huang2025hyperbolicEE}, many-body physics~\cite{Lenggenhager_2025_hsl,Dutkiewicz_2026_BEC,Gluscevich2025dynamic,Saraidaris2025criticalspinmodels,yan2019hfm,yan2019hfm2,yan2025hfm3}, quantum codes~\cite{Breuckmann_2016_hyper_surfacecode,Fahimniya2025faulttolerant,Mahmoud2026QEC}, and holography~\cite{Asaduzzaman_holography,Jahn_2020_holography,Boyle2020conformal,Boyle2025holographic,Santanu2024holography}.
These developments make hyperbolic lattices a natural arena for geometry-dependent phenomena, including subsystem symmetries of quantum states and directed percolation.
However, the same geometric features that make hyperbolic lattices physically rich also create fundamental obstacles: non-Abelian translation symmetry and the finite surface-to-volume ratio in the thermodynamic limit obstruct the direct application of well-established analytical and computational approaches developed for Euclidean lattices~\cite{Crystallography2022,hl_percolation,Margenstern2018,Schrauth_2024_hypertiling}.
Although specific subsystem symmetries have been identified on hyperbolic lattices~\cite{yan2019hfm,yan2019hfm2,yan2025hfm3}, the absence of a general constructive framework limits a systematic understanding of the corresponding quantum states.
Similarly, numerical studies of percolation on hyperbolic lattices face the intrinsic difficulty of labeling lattice sites and storing lattice data as the number of sites grows exponentially~\cite{hl_percolation,Schrauth_2024_hypertiling,Margenstern2018}.
Since DP is a central model of non-equilibrium statistical physics, overcoming these geometric and computational barriers is essential for studying DP-related phenomena on hyperbolic lattices.
It is therefore necessary to develop a local-to-global framework that incorporates the constraints of hyperbolic geometry while retaining physical locality, so that geometry-dependent quantum phases and non-equilibrium dynamics can be treated on the same footing.

\begin{figure*}[t]
	\includegraphics[width=0.95\textwidth]{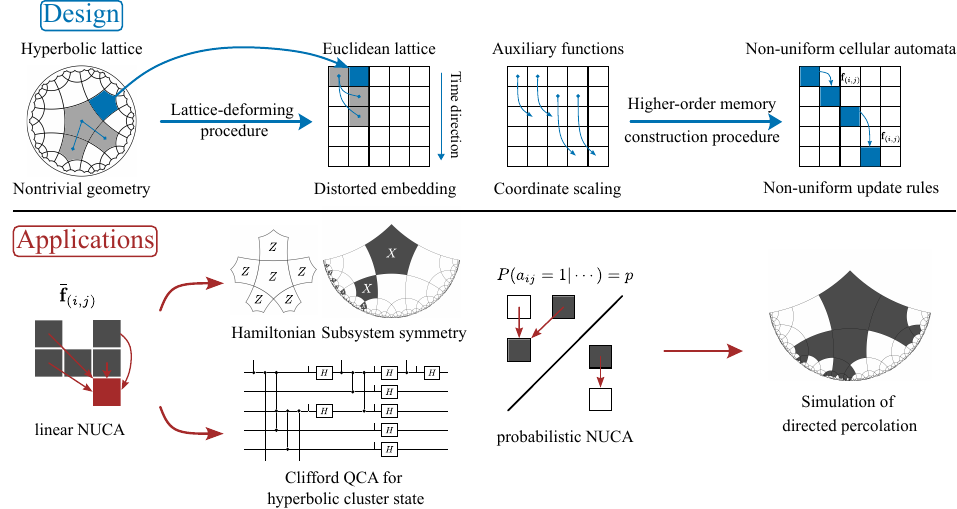}
	\caption{\label{fig_NUCA} 
	Schematic overview of the NUCA framework. By developing a lattice-deforming procedure that embeds hyperbolic lattices into a Euclidean square lattice, we derive coordinate-scaling functions to characterize the geometric distortion. The resulting higher-order non-uniform update rules bridge non-uniformity in CA and physical locality on the hyperbolic lattice. The framework is applied to three geometry-dependent physical settings. Through a linear NUCA, we design and analyze SSPT and SSSB models in Sec.~\ref{sec_NUCA54}, and classify NUCA-generated symmetries in Sec.~\ref{sec_diversity}. As a quantum generalization, we construct non-uniform Clifford QCA for generating subsystem symmetries of the hyperbolic cluster state in Sec.~\ref{sec_QCA}. Using a probabilistic NUCA, we simulate directed percolation on the hyperbolic lattice and estimate thresholds and the phase diagram in Sec.~\ref{sec_DP}.
	} 
\end{figure*}

Cellular automata (CA) are discrete dynamical systems in which simple local update rules can generate rich global structures and long-distance behaviors~\cite{wolfram1983review,wolfram1984,TOFFOLI1977213,ROLLIER2025108362,PCA2018}.
This local-to-global character makes CA and their variants especially suited for physical problems where global phenomena are constrained by microscopic rules, including subsystem symmetries of quantum states~\cite{Devakul2019fractal,Devakul2018fractaluniversal,Devakul2019fractalclass,Stephen2019subsystem,Daniel2020computational,zhang2024hoca,zhang2025set,biswas2022mca} and directed percolation~\cite{DK1984,Pizzi2021,MIPT_DP_2021}.
Their quantum generalizations, quantum cellular automata (QCA), further provide natural frameworks for quantum simulation and universal quantum computation~\cite{Raussendorf_2005_QCA,Osborne_2006_QCA,Farrelly2020reviewofquantum,Arrighi2019qca}.
Conventional CA are usually defined by synchronous evolution under an identical update rule across all sites~\cite{wolfram1984,TOFFOLI1977213,ROLLIER2025108362}, making them naturally adapted to translationally invariant Euclidean lattices.
However, this built-in uniformity also reveals their limitation: conventional one-dimensional CA cannot be directly applied to hyperbolic lattices, where translation symmetry is non-Abelian and physical locality is organized by non-Euclidean geometry.
This motivates a generalization of the uniform CA formalism that can retain the local-to-global principle while incorporating the geometric constraints of hyperbolic lattices.

Non-uniform cellular automata generalize conventional CA by relaxing the requirement of a single uniform update rule, thereby allowing more complex dynamical behavior~\cite{nuca1986,tNUCA1997,tNUCA2023,tNUCA2024,nuca2026} and enabling diverse applications~\cite{Faraoun_2014_NUCA,image_encryption_2017,Kumaravel2013,nuca2022}.
In existing one-dimensional non-uniform CA, however, the variation of update rules is usually independent of the underlying geometry and therefore amounts mainly to a breaking of Euclidean translation invariance.
This is insufficient for physical systems on hyperbolic lattices, where the update dynamics must adapt to the target lattice while preserving physical locality.
In this setting, non-uniformity is not merely a departure from translation invariance, but must carry the geometric constraints imposed by negative curvature.
Since there is no distortion-free two-dimensional Euclidean model of the hyperbolic plane~\cite{Ratcliffe2019hyperbolic}, designing one-dimensional non-uniform CA that preserve locality on hyperbolic lattices poses a fundamental challenge that has remained unresolved.

\begin{figure}
	\centering
	\includegraphics[width=\columnwidth]{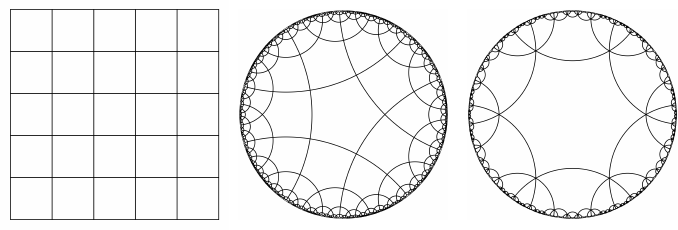}
	\caption{\label{fig_lattices} Visualization of the Euclidean $\{4,4\}$ (square) lattice and the hyperbolic $\{5,4\}$ and $\{6,6\}$ lattices studied in this paper. Physical qubits are located on polygons.}
\end{figure}

In this paper, we develop a \textit{higher-order non-uniform cellular automata} (NUCA) framework for translationally invariant regular $\{p,q\}$ Euclidean and hyperbolic lattices with $p,q\ge4$, as illustrated in Fig.~\ref{fig_NUCA}. 
Here, $\{p,q\}$ denotes a lattice where $q$ regular $p$-gons meet at each vertex, as visualized in Fig.~\ref{fig_lattices}, and NUCA$(p,q)$ denotes the application of our framework to this lattice.
Unlike conventional non-uniform CA whose update rules are independent of geometry~\cite{nuca2026}, our construction introduces position-dependent update rules by embedding a hyperbolic lattice into a Euclidean square lattice with explicitly characterized distortion, thereby encoding hyperbolic geometry while preserving physical locality.
This geometry-dependent NUCA framework allows us to construct deterministic linear NUCA for quantum states with subsystem symmetries~\cite{you2018sspt,Devakul2018classification}, non-uniform Clifford QCA for the hyperbolic cluster state~\cite{Farrelly2020reviewofquantum,Arrighi2019qca}, and stochastic probabilistic NUCA for directed percolation~\cite{DK1984,Hinrichsen2000}.
These constructions provide a unified local-to-global framework for studying subsystem-symmetric quantum matter and non-equilibrium dynamics on hyperbolic lattices.
Moreover, the same geometric construction can in principle be extended to other complex geometries, such as self-similar fractal lattices~\cite{Mandelbrot_fractal,Gefen1980fractal,Kempkes2018,zhou_2024_fractal}.

Concretely, the main results consist of two parts: \textit{Design} and \textit{Applications}, as shown in Fig.~\ref{fig_NUCA}.

In the \textit{Design} part, we formulate the NUCA framework by incorporating nontrivial geometric data into the non-uniform update rules in Eq.~(\ref{eq_update_rule}) and Eq.~(\ref{eq_transposed_update_rule}).
Under these update rules, the discrete dynamics of a one-dimensional NUCA$(p,q)$ generate spatial configurations on the two-dimensional $\{p,q\}$ lattice.
To implement this construction on hyperbolic $\{p,q\}$ lattices, we use the splitting method as a discrete labeling system~\cite{Margenstern2018,margenstern_1999,Margenstern_2000,MARGENSTERN2001,Margenstern2007} and develop a lattice-deforming procedure that embeds the hyperbolic $\{p,q\}$ lattice into a Euclidean square lattice, as shown in Fig.~\ref{fig_54coor}.
The induced distortion is characterized by coordinate-scaling functions on the deformed lattice, which encode the mismatch between Euclidean coordinates and hyperbolic locality.
This construction yields higher-order and position-dependent update rules, thereby connecting the non-uniformity of cellular automata to the physical locality imposed by hyperbolic geometry.

Next, we demonstrate the
above framework in three \textit{Applications} ranging from quantum many-body physics to non-equilibrium statistical physics.  As the first application, we use linear NUCA$(5,4)$ to provide a constructive framework for quantum matter with subsystem symmetries on the hyperbolic $\{5,4\}$ lattice, extending the symmetry-generating mechanisms previously restricted to uniform CA and Euclidean translation invariance~\cite{Devakul2019fractal,Devakul2018fractaluniversal,Devakul2019fractalclass,Sfairopoulos2023,sfairopoulos2025cellularautomataddimensions,zhang2024hoca,zhang2025set}.
Rather than imposing subsystem symmetries on models by hand, the NUCA simultaneously determines subsystem symmetries and the corresponding exactly solvable SSPT models in Eq.~(\ref{eq_54_SSPT}) and spontaneous subsystem symmetry-breaking (SSSB) models in Eq.~(\ref{eq_54_SSB}).
We classify the NUCA-generated symmetries as \textit{regular} or \textit{irregular} according to their growth behavior as shown in Fig.~\ref{fig_sym_fit}, thereby extending the Euclidean classification of subsystem symmetries~\cite{zhang2024hoca} to hyperbolic lattices and revealing the imprint of the exponential expansion of hyperbolic geometry~\cite{Ratcliffe2019hyperbolic,Crystallography2022}.
To diagnose the resulting SSPT order, we construct NUCA-generated \textit{multi-point strange correlators} (MPSC)~\cite{zhou_2022_strangecor,zhang2024hoca,gao_zhang2025ssc,you_xu_2014_strangecor,Wierschem_2014_strangecor} in Eq.~(\ref{eq_MPSC}) to detect the nontrivial SSPT state on the hyperbolic lattice, whose nontrivial support directly reflects the underlying geometric structure.
Furthermore, we derive a sufficient condition on the update rules for NUCA-generated Hamiltonians to be invariant under the non-Abelian translation symmetry.
In this way, the linear NUCA construction provides a systematic route to designing, classifying, and diagnosing quantum matter with subsystem symmetries on hyperbolic lattices.

As a quantum generalization of this framework, we develop a NUCA-guided partition method to construct the underlying non-uniform Clifford quantum cellular automata (CQCA) of the hyperbolic cluster state~\cite{Arrighi2019qca,Farrelly2020reviewofquantum}. 
CQCA map single Pauli operators to tensor products of Pauli operators.
On Euclidean lattices, the classification of translationally invariant one-dimensional CQCA~\cite{CQCA_class_2010,Gtschow2010} correspond to that of subsystem symmetries of cluster states~\cite{Daniel2020computational,Stephen2019subsystem}.
However, the absence of a natural $\mathbb{Z}\times\mathbb{Z}$ spatial-temporal structure makes the Euclidean CQCA construction not directly applicable to hyperbolic lattices.
We overcome this difficulty by partitioning the hyperbolic lattice into disjoint one-dimensional graphs based on the NUCA$(5,4)$ update rule Eq.~(\ref{eq_54_cluster}), and determine the tensor network description of hyperbolic cluster state on the $\{5,4\}$ lattice.
From this tensor network description, we obtain a non-uniform CQCA for which the propagation of Pauli operators generates subsystem symmetries of the hyperbolic cluster state, as visualized in Fig.~\ref{fig_QCA_building_block}.
This construction establishes a hyperbolic extension of the correspondence between CQCA and subsystem symmetries, showing how the dynamics of CQCA and the geometry of subsystem symmetries are reorganized beyond Euclidean translation invariance.

Furthermore, we apply NUCA to non-equilibrium statistical physics by studying directed percolation (DP) on hyperbolic lattices.
On the Euclidean square lattice, the DP process can be simulated by uniform probabilistic CA~\cite{DK1984,Hinrichsen2000}.
To generalize this local dynamical description to the hyperbolic setting, we formulate a probabilistic NUCA$(5,4)$ for the $\{5,4\}$ lattice whose update rules Eq.~(\ref{eq_nuca54_percolation}) encode directed dynamics and treelike structure of the lattice while preserving physical locality.
Within the probabilistic NUCA$(5,4)$, the required neighbor relations are computed on demand from coordinates, avoiding the need to store the full lattice and thereby overcoming a central computational difficulty in numerical studies of percolation on hyperbolic lattices~\cite{hl_percolation}.
Through NUCA simulations and scaling analysis, we numerically estimate several DP thresholds, including those for bond DP and site DP, and obtain the approximate phase diagram in Fig.~\ref{fig_DP_diagram} through interpolation.
In this formulation, the geometric growth of the hyperbolic lattice is incorporated directly into the stochastic evolution, allowing us to study how negative curvature reshapes directed percolation and its phase structure.

This paper is organized as follows.
In Sec.~\ref{sec_preliminaries}, we introduce the NUCA framework and its uniform limit.
In Sec.~\ref{sec_NUCA54}, we construct a linear NUCA$(5,4)$ to design SSPT and SSSB models with subsystem symmetries on the hyperbolic $\{5,4\}$ lattice.
We also construct the multi-point strange correlator for the NUCA-generated SSPT state and derive a sufficient condition for invariance under non-Abelian translations.
In Sec.~\ref{sec_diversity}, we classify these NUCA-generated subsystem symmetries by their growth behavior.
In Sec.~\ref{sec_QCA}, we construct non-uniform CQCA for the hyperbolic cluster state and generate the corresponding symmetries.
In Sec.~\ref{sec_DP}, we construct a probabilistic NUCA$(5,4)$ to simulate directed percolation on the $\{5,4\}$ lattice.
Finally, Sec.~\ref{sec_conclusion} summarizes the results and discusses future directions.
In Appendix~\ref{app_hyperbolic}, we briefly review hyperbolic geometry.
In Appendix~\ref{app_applicability}, we discuss the applicability of the NUCA construction and its consistency with uniform CA.
In Appendix~\ref{app_NUCA54}, we provide mathematical details and derivations for the NUCA$(5,4)$.
In Appendix~\ref{app_NUCA66}, we construct a linear NUCA$(6,6)$ for the $\{6,6\}$ lattice.
In Appendix~\ref{app_hcm}, we discuss physical properties of the hyperbolic cluster state.

\begin{table*}
\caption{Notation for non-uniform cellular automata}\label{tab_notation}
\begin{ruledtabular}
\begin{tabular}{ccl}
Notation & Meaning & First introduced\\
\hline
$\{p,q\}$ & Schl\"afli symbol for regular lattices & Sec.~\ref{sec_introduction}\\
NUCA$(p,q)$ & Non-uniform cellular automata for the $\{p,q\}$ lattice & Sec.~\ref{sec_introduction}\\
$\mathbb{F}_2$ & Alphabet of NUCA & Sec.~\ref{sec_preliminaries}\\
$x^iy^j$ & Coordinate of site in polynomial representation & Sec.~\ref{sec_preliminaries}, Eq.~(\ref{eq_state_def})\\
$a_{ij}$ & State of site $x^iy^j$ & Sec.~\ref{sec_preliminaries}, Eq.~(\ref{eq_state_def})\\
$r_j(x)$ & States of sites at time step $j$ & Sec.~\ref{sec_preliminaries}, Eq.~(\ref{eq_time_step})\\
$\mathscr{F}(x,y)$ & Space-time configuration of NUCA evolution & Sec.~\ref{sec_preliminaries}, Eq.~(\ref{eq_configuration})\\
$\tilde{\mathscr{F}}(x,y)$ & Truncated space-time configuration & Sec.~\ref{sec_preliminaries}, Eq.~(\ref{eq_symmetry})\\
$F^{(k)}_j,\mathbf{F}^{(k)}_j$ & Transition operator and its matrix representation & Sec.~\ref{sec_preliminaries}, Eq.~(\ref{eq_transition_operator})\\
$\mathbf{f}_{(i,j)}(x)$ & Update rule of linear NUCA & Sec.~\ref{sec_preliminaries}, Eq.~(\ref{eq_update_rule})\\
$\bar{\mathbf{f}}_{(i,j)}(x)$ & Transposed update rule of linear NUCA & Sec.~\ref{sec_preliminaries}, Eq.~(\ref{eq_transposed_update_rule})\\
$\mathrm{f}_{(i,j),k}(x),\bar{\mathrm{f}}_{(i,j),k}(x)$ & $k$-th elements of the (transposed) update rule & Sec.~\ref{sec_preliminaries}, Eq.~(\ref{eq_update_rule}), Eq.~(\ref{eq_transposed_update_rule}) \\
$n,\bar{n}$ & Order of the (transposed) NUCA update rule & Sec.~\ref{sec_preliminaries}, Eq.~(\ref{eq_update_rule}), Eq.~(\ref{eq_transposed_update_rule})\\
$\mathbf{q},\mathbf{q}(x),\mathbf{q}(x,y)$ & Initial condition & Sec.~\ref{sec_preliminaries}, Eq.~(\ref{eq_initial_condition_NUCA})\\
$\mathbf{y},\mathbf{y}_{1,n}(y)$ & Function vector $(y^1, y^2, \cdots, y^n)^T$ & Sec.~\ref{sec_preliminaries}, Eq.~(\ref{eq_y_vector})\\
$\bar{\mathbf{y}},\bar{\mathbf{y}}_{1,\bar{n}}(y)$ & Function vector $(y^{-1}, y^{-2}, \cdots, y^{-\bar{n}})^T$ & Sec.~\ref{sec_preliminaries}, Eq.~(\ref{eq_y_vector})\\
$P(a_{ij}=1|\cdots)$ & Update rules of probabilistic NUCA & Sec.~\ref{sec_preliminaries}, Eq.~(\ref{eq_pnuca}) \\
$p_1, p_2$ & Probabilities for probabilistic NUCA & Sec.~\ref{sec_DP}, Eq.~(\ref{eq_nuca54_percolation}) \\
$J(j)$ & Indicator function & Sec.~\ref{sec_NUCA54}, Eq.~(\ref{auxiliaryJ}) \\
$P(j)$ & Parent function & Sec.~\ref{sec_NUCA54}, Eq.~(\ref{auxiliaryP})\\
$C(j)$ & Child function & Sec.~\ref{sec_NUCA54}, Eq.~(\ref{auxiliaryC})\\
$\phi$ & Golden ratio $(1+\sqrt{5})/2$ & Sec.~\ref{sec_NUCA54}
\end{tabular}
\end{ruledtabular}
\end{table*}

\section{Construction of higher-order non-uniform cellular automata}
\label{sec_preliminaries}
We introduce the general construction of NUCA used throughout this work.
Conventionally, CA are defined by uniformity, where the states of all sites evolve synchronously according to an identical local rule~\cite{wolfram1983review,wolfram1984,TOFFOLI1977213}. 
Non-uniform cellular automata allow spatially or temporally varying update rules~\cite{nuca1986,tNUCA1997,tNUCA2023,tNUCA2024,nuca2026}, thereby relaxing the constraint of translation invariance.
This flexibility makes NUCA suitable for encoding geometric constraints in lattice systems without Abelian translation symmetry, such as hyperbolic lattices.

In this work, we focus on one-dimensional higher-order non-uniform CA.
Here, one-dimensional means that the NUCA configuration at each time step is supported on a line, while higher-order means that a NUCA has memory of multiple preceding time steps~\cite{wolfram1984,TOFFOLI1977213}.
Our construction applies to both deterministic and probabilistic NUCA.
For deterministic and linear NUCA, the update rules uniquely determine the states of sites from preceding configurations.
For probabilistic NUCA, the states are updated from probability distributions specified by preceding configurations.
All notation used for NUCA throughout this paper is summarized in Table~\ref{tab_notation} for convenience.

\subsection{Constructing procedures}
Consider a two-dimensional square lattice where the rectangles are labeled as sites $(i,j),\,j\ge0$.
During the evolution of NUCA, each lattice site takes a state from a finite set $\mathbb{F}_2 = \{0,1\}$ denoted as the \textit{alphabet} of the NUCA. 
Now we use the \textit{polynomial representation} to represent a site $(i,j)$ as $x^iy^j$, and the state of the site $a(i,j)$ is:
\begin{equation}
	\label{eq_state_def}
	a(i,j)\rightarrow a_{ij}x^iy^j\,, \quad a(i,j)\equiv a_{ij}\in \mathbb{F}_2 \,.
\end{equation}
The coordinate $j$ is interpreted as a time step of NUCA evolution, and the states of sites at time $j$ are represented by
\begin{equation}
	\label{eq_time_step}
	r_j(x)= \sum_{i=-\infty}^{\infty} a_{ij}x^{i},\quad a_{ij}\in \mathbb{F}_2.
\end{equation}
Then the space-time configuration of the whole NUCA evolution is represented by
\begin{equation}
	\label{eq_configuration}
	\mathscr{F}(x,y) = \sum_{j=0}^{\infty} r_j(x)y^j =  \sum_{i=-\infty}^{\infty} \sum_{j=0}^{\infty} a_{ij}x^{i}y^{j}\, .
\end{equation}
The summation ranges of $i,j$ are subject to constraints if the lattice is finite.

We first consider the linear NUCA\footnote{In the following sections, we refer to linear NUCA simply as NUCA when there is no ambiguity.} without introducing any randomness. 
For the linear NUCA, the states of sites $r_j(x)$ at time $j$ are uniquely determined by the states of sites in preceding time steps:
\begin{equation}
	\label{eq_time_step_evo}
	r_j(x)= \sum_{k=1}^{j} F^{(k)}_{j-k}[r_{j-k}(x)]+q_j(x)\,.
\end{equation}
Here $F^{(k)}_{j-k}$ is a linear transition operator representing how $r_{j-k}(x)$ contributes to $r_{j}(x)$ after $k$ time steps.
The term $q_j(x)$ represents the initial condition specified at time $j$ for specific sites. 
For a general time step $j$, the action of $F^{(k)}_j$ on the spatial basis $\{x^i\}$ is represented by a finite Laurent polynomial $\mathrm{f}_{(i,j),k}(x)$:
\begin{equation}
	\label{eq_transition_operator}
	F^{(k)}_j[x^i] = \sum_{m} c_{j,k;i,m} x^m = x^i \mathrm{f}_{(i,j),k}(x)\,,\quad \forall i,j\,,
\end{equation}
where $c_{j,k;i,m}\in \mathbb{F}_2$ are coefficients.

Consequently, we define a NUCA \textit{update rule} $\mathbf{f}_{(i,j)}(x)$ as a vector function:
\begin{equation}
	\label{eq_update_rule}
	\mathbf{f}_{(i,j)}(x) \equiv (\mathrm{f}_{(i,j),1}(x),\mathrm{f}_{(i,j),2}(x),\cdots ,\mathrm{f}_{(i,j),n}(x))^T \,,
\end{equation}
where $T$ denotes the transpose of a vector.
The update rule specifies how the state $a_{ij}$ influences states of other sites during the evolution.
The order $n\equiv n(i,j)$ is defined as the maximal integer $k$ such that $\mathrm{f}_{(i,j),k}(x)\ne 0$. 
In general, $\mathbf{f}_{(i,j)}(x)$ is position-dependent.
By using the update rule, the evolution of NUCA can be written as:
\begin{equation}
	\label{state_evolution}
	r_j(x) = \sum_{k=1}^{n_{\max}} \sum_{i}  a_{i,j-k}x^i \mathrm{f}_{(i,j-k),k}(x) +q_j(x)\, ,
\end{equation}
where $n_{\max}$ is the maximal integer $k$ determined by $F$ such that $\exists i,\mathrm{f}_{(i,j-k),k}(x)\ne 0$.
The evolution equation can be rewritten in terms of matrix operations on the coefficient vectors $\vec{a}_j=(\cdots, a_{i-1,j}, a_{i,j}, a_{i+1,j},\cdots)$ and similarly for $\vec{q}_j$ as:
\begin{equation}
	\label{eq_coeff_vec}
	\vec{a}_j = \sum_{k=1}^{n_{\max}} \vec{a}_{j-k} \mathbf{F}^{(k)}_{j-k} +\vec{q}_j\, ,
\end{equation}
where $\mathbf{F}^{(k)}_j$ is the matrix representation of $F^{(k)}_j$ such that $(\mathbf{F}^{(k)}_j)_{i,m} = c_{j,k;i,m}$.

To determine which preceding sites contribute to the state of a given site, we introduce the transposed transition operator $\bar{F}^{(k)}_j$ whose matrix representation is the transpose of $\mathbf{F}^{(k)}_{j-k}$.
Accordingly, a finite Laurent polynomial $\bar{\mathrm{f}}_{(i,j),k}(x)$ represents the action of $\bar{F}^{(k)}_j$ on $x^iy^j$ as:
\begin{equation}
	\bar{F}^{(k)}_j [x^i] = \sum_{m} \bar{c}_{j,k;i,m} x^m = x^i \bar{\mathrm{f}}_{(i,j),k}(x)\,,\quad \forall i,j\,,
\end{equation}
where $\bar{c}_{j,k;i,m} \equiv [(\mathbf{F}^{(k)}_{j-k})^T]_{i,m} = c_{j-k,k;m,i}$.
Therefore, we define the transposed update rule as:
\begin{equation}
	\label{eq_transposed_update_rule}
	\bar{\mathbf{f}}_{(i,j)}(x) = \left( \bar{\mathrm{f}}_{(i,j),1}(x),\bar{\mathrm{f}}_{(i,j),2}(x),\cdots, \bar{\mathrm{f}}_{(i,j),\bar{n}}(x) \right)^T \,,
\end{equation}
which specifies how the state $a_{ij}$ of the site $x^iy^j$ is determined during the evolution.
The order $\bar{n}\equiv \bar{n}(i,j)$ is defined as the maximal integer $k$ such that $\bar{\mathrm{f}}_{(i,j),k}(x)\ne 0$. 
The update rule is uniquely determined by its transposed counterpart and vice versa.
If $x^a$ is a term in $x^b\mathrm{f}_{(b,j-k),k}(x)$, then $x^b$ is a term in $x^a\bar{\mathrm{f}}_{(a,j),k}(x)$.

If the update rule is independent of the position as $\mathbf{f}_{(i,j)}(x)\equiv\mathbf{f}(x)$ and $\bar{\mathbf{f}}_{(i,j)}(x)\equiv\bar{\mathbf{f}}(x)$, the NUCA reduces to the uniform higher-order CA.
Furthermore, if the orders are $n=\bar{n}=1$ for $\mathbf{f}(x)$ and $\bar{\mathbf{f}}(x)$, the uniform higher-order CA reduces to order-one CA.
Consequently, by choosing uniform update rules, our NUCA can naturally implement uniform CA.

The position-dependent update rules $\mathbf{f}_{(i,j)}(x)$, or equivalently $\bar{\mathbf{f}}_{(i,j)}(x)$, are designed to incorporate the geometry induced by the embedding.
Through evolution under the update rules, the discrete dynamics of a one-dimensional NUCA are mapped to spatial configurations on lattices without Abelian translation symmetry, enabling us to investigate hyperbolic physics directly through NUCA.

The \textit{initial condition} $\mathbf{q}$ is defined as a vector function composed of the polynomials $q_j(x)$:
\begin{equation}
	\label{eq_initial_condition_NUCA}
	\mathbf{q}(x)\equiv (q_0(x),q_1(x),\cdots)^T\,.
\end{equation}
The initial condition can be specified for sites $x^iy^j$ with trivial transposed update rule $\bar{\mathbf{f}}_{(i,j)}(x) = \mathbf{0}$, since states of these sites are not determined by preceding time steps during the evolution.
The space-time configuration $\mathscr{F}(x,y)$ in Eq.~(\ref{eq_configuration}) is completely determined by the update rule and the initial condition.

In addition to the above deterministic and linear NUCA, we also consider probabilistic NUCA~\cite{ROLLIER2025108362,PCA2018}.
A probabilistic NUCA is specified by a function $P$ that defines the probability of the state of a site $a_{ij}$ transitioning to a state in the alphabet $\mathbb{F}_2$ based on the evolution configuration of preceding $\bar{n}\equiv \bar{n}(i,j)$ time steps:
\begin{equation}
	\label{eq_pnuca}
	\begin{aligned}
		&P(a_{ij}=1|\{r_{j-k}(x)\}_{k=1,\cdots,\bar{n}})= p\,,\\
		&P(a_{ij}=0|\{r_{j-k}(x)\}_{k=1,\cdots,\bar{n}})= 1-p\,.
	\end{aligned}
\end{equation}
Here, $p\in [0,1]$ is a probability and should not be confused with the integer $p$ in the Schl\"afli symbol $\{p,q\}$. 
In general, this probability $p\equiv p_{ij}(\{r_{j-k}(x)\})$ depends on both the position $x^iy^j$ and the specific configuration of preceding time steps, and is therefore non-uniform. 
Assuming independence of evolution at the same time step given preceding configurations, the evolution of the probabilistic NUCA is written as:
\begin{equation}
	\begin{aligned}
		P(\{r_j(x)\}|&\{r_{j-k}(x)\}_{k=1,\cdots,\bar{n}_{\max}}) = \\
		&\prod_i P(a_{ij}|\{r_{j-k}(x)\}_{k=1,\cdots,\bar{n}}),
	\end{aligned}
\end{equation}
where $\{r_j(x)\}$ is a specific configuration of $r_j(x)$ and $\bar{n}_{\max}$ is the maximum of all $\bar{n}(i,j)$.
The probabilistic NUCA reduces to the deterministic case when $p\in\{0,1\}$ for all sites and possible preceding configurations, where the discrete dynamics are governed by the update rules and the initial conditions without randomness.

\subsection{Polynomial representation of operators}
The polynomial representation on $\mathbb{F}_2$ also provides a convenient notation for qubit states and Pauli operators supported on sites $x^iy^j$.
Its physical interpretation is as follows:
\begin{enumerate}
	\item Each site $(i,j)$ is represented by $x^iy^j$ as defined previously, with $a_{ij} \in \mathbb{F}_2$ representing the state of the qubit at this site. $a_{ij}=0$ represents that the qubit at this site is in the state $\left| 0 \right>$, while $a_{ij}=1$ represents $\left| 1 \right>$.
	\item A Pauli operator $\hat{O}\in\{\hat{X},\,\hat{Y},\,\hat{Z}\}$ acting nontrivially on a single qubit $x^iy^j$ is expressed as $\hat{O}(x^iy^j)$\footnote{In the case where there is no ambiguity, we will omit the hat notation of physical operators acting on the Hilbert space for convenience.}.  Consequently, a many-body Pauli operator is defined as $O(x^{i_1}y^{j_1}+x^{i_2}y^{j_2}+\cdots+x^{i_k}y^{j_k})$.
	\item Because the coefficients $a_{ij}\in\mathbb{F}_2$, the product of Pauli operators $O_1(\alpha(x,y))$ and $O_2(\beta(x,y))$ is defined as:
	\begin{equation}
		O_1O_2=O(\alpha(x,y) + \beta(x,y)) \, ,
	\end{equation}
	where $O=O_1=O_2\in\{X,Y,Z\}$. $\alpha(x,y)$ and $\beta(x,y)$ are Laurent polynomials representing sites on which the operators have nontrivial action. The product of two operators is represented by the sum of two Laurent polynomials.
\end{enumerate}

Then we introduce the \textit{commutation polynomial} for examining the commutation relation of operators~\cite{Devakul2019fractal,zhang2024hoca}.
For two operators $X(\alpha)$ and $Z(\beta)$ represented by polynomials $\alpha(x,y)$ and $\beta(x,y)$, the commutation polynomial with respect to them is defined as 
\begin{equation}
	\mathrm{C}(\alpha,\beta) \equiv \alpha(x,y)\beta(x^{-1},y^{-1})\,.
\end{equation}
If $[\mathrm{C}(\alpha,\beta)]_{x^0y^0}=0$, $X(\alpha)$ and $Z(\beta)$ commute with each other.
Here, $[\,\cdot\,]_{x^iy^j}$ denotes the coefficient of $x^iy^j$ in the polynomial.

\begin{figure}
	\centering
	\includegraphics[width=0.9\columnwidth]{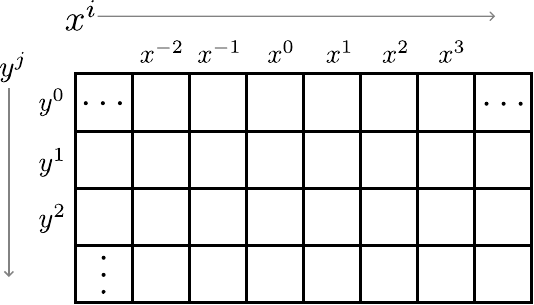}
	\caption{\label{fig_euclicoor} Visualization of a semi-infinite Euclidean square lattice on which NUCA$(4,4)$ is performed. Qubits are located on squares.
	}
\end{figure}

\subsection{Uniform limit of NUCA and subsystem symmetries on Euclidean lattices}
\label{ssec_nuca44}
Following the general construction, we construct a linear NUCA$(4,4)$ to generate many-body Hamiltonians with subsystem symmetries on the Euclidean square lattice as a consistency check.
An onsite subsystem symmetry acts only on a rigid, non-deformable and subextensive subsystem of the whole system, while an onsite global symmetry acts uniformly on the whole system~\cite{you2018sspt,Devakul2018classification}. 
In Ref.~\cite{zhang2024hoca,zhang2025set}, it is proved that all possible subsystem symmetry patterns defined on the square lattice can be generated by uniform higher-order CA, with a complete classification of them into four classes. 
The NUCA reduces to uniform higher-order CA if we adopt uniform update rules, and now we demonstrate the design of these symmetries~\cite{Devakul2019fractal,zhang2024hoca}.

We consider a semi-infinite Euclidean square lattice $(-\infty<i<\infty,\, j\ge0)$ as shown in Fig.~\ref{fig_euclicoor}. 
According to our convention, sites with the same $y^j$ belong to the same time step $j$ such that $r_j(x) =\sum_{i} a_{ij} x^i$.
In the Euclidean case, we are interested in the uniform update rules 
\begin{equation}
	\mathbf{f}_{(i,j)}(x)\equiv \mathbf{f}(x)\,,\quad \bar{\mathbf{f}}_{(i,j)}(x)\equiv \bar{\mathbf{f}}(x)\,,
\end{equation}
independent of the site position, and thus $\bar{\mathbf{f}}(x) = \mathbf{f}(x^{-1}),\,n=\bar{n}$ as shown in Appendix~\ref{app_applicability}.
When adopting uniform update rules, the evolution is simplified to:
\begin{equation}
	r_j(x) = \sum_{k=1}^{n} r_{j-k}(x) \mathrm{f}_{k}(x)\, ,
\end{equation}
with the initial condition consisting of sites on the first $n$ rows of the boundary:
\begin{equation}
	\label{Euclidean_initial_condition}
	\mathbf{q}(x)= (q_0(x),q_1(x),\cdots,q_{n-1}(x))^T\,.
\end{equation}

The uniform update rules lead to translationally invariant NUCA-generated Hamiltonians on the Euclidean lattice. 
For the SSSB models, the Hamiltonians read
\begin{equation}
	\label{eq_44_SSB}
	\mathscr{H}=-\sum_{i,j}Z(x^iy^j(1+\bar{\mathbf{f}}\cdot\bar{\mathbf{y}}))\,,
\end{equation}
which can be appended by a transverse field $-\sum_{ij}X(x^iy^j)$. 
Here, we introduce $\bar{\mathbf{y}}$ and the corresponding $\mathbf{y}$ as function vectors of $y$ 
\begin{equation}
	\label{eq_y_vector}
	\begin{aligned}
		&\mathbf{y} \equiv \mathbf{y}_{1,n}(y) = (y^{1},y^{2},\cdots,y^{n})^T\,,\\
		&\bar{\mathbf{y}} \equiv \bar{\mathbf{y}}_{1,\bar{n}}(y) = (y^{-1},y^{-2},\cdots,y^{-\bar{n}})^T\,,
	\end{aligned}
\end{equation}
which are composed of monomials of $y$ for notational simplicity.
In the Hamiltonian Eq.~(\ref{eq_44_SSB}), each site determines a Hamiltonian term through the update rule $\bar{\mathbf{f}}$.
Under open boundary conditions on a finite lattice, all terms defined at $x^iy^j$ with support extending beyond the lattice are excluded by specifying $\bar{\mathbf{f}}=\mathbf{0}$ for these sites.
Through the examination of the commutation polynomial, $\mathscr{H}$ in Eq.~(\ref{eq_44_SSB}) commutes with linear NUCA-generated symmetries
\begin{equation}
	\label{eq_symmetry}
	S(\mathbf{q}) = X(\tilde{\mathscr{F}}(x,y))\,,
\end{equation}
where $\tilde{\mathscr{F}}$ is the truncated configuration $\mathscr{F}$ specified by the update rule $\mathbf{f}$ and the initial condition $\mathbf{q}(x)$, while all terms which are not fully supported in the lattice are excluded.
Here, $a_{ij} = 1$ in a configuration represents nontrivial action of a symmetry operator at site $x^iy^j$. 
By specifying different initial conditions, we can enumerate symmetry elements of the NUCA-generated models.

There are two sublattices $\{x^iy^j\}_{i+j\equiv0\pmod2}$ and $\{x^iy^j\}_{i+j\equiv1\pmod2}$ for the $\{4,4\}$ lattice, and the update rule for the SSPT model should be designed to reflect the bipartite structure.
Consequently, a general SSPT Hamiltonian generated by the uniform update rules is written as:
\begin{equation}
	\label{eq_44_SSPT}
	\begin{aligned}
		\mathscr{H}=&-\sum_{i+j\equiv0\, (\text{mod }2)}Z(x^iy^j(1+\bar{\mathbf{f}}\cdot\bar{\mathbf{y}}))Z(x^iy^{j-1})\\
		&-\sum_{k+l\equiv1\, (\text{mod }2)}X(x^ky^l(1+\mathbf{f}\cdot\mathbf{y}))X(x^ky^{l+1}) \, ,
	\end{aligned}
\end{equation}
and we exclude terms with support extending outside the lattice.
Here, we decompose operators belonging to different sublattices in each Hamiltonian term explicitly.
The Hamiltonian Eq.~(\ref{eq_44_SSPT}) describes an exactly solvable SSPT model with a short-range entangled unique ground state on a torus~\cite{you2018sspt,Devakul2019fractal,zhang2024hoca}.
The exact solvability of the model can be proved by noticing that there are always $0$ or $2$ overlapping sites with nontrivial support between two Hamiltonian terms, such that all Hamiltonian terms commute with each other.

Each Hamiltonian term in Eq.~(\ref{eq_44_SSPT}) commutes with the following subsystem symmetries:
\begin{equation}
	\label{eq_symmetry_44}
	S^{(a)}(\mathbf{q}^{(a)}) = X(\tilde{\mathscr{F}}^{(a)}(x,y))\, ,\quad  S^{(b)}(\mathbf{q}^{(b)}) = Z(\tilde{\bar{\mathscr{F}}}^{(b)}(x,y))\, ,
\end{equation}
where $\bar{\mathscr{F}}(x,y)=\mathscr{F}(x^{-1},y^{-1})$.
The commutation relation is checked in Appendix~\ref{app_applicability}.
The evolution configuration $\tilde{\mathscr{F}}^{(a)/(b)}$ restricted to one sublattice is generated by the initial condition $\mathbf{q}^{(a)}(x)/\mathbf{q}^{(b)}(x)$ defined on one sublattice $(a)/(b)$.

\begin{figure*}[t]
	\includegraphics[width=0.8\textwidth]{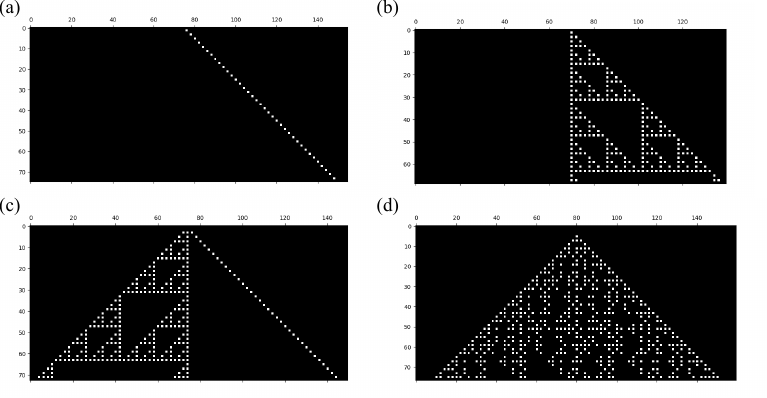}
	\caption{\label{fig_Euclid_sym} Visualization of four classes of subsystem symmetries for translationally invariant SSPT models on the Euclidean lattice.
	(a) Regular symmetry pattern generated by $\mathbf{f}_1,\,\mathbf{q}_1$ in Eq.~(\ref{eq_regular}).
	(b) Fractal symmetry pattern generated by $\mathbf{f}_2,\,\mathbf{q}_2$ in Eq.~(\ref{eq_fractal}).
	(c) Mixed symmetry pattern generated by $\mathbf{f}_3,\,\mathbf{q}_3$ in Eq.~(\ref{eq_mixed}).
	(d) Chaotic symmetry pattern generated by $\mathbf{f}_4,\,\mathbf{q}_4$ in Eq.~(\ref{eq_chaotic}).}
\end{figure*}

The subsystem symmetries for translationally invariant SSPT models on the Euclidean lattice are well-studied in the literature~\cite{zhang2024hoca,zhang2025set,Devakul2019fractal,Devakul2018fractaluniversal,Devakul2019fractalclass,Stephen2019subsystem,Daniel2020computational}. 
These symmetry patterns can be generated and classified into regular (e.g., linelike and checkerboard-like structures) and fractal (e.g., Sierpinski triangle), mixed (characterized by the coexistence of distinct patterns) and chaotic classes by certain update rules and initial conditions~\cite{zhang2024hoca}.
The following update rules and initial conditions 
\begin{subequations}
	\begin{align}
		&\mathbf{f}_1(x) = (0,x^{-2}+x^2,0,1)^T \,,\notag\\ &\mathbf{q}_1(x) = (0,1,0,x^2)^T\,; \label{eq_regular} \\
		&\mathbf{f}_2(x) = (0,1+x^2)^T\,,\notag\\
		&\mathbf{q}_2(x) = (1,0)^T\,;  \label{eq_fractal}\\
		&\mathbf{f}_3(x) = (0,x^{-2}+1+x^2,0,1+x^2)^T \,,\notag\\
		&\mathbf{q}_3(x) = (0,0,0,x^{-2}+1+x^2)^T\,; \label{eq_mixed}\\
		&\mathbf{f}_4(x) = (0,x^{-2}+1+x^2,0,1,0,x^2)^T \,,\notag\\
		&\mathbf{q}_4(x) = (0,0,0,0,0,1)^T \label{eq_chaotic}
	\end{align}
\end{subequations} 
correspond to regular, fractal, mixed and chaotic SSPT models and subsystem symmetry patterns respectively.
In Fig.~\ref{fig_Euclid_sym}, we visualize these symmetry patterns generated by NUCA$(4,4)$.
We explicitly present the symmetries on one sublattice, as those for the other one are analogous.
The linear NUCA$(4,4)$ therefore reproduces the known Euclidean classification and serves as a consistency check of the NUCA framework in the uniform limit.

\section{Linear NUCA and quantum states with subsystem symmetries on the hyperbolic lattice}
\label{sec_NUCA54}

\subsection{Overview}
A central difficulty in applying CA-type constructions to hyperbolic lattices is the locating problem, namely the labeling of polygons and the computation of their neighborhoods~\cite{hl_percolation,Margenstern2018}.
In this section, we overcome this obstruction by developing a lattice-deforming procedure and construct a coordinate system for one dimensional NUCA.
Using M. Margenstern's splitting method~\cite{margenstern_1999,Margenstern_2000,MARGENSTERN2001,Margenstern2007} which provides a labeling system for polygons on regular lattices $\{p\ge4,q\ge4\}$, our lattice-deforming procedure embeds hyperbolic lattices into a Euclidean square lattice, allowing the construction of NUCA.
The resulting coordinate-scaling auxiliary functions characterize the embedding distortion and allow non-uniform update rules to remain local with respect to the target hyperbolic geometry.

Although several specific subsystem symmetries have been identified on hyperbolic lattices~\cite{yan2019hfm,yan2019hfm2,yan2025hfm3}, the lack of a general framework for constructing Hamiltonians with such symmetries limits a systematic understanding of the corresponding quantum states.
Based on our construction above, we use linear NUCA$(5,4)$ to provide such a framework and realize explicit examples of exotic quantum matter with NUCA-generated subsystem symmetries on the hyperbolic $\{5,4\}$ lattice.
In particular, we construct exactly solvable SSPT models and SSSB models whose symmetries are generated by NUCA evolution rather than imposed by hand.
We demonstrate that the supports of subsystem symmetries and the associated multi-point strange correlator are organized by the negative curvature and exponential growth of the underlying geometry.
The NUCA therefore provides a controllable framework for designing and analyzing hyperbolic quantum matter protected or organized by subsystem symmetries.

We develop the construction in steps.
We first introduce the splitting method, then define the lattice-deforming procedure and auxiliary functions. 
We next construct NUCA$(5,4)$-generated SSPT and SSSB Hamiltonians, diagnose the nontrivial SSPT state using NUCA-generated multi-point strange correlator, and discuss a sufficient condition for translation invariance under non-Abelian translation symmetry.
We discuss the applicability of our construction for $\{p\ge4,q\ge4\}$ lattices in Appendix~\ref{app_applicability}.
We construct another example of a linear NUCA$(6,6)$ for the $\{6,6\}$ lattice in Appendix~\ref{app_NUCA66} to show the generality of the NUCA framework to hyperbolic lattices.

\subsection{Hyperbolic lattice labeling from the splitting method}
\label{ssec_splitting}
\begin{figure*}[t]
	\includegraphics[width=0.95\textwidth]{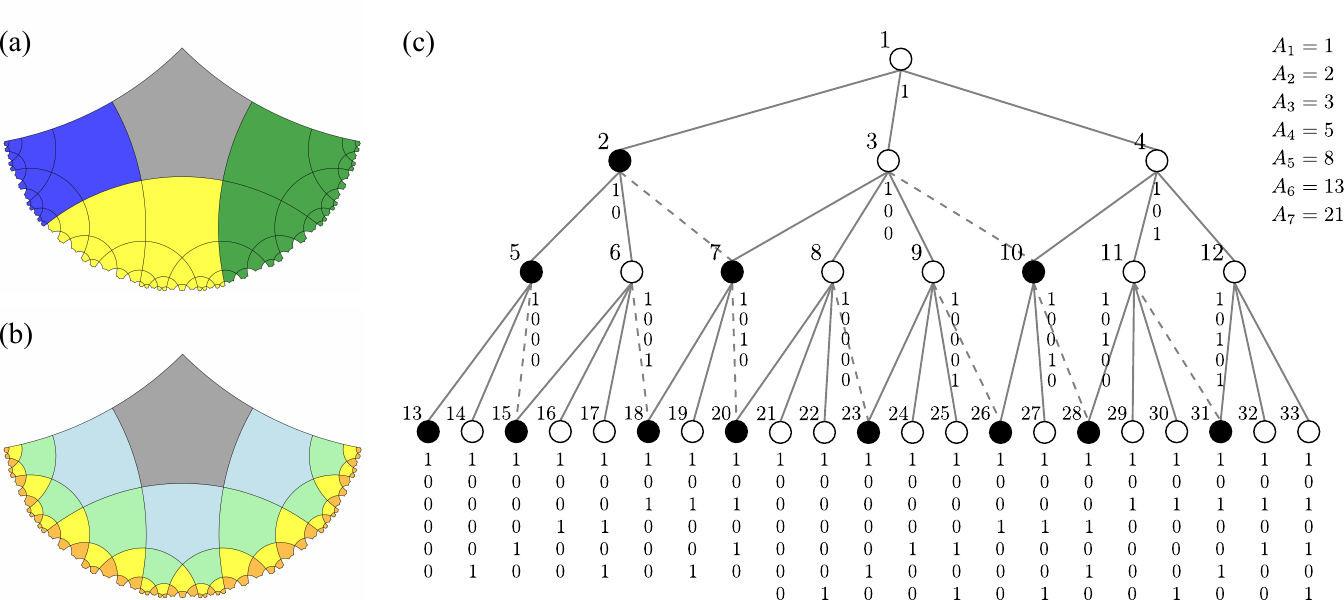}
	\caption{\label{fig_los54} 
	(a) The splitting method applied to a quarter of the $\{5,4\}$ lattice. The gray region denotes the leading pentagon of a quarter, the green and yellow regions denote two copies of this quarter, while the blue region denotes a strip.
	(b) Lattice generated by the splitting method. The pentagons with the same color correspond to the same level.
	(c) The spanning tree of the $\{5,4\}$ lattice. 
	The nodes (pentagons) are uniquely assigned discrete indices in terms of the Fibonacci representation. The neighboring relation is represented by lines (offspring) and dashed lines (non-offspring).}
\end{figure*}

Considering the hyperbolic plane $\mathbb{H}^2$ equipped with a regular tessellation $\{p,q\}$, we construct a bijection between a quarter of this tessellation and a spanning tree:
\begin{enumerate}
	\label{generating_steps}
	\item \textit{Initial step:} For a quarter $\mathcal{Q}_0$ of the lattice, specify its leading $p$-gon and assign it as the root node of the spanning tree.
	\item \textit{Iterative steps:} If the current $p$-gon $\mathcal{P}$ is the leading polygon of a quarter $\mathcal{Q}_k$, its complement in $\mathcal{Q}_k$ decomposes into $(p-3)(\lfloor q/2 \rfloor-1)$ isometric images (copies) of $\mathcal{Q}_0$ for even $q$, or $(p-3)\lfloor q/2 \rfloor$ copies of $\mathcal{Q}_0$ for odd $q$, together with a remaining strip region $\mathcal{S}_k$. 
	Here, $\lfloor \cdot \rfloor $ is the floor function.
	In the spanning tree, $\mathcal{P}$ is represented by a white node \begin{tikzpicture}[] \node[wcirc] {}; \end{tikzpicture}.
	
	If $\mathcal{P}$ is the leading polygon of a strip $\mathcal{S}_k$, its complement in $\mathcal{S}_k$ decomposes into $(p-2)(\lfloor q/2 \rfloor-1)-2$ copies of $\mathcal{Q}_0$ for even $q$, or $(p-2)\lfloor q/2 \rfloor-3$ copies of $\mathcal{Q}_0$ for odd $q$, together with a remaining strip $\mathcal{S}_{k+1}$. In the spanning tree, $\mathcal{P}$ is represented by a black node \begin{tikzpicture}[] \node[bcirc] {}; \end{tikzpicture}.
\end{enumerate}

The iterative process generates nodes of the spanning tree level by level. 
As proved by M. Margenstern, this bijection maps $p$-gons of a quarter of $\{p,q\}$ lattice to nodes of the spanning tree~\cite{margenstern_1999,Margenstern_2000,MARGENSTERN2001,Margenstern2007}. 
An intuitive example of splitting the $\{5,4\}$ lattice is shown in Fig.~\ref{fig_los54}, whose generating rule is 
\begin{tikzpicture}[baseline=-0.5ex, node distance=0.3ex]
    \node[wcirc] (c1) {};
    \node[right=of c1] (arrow) {$\rightarrow$};
    \node[bcirc, right=of arrow] (c2) {};
    \node[wcirc, right=of c2] (c3) {};
    \node[wcirc, right=of c3] (c4) {};
\end{tikzpicture} and 
\begin{tikzpicture}[baseline=-0.5ex, node distance=0.3ex]
    \node[bcirc] (c1) {};
    \node[right=of c1] (arrow) {$\rightarrow$};
    \node[bcirc, right=of arrow] (c2) {};
    \node[wcirc, right=of c2] (c3) {};
\end{tikzpicture}. 
Therefore, the recursive decomposition of a quarter of the lattice is described by the splitting matrix:
\begin{equation}
	\label{splittingmatrix}
	S_{p,q}=\begin{pmatrix}
	(p-3)(\lfloor q/2 \rfloor-1+\epsilon)&		1\\
	(p-2)(\lfloor q/2 \rfloor-1+\epsilon)-2-\epsilon&		1\\
	\end{pmatrix}\,,
\end{equation}
where $\epsilon = q \bmod 2$. The first row represents how a quarter splits into copies of a quarter together with a strip, while the second row represents how a strip splits into copies of a quarter together with a strip.
The characteristic polynomial of the splitting matrix gives the recursive relation for the number of nodes $u_k$ on each level:
\begin{equation}
	u_{k+2} = ((p-3)(\lfloor q/2 \rfloor-1+\epsilon)+1)u_{k+1} + (\lfloor q/2 \rfloor-3)u_k \,,
\end{equation}
where $k=0,1,2\cdots$ denotes levels of the tree, with initial values determined by the root and the first splitting step.

Solving the characteristic polynomial of the splitting matrix Eq.~(\ref{splittingmatrix}) for the $\{5,4\}$ lattice, whose greatest root is the square of the golden ratio $\phi=(1+\sqrt{5})/2$, gives the number of nodes on each level $u_k = A_{2k+1},\, k =0,1,2,\cdots$. 
Here, $\{A_k\}$ is the Fibonacci sequence defined by $A_1=1$, $A_2=2$, and $A_{k+2}=A_{k+1}+A_k$.
After the construction of the spanning tree, each node is labeled by a unique natural number $v=1,2,3\cdots$ from top to bottom and from left to right, which equivalently labels polygons on the hyperbolic lattice. 
Now we introduce the Fibonacci representation for the nodes.
By Zeckendorf's representation of natural numbers, a positive integer $v$ is uniquely represented as a sum of distinct terms of the Fibonacci sequence $\{A_k\}$:
\begin{equation}
	\label{eq_Fibonacci_Rep}
	v = \sum_{i=1}^{i_{\max}} \alpha_iA_i \, ,
\end{equation}
with $\alpha_i\in\{0,1\}$ satisfying $\alpha_i\alpha_{i+1}=0,\forall i$. 
The sequence $\alpha_{i_{\max}}\alpha_{i_{\max}-1}\cdots \alpha_2\alpha_1$ with largest $i_{\max}$ such that $\alpha_{i_{\max}}\ne 0$ is the Fibonacci representation $\mathcal{A}(v)$ corresponding to node $v$ in the spanning tree. 
The representation of nodes in the $\{5,4\}$ spanning tree is shown in Fig.~\ref{fig_los54}(c). 

By using a greedy algorithm, the Fibonacci representation of any $v$ is uniquely and readily computed. 
It is important that the neighboring relation of nodes can be calculated efficiently in the Fibonacci representation~\cite{margenstern_1999,Margenstern_2000,MARGENSTERN2001} with details in Appendix~\ref{app_NUCA54}. 
In Fig.~\ref{fig_los54}(c), the lines represent the neighboring relation and the offspring relation of nodes, while the dashed lines represent the neighboring relation only.
The splitting method can be applied to other regular Euclidean and hyperbolic lattices $\{p\ge 4,q\ge 4\}$, and we recommend readers to refer to Ref.~\cite{margenstern_1999,Margenstern_2000,MARGENSTERN2001,Margenstern2007,Margenstern2018} for more details and mathematical proofs.

\subsection{Lattice-deforming procedure for NUCA}
\label{ssec_54lattice}
As introduced in Sec.~\ref{sec_preliminaries}, the NUCA is performed on the square lattice, so both the update rules and the direction of its dynamics must respect hyperbolic geometry.
To address these issues, we develop a lattice-deforming procedure based on the splitting method, which maps each polygon on the hyperbolic lattice to a site $x^iy^j$ on the square lattice and simultaneously defines the NUCA time direction.
As there is no distortion-free two-dimensional Euclidean model of the hyperbolic plane~\cite{Ratcliffe2019hyperbolic}, one-dimensional CA designed for the hyperbolic lattice are intrinsically non-uniform.
The universal applicability of the lattice-deforming procedure for hyperbolic lattices is discussed in Appendix~\ref{app_applicability}.

\begin{figure}[b]
	\centering
	\includegraphics[width=0.9\columnwidth]{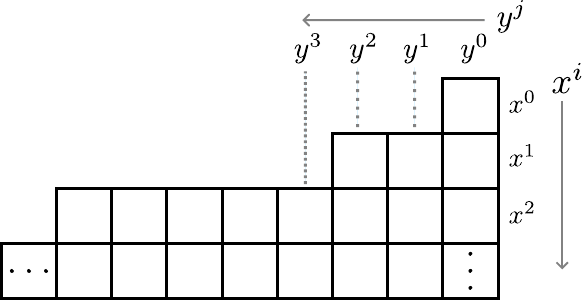}
	\caption{\label{fig_54coor} Visualization of the deformed lattice on which NUCA$(5,4)$ is performed. Sites $x^iy^j$ with $j<A_{2i+1}$ are mapped to pentagons on the $\{5,4\}$ lattice. Nearest-neighbor relations on the deformed lattice do not represent physical locality of the corresponding pentagons.	}
\end{figure}

The deformed lattice is a square lattice with coordinate $x^iy^j$ in terms of the polynomial representation, as visualized in Fig.~\ref{fig_54coor}.
This deformed lattice defines the time direction for the NUCA dynamics, where sites sharing the same $y^j$ belong to the same time step $j$.
A finite deformed lattice is obtained by specifying the maximal $i_{\max}$ of $x^i$.
On the deformed lattice, a site $x^iy^j,j<A_{2i+1}$ corresponds one-to-one to a node $v$ on the $i$-th level in the spanning tree through $v= \sum_{l=0}^{i} A_{2l+1}-j$.
Consequently, we denote it as a physical site since it represents a pentagon of the $\{5,4\}$ lattice.
The neighboring sites on the deformed lattice do not represent real physical locality of qubits on the hyperbolic $\{5,4\}$ lattice. 
Therefore, the update rules are generally nonlocal with respect to the neighboring relation on the deformed lattice, but remain local with respect to the target hyperbolic lattice.

Now we define several auxiliary functions for designing update rules, reflecting physical locality on the hyperbolic lattice. 
The auxiliary functions are directly defined on the deformed lattice as functions of the coordinate.
First, the indicator function $J(j)$ for $j>0$ is:
\begin{equation}
	\label{auxiliaryJ}
	J\left( j \right) =\begin{cases}
	1, \quad \text{if }j=\lfloor \phi^2 \lceil j/\phi^2 \rceil \rfloor\\
	0, \quad \text{if }j\ne \lfloor \phi^2 \lceil j/\phi^2 \rceil \rfloor\\
	\end{cases}\,,
\end{equation}
where $\lceil \cdot \rceil $ is the ceiling function and we define $J(0)=0$. 
If a site $x^iy^j$ is mapped to a black node in the spanning tree then $J(j)=1$.
Moreover, we define two functions for physical locality.
The parent function $P(j)$ is written as:
\begin{equation}
	\label{auxiliaryP}
	P(j) = \lfloor j/\phi^2 \rfloor \, ,
\end{equation}
where $\phi$ is the golden ratio. The function $P(j)$ identifies that the pentagon corresponding to $x^iy^j$ is connected to that of $x^{i-1}y^{P(j)}$. 
The corresponding sites are not neighboring sites on the deformed lattice.
Analogously, the child function $C(j)$ is defined as:
\begin{equation}
	\label{auxiliaryC}
	C(j) = \begin{cases}
		\lfloor \phi^2(j+1) \rfloor,\qquad \text{if }J(j)=1\\
		\lfloor \phi^2(j+1) \rfloor - 1,\quad \text{if }J(j)=0
	\end{cases}\,.
\end{equation}
The function $C(j)$ identifies that the pentagon corresponding to $x^iy^j$ is connected to that of $x^{i+1}y^{C(j)}$. 
The complete local neighborhood can be reconstructed from these functions, and we derive the above properties and functions in Appendix~\ref{app_NUCA54}.

These auxiliary functions stem from the distortion of embedding a  hyperbolic lattice into a Euclidean lattice.
Because the number of lattice sites grows exponentially with the radius, the coordinate scaling in these functions is an intrinsic property of NUCA$(p,q)$ for hyperbolic lattices with no counterpart in the Euclidean case.
On the hyperbolic plane, the locality is governed by the geodesic distance associated with a nontrivial metric~\cite{Ratcliffe2019hyperbolic}.
Although this geometric feature cannot be directly realized on a square lattice where a CA is defined, it is accounted for by the auxiliary functions.
Update rules designed with these auxiliary functions bridge the non-uniformity in CA and physical locality on the hyperbolic lattice, allowing us to investigate physical systems through NUCA.

\subsection{Construction of Linear NUCA$(5,4)$ for SSPT and SSSB models on the $\{5,4\}$ lattice}
\label{ssec_NUCA54}
We now construct the linear NUCA$(5,4)$, which allows us to design SSPT and SSSB models and to detect SSPT states on the $\{5,4\}$ lattice.
Starting from the general construction introduced in Sec.~\ref{sec_preliminaries}, we specialize the formalism to the hyperbolic case.

The alphabet of NUCA is $\mathbb{F}_2$ and the state of a site $x^iy^j$ is expressed as $a_{ij}x^iy^j,\,a_{ij}\in \mathbb{F}_2$.
The time direction of the NUCA$(5,4)$ is defined as the $y^+$-direction, thus sites sharing the same $j$ are updated synchronously.
The evolution configuration, including states determined by time evolution and the initial condition, is represented by a polynomial:
\begin{equation}
	\label{configuration}
	\mathscr{F}(x,y) = \sum_{i=0}^{\infty}\sum_{j=0}^{\infty} a_{ij}x^{i}y^{j}\, .
\end{equation} 
As only sites $x^iy^j$ with $A_{2i+1}>j$ are mapped to the pentagons on the hyperbolic $\{5,4\}$ lattice, we only need to consider the configuration of NUCA$(5,4)$ evolution restricted to these sites.
The states of sites at a time step $j$ are represented by
\begin{equation}
	r_{j}(x)\equiv\sum_{i=0}^{\infty} a_{ij}x^{i}\, .
\end{equation}
The synchronous evolution of sites does not imply locality of the corresponding pentagons on the hyperbolic $\{5,4\}$ lattice.
A general NUCA$(p,q)$ update rule $\mathbf{f}_{(i,j)}(x)$ depends on the site position $x^iy^j$ to preserve physical locality.

In the deformed lattice, the locality of qubits is specified by the auxiliary functions $P(j)$ and $C(j)$ in Eq.~(\ref{auxiliaryP}) and Eq.~(\ref{auxiliaryC}), which are functions of $j$.
Thus the update rules $\mathbf{f}_{(i,j)}(x)$ and $\bar{\mathbf{f}}_{(i,j)}(x)$ depend on $j$ only, i.e.:
\begin{equation}
	\mathbf{f}_{(i,j)}(x)\equiv \mathbf{f}_{j}(x)\,,\quad \bar{\mathbf{f}}_{(i,j)}(x) \equiv \bar{\mathbf{f}}_{j}(x)\,.
\end{equation}
The update rules of NUCA$(5,4)$ vary temporally during the evolution, and thus belong to a subclass of NUCA specifically studied in the literature~\cite{nuca2026,tNUCA1997,tNUCA2023,tNUCA2024}.
The elements of $\mathbf{f}_{j}(x)$ and $\bar{\mathbf{f}}_{j}(x)$ are expressed as
\begin{equation}
	\label{update_rule}
	\begin{aligned}
		&\mathbf{f}_{j}(x) \equiv (\mathrm{f}_{j,1}(x),\mathrm{f}_{j,2}(x),\cdots ,\mathrm{f}_{j,n}(x))^T\, ,\\
		&\bar{\mathbf{f}}_j(x) \equiv (\bar{\mathrm{f}}_{j,1}(x),\bar{\mathrm{f}}_{j,2}(x),\cdots ,\bar{\mathrm{f}}_{j,\bar{n}}(x))^T\,,
	\end{aligned}
\end{equation}
where the order $n\equiv n(j)$ and $\bar{n}\equiv \bar{n}(j)$ vary for different $j$. 

By using the update rule Eq.~(\ref{update_rule}), the time evolution of NUCA$(5,4)$ is represented as
\begin{equation}
	\label{eq_state_evolution_54}
	r_j(x) = \sum_{k=1}^{n_{\max}(j)} \sum_{i} a_{i,j-k}x^i \mathrm{f}_{j-k,k}(x) +q_j(x)\, ,
\end{equation}
where $n_{\max}(j)$ is the maximal integer $k$ determined by $F$ such that $\exists i,\mathrm{f}_{j-k,k}(x)\ne 0$.
By specifying the initial condition $\mathbf{q}$ defined in Eq.~(\ref{eq_initial_condition_NUCA}):
\begin{equation}
	\mathbf{q}(x)\equiv (q_0(x),q_1(x),\cdots)^T\,,
\end{equation}
the whole space-time configuration $\mathscr{F}(x,y)$ is uniquely determined.
In general, states of sites $r_j(x)$ at a time step $j$ are either determined by time evolution under a nontrivial transposed update rule $\bar{\mathbf{f}}_{j}(x)$, or determined by the initial condition $q_j(x)$. 
The initial condition $\mathbf{q}(x)$ specified for sites in the bulk is a property of NUCA that is absent in the uniform CA.

Now we turn to define NUCA$(5,4)$-generated Hamiltonians. 
On the deformed lattice, there are two sublattices $\{x^iy^j\}_{i\equiv0\pmod2}$ and $\{x^iy^j\}_{i\equiv1\pmod2}$.
The general form of NUCA$(5,4)$-generated SSPT Hamiltonian can be written as:
\begin{equation}
	\label{eq_54_SSPT}
	\begin{aligned}
		\mathscr{H}=&-\sum_{\substack{i,j\\i\equiv0\, (\text{mod }2)\\ \bar{\mathbf{f}}_j\neq \mathbf{0}}}Z(x^iy^j(1+\bar{\mathbf{f}}_j\cdot\bar{\mathbf{y}}))Z(x^{i-u(j)}y^{m(j)} )\\
		&-\sum_{\substack{k,l\\k\equiv1\, (\text{mod }2)\\ \bar{\mathbf{f}}_l\neq \mathbf{0}}}X(x^ky^l(1+\bar{\mathbf{f}}_l\cdot\bar{\mathbf{y}}))X(x^{k-u(l)}y^{m(l)})\,,
	\end{aligned}
\end{equation}
where $\bar{\mathbf{y}}\equiv \bar{\mathbf{y}}_{1,\bar{n}(j)}(y)$ is similar to the definition in Eq.~(\ref{eq_y_vector}) but depends on the position.
Here, $x^{i-u(j)}y^{m(j)}$ and $x^iy^j$ are defined on distinct sublattices, and we use $u(j)$ and $m(j)$ to specify the relative offset between them.
To make the Hamiltonian terms commute with each other, the update rule is subject to the following constraint:
\begin{equation}
	\begin{aligned}
		\label{constraint}
		[(1+&\bar{\mathbf{f}}_j\cdot\bar{\mathbf{y}})x^{i-k+u(l)}y^{j-m(l)} \\
		&+ (1+\bar{\mathbf{f}}_l \cdot\bar{\mathbf{y}}) x^{k-i+u(j)}y^{l-m(j)}]_{x^0y^0}=0\,,
	\end{aligned}
\end{equation}
for any $i,j,k,l$ in the summation range.
The constraint is obtained by requiring that there are always even overlapping sites with nontrivial support between two Hamiltonian terms.
If a term $h_{ij}$ defined by $\bar{\mathbf{f}}_{j}$ cannot be fully supported in a finite deformed lattice, $h_{ij}$ is excluded from the Hamiltonian by specifying trivial $\bar{\mathbf{f}}_{j}$ for the site $x^iy^j$.

After specifying physical update rules which define local Hamiltonians, the evolution of linear NUCA$(5,4)$ generates subsystem symmetries.
For the Hamiltonians Eq.~(\ref{eq_54_SSPT}), the following two sets of subsystem symmetries hold:
\begin{equation}
	\label{symmetry}
	S^{(a)}(\mathbf{q}^{(a)}) = X(\tilde{\mathscr{F}}^{(a)}(x,y))\, ,\quad  S^{(b)}(\mathbf{q}^{(b)}) = Z(\tilde{\mathscr{F}}^{(b)}(x,y))\, ,
\end{equation}
where $\mathbf{q}^{(a)/(b)}$ is the initial condition on one sublattice, similar to Eq.~(\ref{eq_symmetry_44}). 
$\tilde{\mathscr{F}}^{(a)/(b)}$ is the truncated NUCA$(5,4)$ evolution configuration specified by $\mathbf{f}_{j}$ and the initial condition, and all terms which are not fully supported in a finite deformed lattice with the specified $x^{i_{\max}}$ are excluded.
We check the commutation polynomial in Appendix~\ref{app_NUCA54}. The update rules $\mathbf{f}_{j}$ determine the geometry and growth behavior of the corresponding subsystem symmetries.
By choosing different $\mathbf{q}$, we obtain distinct NUCA$(5,4)$-generated symmetries associated with the NUCA$(5,4)$-generated SSPT models.

The edge modes of an SSPT model are protected by the corresponding subsystem symmetries~\cite{you2018sspt,Devakul2018classification}.
For a general NUCA-generated SSPT model, the edge operator is defined by truncated operator on an open finite lattice.
For a site $x^{i-u(j)}y^{m(j)}$ in sublattice $(a)$ on the boundary, its edge Pauli operator is written as:
\begin{equation}
	\label{eq_edge_operators}
	\begin{aligned}
		\mathscr{X}^{(a)} &= X(x^{i-u(j)}y^{m(j)}+x^iy^j(1+\bar{\mathbf{f}}_j\cdot\bar{\mathbf{y}})) \,,\\
		\mathscr{Y}^{(a)} &= Y(x^{i-u(j)}y^{m(j)}) X(x^iy^j(1+\bar{\mathbf{f}}_j\cdot\bar{\mathbf{y}}))\,,\\
		\mathscr{Z}^{(a)} &= Z(x^{i-u(j)}y^{m(j)})\,,
	\end{aligned}
\end{equation}
where these operators are truncated to a finite lattice and the case for the other sublattice is obtained by exchanging $X$ and $Z$. 
These edge operators all commute with the remaining Hamiltonian terms and form a Pauli algebra.

For the SSSB models, the general form of the Hamiltonian is simplified as:
\begin{equation}
	\label{eq_54_SSB}
	\mathscr{H}=-\sum_{i,j;\bar{\mathbf{f}}_j\neq \mathbf{0}}Z(x^iy^j(1+\bar{\mathbf{f}}_j\cdot\bar{\mathbf{y}}))\,,
\end{equation}
and the following NUCA-generated configuration gives the subsystem symmetries:
\begin{equation}
	\label{eq_SSB_symmetry}
	S(\mathbf{q}) = X(\tilde{\mathscr{F}}(x,y))\, .
\end{equation}
In the thermodynamic limit, the zero-temperature ground state of Eq.~(\ref{eq_54_SSB}) is degenerate and spontaneously breaks the NUCA-generated symmetries.

\begin{figure*}[t]
	\includegraphics[width=0.9\textwidth]{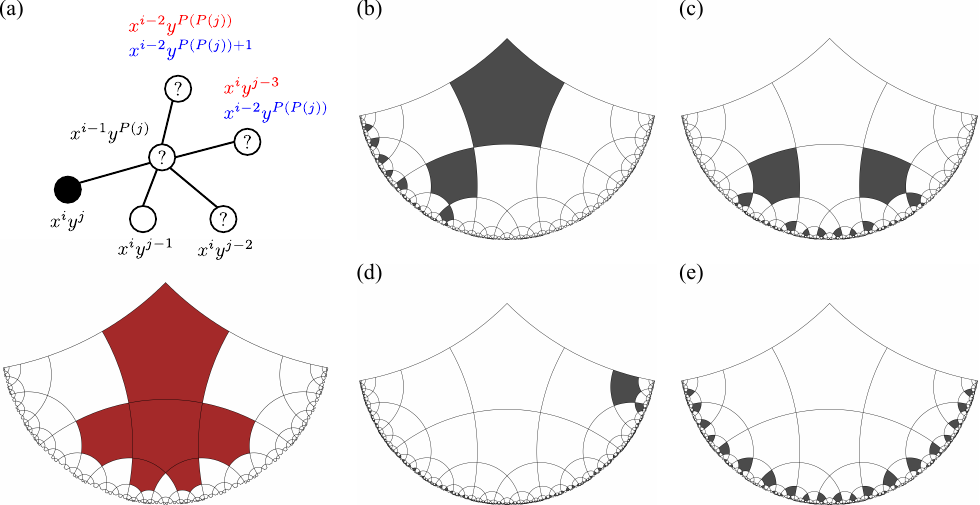}
	\caption{\label{fig_54_model_1} Hamiltonian and symmetry patterns of the cluster model on the $\{5,4\}$ lattice generated by the update rule Eq.~(\ref{eq_54_cluster}). (a) Coordinate of a general Hamiltonian term Eq.~(\ref{eq_54clusterH}) on the deformed lattice. The red and blue terms correspond to $J(P(j))=0$ and $1$, respectively, and are physically equivalent. The symbol ``?'' indicates that $J(\cdot)$ at this site can be $0$ or $1$. The red region below illustrates the support of a Hamiltonian term on the hyperbolic lattice. Panels (b)--(e) show symmetry patterns generated by initial conditions (b) $\mathbf{q}_1(x,y)$ in Eq.~(\ref{eq_54_cluster_ic1}), (c) $\mathbf{q}_2(x,y)$ in Eq.~(\ref{eq_54_cluster_ic2}), (d) $\mathbf{q}_3(x,y)$ in Eq.~(\ref{eq_54_cluster_ic3}), and (e) $\mathbf{q}_4(x,y)$ in Eq.~(\ref{eq_54_cluster_ic4}). Nontrivial Pauli $X$ actions of the symmetry are represented by black polygons.}
\end{figure*}

After construction of the linear NUCA$(5,4)$, we explicitly design some NUCA$(5,4)$-generated subsystem symmetry-protected models. 
The cluster model is a well-studied nontrivial SSPT model~\cite{Raussendorf_2001_MBQC,you2018sspt,Devakul2018classification,Daniel2020computational,Stephen2019subsystem,zhou_2022_strangecor,zhang2024hoca,li2026SEE} and can be directly defined on the hyperbolic lattice. 
Considering the cluster model on the $\{5,4\}$ lattice, the translationally invariant Hamiltonian in standard NUCA form Eq.~(\ref{eq_54_SSPT}) is written as:
\begin{widetext}
\begin{equation}
	\label{eq_54clusterH}
	\begin{aligned}
		\mathscr{H}_{\text{cluster}} =&-\sum_{\substack{(i,j)\in \mathrm{S}_1 \\i\equiv0\, (\text{mod }2)}}Z\left( x^iy^j\left[ 1+y^{-1}+y^{-2}+\left( 1-J\left( P(j) \right) \right) y^{-3}+x^{-2}y^{P\left( P\left( j \right) \right) -j}+J\left( P(j) \right)x^{-2} y^{P\left( P\left( j \right) \right) -j+1} \right] \right) \\ &\qquad \qquad \qquad \qquad \qquad \qquad \qquad \qquad \qquad \qquad \qquad \qquad\qquad \qquad \qquad \qquad \qquad \qquad \quad Z\left( x^{i-1}y^{P\left( j \right)} \right)  \\
		& -\sum_{\substack{(k,l)\in \mathrm{S}_1 \\k\equiv1\, (\text{mod }2)}}X\left( x^k y^l\left[ 1+y^{-1}+y^{-2}+\left( 1-J\left( P(l) \right) \right) y^{-3}+x^{-2}y^{P\left( P\left( l \right) \right) -l}+J\left( P(l) \right)x^{-2} y^{P\left( P\left( l \right) \right) -l+1} \right] \right) \\ &\qquad \qquad \qquad \qquad \qquad \qquad \qquad \qquad \qquad \qquad \qquad \qquad\qquad \qquad \qquad \qquad \qquad \qquad \quad  X\left( x^{k-1}y^{P\left( l \right)} \right) \,,
	\end{aligned}
\end{equation}
\end{widetext}
where $\mathrm{S}_1=\{(i,j)\mid i\ge2,3\le j\le A_{2i+1}-2,J(j)=1\}$ as we exclude terms that are not fully supported on the lattice.
We visualize a general Hamiltonian term on the hyperbolic lattice and its explicit coordinate on the deformed lattice in Fig.~\ref{fig_54_model_1}(a).
We discuss the properties of the hyperbolic cluster state as an SSPT state in Appendix~\ref{app_hcm}.
The position-dependent transposed update rule of Hamiltonian Eq.~(\ref{eq_54clusterH}) is explicitly written as
\begin{equation}
	\label{eq_54_cluster}
	\begin{aligned}
	\bar{\mathbf{f}}_j&\cdot\bar{\mathbf{y}} = y^{-1}+y^{-2}+(1-J(P(j)))y^{-3}\\
	&+J(P(j))x^{-2}y^{P(P(j))-j+1}+x^{-2}y^{P(P(j))-j},\,\, (i,j)\in \mathrm{S}_1\,,
	\end{aligned}
\end{equation}
and the rule is trivial otherwise.
Here, we equivalently represent $\bar{\mathbf{f}}_j(x)$ as $\bar{\mathbf{f}}_j\cdot\bar{\mathbf{y}}$ to simplify the expression.

\begin{figure}[b]
	\includegraphics[width=0.6\columnwidth]{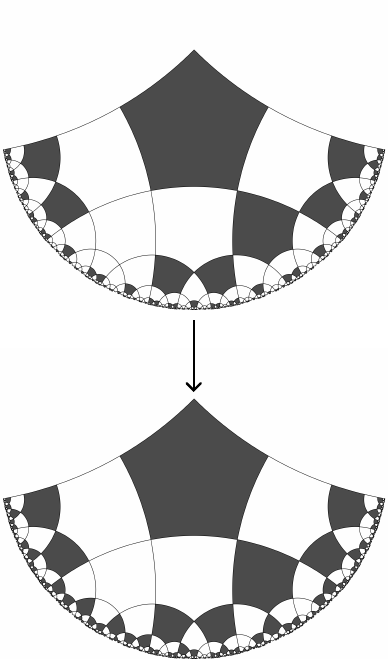}
	\caption{\label{fig_initial_condition} 
	Visualization of nontrivial initial conditions for the update rule of the cluster model Eq.~(\ref{eq_54_cluster}). Sites belonging to one sublattice with trivial $\bar{\mathbf{f}}_j$ can be assigned initial conditions and are shown as black pentagons above. If all these sites are assigned $a=1$, the resulting symmetry pattern generated by NUCA$(5,4)$ is shown below.}
\end{figure}

The position-dependent update rules can lead to a nontrivial initial condition that is absent in the uniform CA.
By the definition of the initial condition in Eq.~(\ref{eq_initial_condition_NUCA}), the site $x^iy^j$ with $\bar{\mathrm{f}}_{j,k}(x) = 0,\forall k$ can be assigned $a_{ij}\in \mathbb{F}_2$ as the initial condition.
Then a subsystem symmetry Eq.~(\ref{symmetry}) is obtained by the NUCA$(5,4)$ evolution configuration on one set of sublattices.
For instance, in Fig.~\ref{fig_54_model_1}(b)--(e) we show symmetry patterns with different initial conditions:
\begin{subequations}
	\begin{align}
		\mathbf{q}_1(x,y) = x^0y^0\,, \label{eq_54_cluster_ic1}\\
		\mathbf{q}_2(x,y) = x^2y^2\,, \label{eq_54_cluster_ic2}\\
		\mathbf{q}_3(x,y) = x^2y^0\,, \label{eq_54_cluster_ic3}\\
		\mathbf{q}_4(x,y) = x^4y^2\,, \label{eq_54_cluster_ic4}
	\end{align}
\end{subequations} 
which generate subsystem symmetries that have nontrivial $X$-action on the black polygons.
In Fig.~\ref{fig_initial_condition}, we visualize the symmetry pattern generated by the maximal initial condition, where all sites with trivial $\bar{\mathbf{f}}_{j}(x)$ are assigned $a=1$ as the initial condition.

\begin{figure*}[t]
	\includegraphics[width=0.9\textwidth]{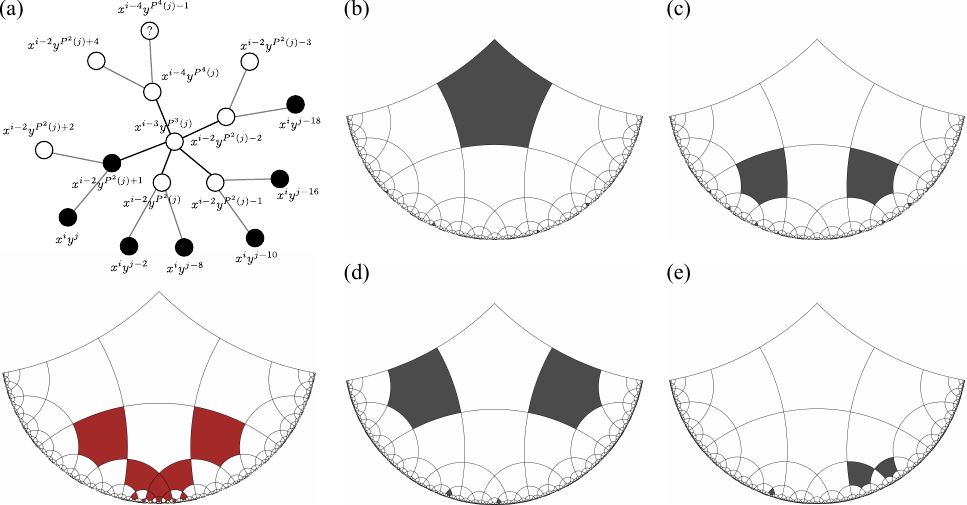}
	\caption{\label{fig_54_model_2} Hamiltonian and symmetry patterns of an SSPT model on the $\{5,4\}$ lattice with the explicit update rule given in Appendix~\ref{app_update_rule_forms}. (a) Coordinate of a specific Hamiltonian term on the deformed lattice. The symbol ``?'' indicates that $J(\cdot)$ at this site can be $0$ or $1$. The red region below illustrates the support of this Hamiltonian term on the hyperbolic lattice. Panels (b)--(e) show symmetry patterns generated by initial conditions (b) $\mathbf{q}_1(x,y)$ in Eq.~(\ref{eq_54_model2_ic1}), (c) $\mathbf{q}_2(x,y)$ in Eq.~(\ref{eq_54_model2_ic2}), (d) $\mathbf{q}_3(x,y)$ in Eq.~(\ref{eq_54_model2_ic3}), and (e) $\mathbf{q}_4(x,y)$ in Eq.~(\ref{eq_54_model2_ic4}). Nontrivial Pauli $X$ actions of the symmetry are represented by black polygons.}
\end{figure*}

Another example of translationally invariant NUCA$(5,4)$-generated SSPT model is shown in Fig.~\ref{fig_54_model_2}.
The update rule and commutation relation corresponding to this model are complex, and we present the explicit form in Eq.~(\ref{eq_54_sspt_2}) of Appendix~\ref{app_NUCA54}.
By specifying the initial condition as 
\begin{subequations}
	\begin{align}
		&\mathbf{q}_1(x,y) = x^0y^0  \,, \label{eq_54_model2_ic1}\\
		&\mathbf{q}_2(x,y) = x^2y^2+x^2y^5  \label{eq_54_model2_ic2}\,,\\
		&\mathbf{q}_3(x,y) = x^1y^0+x^1y^2\,, \label{eq_54_model2_ic3}\\
		&\mathbf{q}_4(x,y) = x^3y^6+x^3y^7\,, \label{eq_54_model2_ic4}
	\end{align}
\end{subequations} 
we obtain symmetry patterns with nontrivial action of Pauli $X$ operators on the black pentagons in Fig.~\ref{fig_54_model_2}(b)--(e).

\subsection{NUCA-generated multi-point strange correlator for SSPT states}
The strange correlator is a useful tool to diagnose nontrivial short-range entangled states, especially for (subsystem) symmetry-protected topological states~\cite{zhou_2022_strangecor,zhang2024hoca,gao_zhang2025ssc}.
A nonlocal multi-point strange correlator (MPSC) is defined as: 
\begin{equation}
	\mathbf{C}\left(\left|\Psi \right>; r_1,r_2,\cdots ,r_{\mathrm{m}} \right) =\frac{\left< \Omega \right|O \left( r_1 \right) O \left( r_2 \right) \cdots O \left( r_{\mathrm{m}} \right) \left| \Psi \right>}{\left< \Omega |\Psi \right>}\,,
\end{equation}
where $\left|\Psi \right>$ is a short-range entangled state to be diagnosed and $\left|\Omega \right>$ is a trivial (product) state in the same Hilbert space as $\left|\Psi \right>$ preserving the same symmetry. 
For a given local operator $O$ and spatial configuration $\{r_i\}$,
an MPSC is nontrivial if $C(\left|\Psi \right>,\{r_i\}) - C(\left|\Omega \right>,\{r_i\})$ saturates to a constant or decays algebraically, thus it can be used to detect nontrivial SPT order. 
Otherwise, the chosen MPSC cannot distinguish the nontrivial SPT state from the trivial symmetric state.

One can construct nonlocal MPSC to detect SSPT orders with $O$ being the onsite Pauli operators generated by NUCA evolution as:
\begin{equation}
	\mathbf{C}\left(\left|\Psi \right>; r_1,r_2,\cdots ,r_{\mathrm{m}} \right) =\frac{\left< \Omega \right|O\left( D_L\left( \mathbf{q},\mathbf{f}_{(i,j)};x,y \right) \right) \left| \Psi \right>}{\left< \Omega |\Psi \right>} \,.
\end{equation}
Here, $D_L\left( \mathbf{q},\mathbf{f}_{(i,j)};x,y \right)$ denotes the support configuration generated by NUCA from the initial condition $\mathbf{q}$, with characteristic distance $L$ among its support, and $O(D_L)$ denotes the corresponding product of onsite Pauli operators on this support.
The trivial state is typically chosen as $\left|\Omega \right> = \left|\hat{X}^{\text{(a)}} = \hat{Z}^{\text{(b)}} = 1 \right>$, which respects the subsystem symmetry of the system. 

For the Euclidean case, the existence of an MPSC with constant and minimal ${\mathrm{m}}$ that does not grow with $L$ is a universal feature for all translationally invariant Euclidean SSPT models~\cite{zhang2024hoca}.
However, this is not the case on the hyperbolic lattice due to the underlying geometry.
For instance, we consider the following update rule:
\begin{equation}
	\label{hcmrule_g}
	\mathbf{g}_j(x) {=} (J(j),\underbrace{0,0,\cdots,0,0}_{C(C(j)){-}j {-}4},x^2,x^2,x^2,(1-J(j))x^2)^T \,,
\end{equation}
which generates a subset of all the cluster Hamiltonian terms on the $\{5,4\}$ lattice. 
The update rule $\mathbf{g}_j(x)$ can be directly constructed from the SSPT Hamiltonians.

On the deformed lattice, two sublattices $(a)$ and $(b)$ are distinguished by even and odd power $i$ of $x^i$.
Now we consider the operator $O_{ij} = Z(x^iy^j(1+\mathbf{g}_j\cdot\mathbf{y})),i\equiv0\, (\text{mod }2)$, which commutes with the symmetry. 
With $O^1_{ij}=O_{ij}$, the product of $O$
\begin{equation}
	\label{eq_MPSC}
	\begin{aligned}
		O^\mathrm{n}_{ij} &\coloneqq \prod_{x^ly^m\in \text{Supp}\{O^{\mathrm{n}-1}\}} {O_{lm}} \\
		&= \prod_{x^ly^m\in \text{Supp}\{O^{\mathrm{n}-1}\}} Z((1+\mathbf{g}_m\cdot\mathbf{y})x^ly^m) \,,
	\end{aligned}
\end{equation} 
acts equivalently as products of Hamiltonian terms in the given sublattice.
The MPSC corresponding to $O^\mathrm{n}_{ij}$
\begin{equation}
	\begin{aligned}
		&\mathbf{C}\left(\left|\Psi \right>; r_1,\cdots ,r_{\mathrm{m}} \right) \\
		&=\frac{\left< \Omega \right|\prod_{x^ly^m\in \text{Supp}\{O^{\mathrm{n}-1}\}} Z((1+\mathbf{g}_m\cdot\mathbf{y})x^ly^m) \left| \Psi \right>}{\left< \Omega |\Psi \right>} \\
		& =\frac{1}{\left< \Omega |\Psi \right>} \left< \Omega \right|(\prod_{\substack{x^ly^m\in \text{Supp}\{O^{\mathrm{n}-1}\} \\J(m)=1}}  Z(x^{l+1}y^{C(m)}) h_{l+2,C(C(m))} \\
		&\qquad \times \prod_{\substack{x^ly^m\in \text{Supp}\{O^{\mathrm{n}-1}\} \\J(m)=0}}  Z(x^{l+1}y^{C(m)}) h_{l+2,C(C(m))+1}) \left| \Psi \right> \\
		&=1
	\end{aligned}
\end{equation}
gives a nontrivial result, where $h_{i,j}$ is a Hamiltonian term defined in Eq.~(\ref{eq_54clusterH}).
The corresponding multi-point normal correlator for the trivial symmetric state $C(\left|\Omega \right>,\{r_i\})$ vanishes, thus this nontrivial MPSC distinguishes the hyperbolic cluster state from the trivial symmetric state.

To count the nontrivial support of the MPSC, we define a support map for the supporting region of $O_{ij}$:
\begin{equation}
	\hat{M}[x^iy^j] \coloneqq x^iy^j(1 + \mathbf{g}_j\cdot\mathbf{y})\,,
\end{equation}
such that the supporting region of $O^\mathrm{n}_{ij}$ is given by:
\begin{equation}
	\text{Supp}\{O^\mathrm{n}_{ij}\} = \hat{M}^\mathrm{n}[x^iy^j]\,,
\end{equation}
with coefficients mod $2$.
By substituting the explicit form Eq.~(\ref{hcmrule_g}) of $\mathbf{g}_j$ into $\hat{M}^2[x^iy^j]$:
\begin{equation}
	\begin{aligned}
		\hat{M}^2&[x^iy^j] = x^iy^j+J(j)x^iy^{j+1}\mathbf{g}_{j+1}\cdot\mathbf{y} \\
		&+x^{i+2}y^{C(C(j))}(y^{-2}\mathbf{g}_{C(C(j))-2}\cdot\mathbf{y}+y^{-1}\mathbf{g}_{C(C(j))-1}\cdot\mathbf{y}\\
		&\quad +\mathbf{g}_{C(C(j))}\cdot\mathbf{y} +(1-J(j))y\mathbf{g}_{C(C(j))+1}\cdot\mathbf{y}) 
	\end{aligned}
\end{equation}
for $x^iy^j$, we obtain a nontrivial and nonlocal MPSC that detects the cluster state from the trivial product state.
From the $C(j)$ function in Eq.~(\ref{auxiliaryC}) we directly find that $\mathbf{g}_j\ne \mathbf{g}_{j'}$ if $j\ne j'$.
Through the sequential action of $\hat{M}$ which can be regarded as generated by NUCA$(5,4)$, the supporting region of this MPSC is extended to the boundary.
$\text{Supp}\{O^\mathrm{n}_{ij}\}$ grows with $\mathrm{n}$ according to the update rule $\mathbf{g}_j(x)$, i.e., the supporting points $\mathrm{m}$ of a nonlocal MPSC grow exponentially.
Unlike the MPSC for Euclidean SSPT models for which $\mathrm{m}$ can be independent of the distance $L\sim\mathrm{n}$, the growing behavior intrinsically reflects the exponentially growing lattice size of hyperbolic geometry.

\subsection{Non-uniformity of NUCA and Non-abelian translation invariance}
\label{ssec_translation}
Conventional CA, based on synchronous evolution and the uniform update rule $\mathbf{f}(x)$, are widely applied to simulate Euclidean physics.
In general, the non-uniformity of CA is summarized into three major variants~\cite{nuca2026}: asynchronous CA with non-simultaneous evolution of sites, network CA with complex neighborhood structures, and non-uniform CA with site-specific update rules.
For NUCA defined on the deformed lattice, the non-uniformity is intrinsically manifested by the position-dependent update rules, as exemplified by NUCA$(5,4)$ discussed in this section. 

We further discuss the translation invariance of NUCA-generated Hamiltonians, leading to constraints for the update rules.
As the lattice-deforming procedure for hyperbolic lattices induces distortion, the translation symmetry is not manifest on the deformed lattice.
Therefore, a NUCA-generated Hamiltonian may or may not be translationally invariant and we construct a sufficient condition to ensure this symmetry.
As reviewed in Appendix~\ref{app_hyperbolic}, translations in hyperbolic space are non-Abelian, leading to nontrivial periodic boundary conditions (PBC) of the hyperbolic lattice~\cite{Crystallography2022,Lux_2023_PBC,Maciejko2020HBT,Maciejko2022AHBT,Lenggenhager2023supercell}.
A PBC lattice is constructed by the factor group $\Gamma /\Gamma_{\text{PBC}}$, where $\Gamma_{\text{PBC}}$ is a normal subgroup of the non-Abelian translation group $\Gamma$.
The specific lattice topology depends on $\Gamma_{\text{PBC}}$.
We denote the translation operator for $[g_a] \in \Gamma /\Gamma_{\text{PBC}}$ as $\mathcal{T}_{[g_a]}$, and the translation symmetry of the Hamiltonian is defined as $\mathcal{T}_{[g_a]} \mathscr{H} \mathcal{T}_{[g_a]}^{-1} = \mathscr{H},\,\forall [g_a]\in \Gamma /\Gamma_{\text{PBC}}$.

We first consider the general SSPT Hamiltonian Eq.~(\ref{eq_54_SSPT}), and recall that it can be transformed to a uniform form by applying Hadamard gates on one sublattice.
In the Hamiltonian Eq.~(\ref{eq_54_SSPT}), we identify the center of a term $h_{ij}$ defined by $\bar{\mathbf{f}}_{j}$ as site $x^{i-u(j)}y^{m(j)}$ which is the unique nontrivial support on the complementary sublattice.
Under a translation $[g_a]$ mapping this center to $x^{i'-u(j')}y^{m(j')}$, the Hamiltonian term $h_{i'j'}$ does not necessarily coincide with $\mathcal{T}_{[g_a]} h_{ij} \mathcal{T}_{[g_a]}^{-1}$ as it is defined by the update rule $\bar{\mathbf{f}}_{j'}$ at site $x^{i'}y^{j'}$.
A sufficient condition for $h_{i'j'}=\mathcal{T}_{[g_a]} h_{ij} \mathcal{T}_{[g_a]}^{-1}$ is that the action of the Hamiltonian term must be isotropic with respect to the geodesic distance from its center.
Specifically, for qubits with the same geodesic distance from the center, the term has either nontrivial or trivial action on all of them simultaneously.
This condition can be incorporated into the design of update rules, and is not satisfied by a uniform $\bar{\mathbf{f}}$. 
The SSPT models we designed satisfy this sufficient condition and thereby are translationally invariant.

The analysis for SSSB models Eq.~(\ref{eq_54_SSB}) is more complex as it does not rely on a sublattice structure. 
In this case, the center of a Hamiltonian term can be a polygon or a vertex.
For a Hamiltonian term centered on a polygon such as Eq.~(\ref{eq_rule_reg_2}), the condition is similar to the SSPT case.
For qubits with the same geodesic distance from the center, the term has either nontrivial or trivial action on all of them simultaneously.
For a Hamiltonian term centered on a vertex such as Eq.~(\ref{eq_rule_reg_1}), a similar condition holds.
For qubits with the same geodesic distance from the central vertex, the term has either nontrivial or trivial action on all of them simultaneously.
These conditions are both incorporated into the design of update rules. 
A NUCA-generated Hamiltonian is translationally invariant on the hyperbolic lattice if it satisfies this sufficient condition.
In Appendix~\ref{app_NUCA54}, we provide a more comprehensive discussion of translation invariance on the hyperbolic lattice.

\begin{figure*}[t]
	\includegraphics[width=0.95\textwidth]{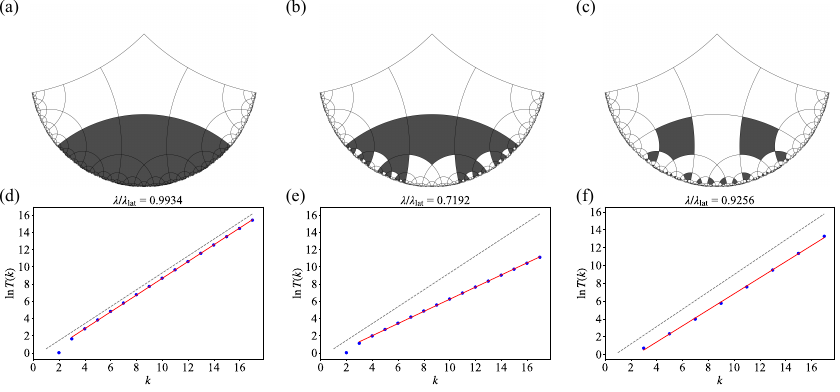}
	\caption{\label{fig_sym_fit} Numerical growth of NUCA-generated subsystem symmetries. (a,d) Regular symmetry pattern generated by the update rule Eq.~(\ref{eq_rule_reg_1}) with initial condition $\mathbf{q}_1(x,y)$ in Eq.~(\ref{eq_reg1_ic1}), whose growth rate satisfies $\lambda = \lambda_{\text{lat}}$. (b,e) Irregular symmetry pattern generated by the update rule Eq.~(\ref{eq_rule_tree}) with initial condition $\mathbf{q}_1(x,y)$ in Eq.~(\ref{eq_rule_tree_ic1}), whose growth rate satisfies $\lambda < \lambda_{\text{lat}}$. (c,f) Irregular symmetry pattern generated by the update rule Eq.~(\ref{eq_54_cluster}) with initial condition $\mathbf{q}_2(x,y)$ in Eq.~(\ref{eq_54_cluster_ic2}), whose growth rate satisfies $\lambda < \lambda_{\text{lat}}$. The gray dashed lines denote the growth of the lattice size.}
\end{figure*}

\section{Diversity of NUCA-generated subsystem symmetries}
\label{sec_diversity}
Subsystem symmetries are intimately related to the physical properties of quantum systems.
The geometric constraints of subsystem symmetries make the determination and classification of them challenging even in the Euclidean case~\cite{Devakul2019fractal,Devakul2018fractaluniversal,Devakul2019fractalclass,zhang2024hoca,Stephen2019subsystem,Daniel2020computational,Sfairopoulos2023,sfairopoulos2025cellularautomataddimensions}.
To date, only a few subsystem symmetries associated with specific classical spin models on hyperbolic lattices have been discovered in the literature~\cite{yan2019hfm,yan2019hfm2,yan2025hfm3}, without a general framework for designing and classifying them. 
In the previous section, we have designed SSPT and SSSB models by NUCA$(5,4)$, showing that the hyperbolic geometry leads to new classes of subsystem symmetries.
In this section, we take NUCA$(5,4)$-generated symmetries as examples to classify NUCA-generated subsystem symmetries into \textit{regular} and \textit{irregular} symmetries based on their growth properties, extending the classification of subsystem symmetries beyond Euclidean lattices and revealing exponential expansion of hyperbolic geometry. 

The lattice size of a hyperbolic lattice grows exponentially with the level $k$ of $x^{k-1}$ as $N(k)\sim \exp (\lambda_{\text{lat}}k)$, allowing the symmetry pattern $\mathscr{F}$ to grow exponentially as $T(\tilde{\mathscr{F}},k)\sim \exp (\lambda k)$. 
Here, $T(\tilde{\mathscr{F}},k)$ is the number of sites with state $a=1$ of the truncated configuration $\tilde{\mathscr{F}}$ on a finite lattice.
To characterize these symmetries, we numerically compute the growth of a symmetry configuration $\tilde{\mathscr{F}}$ through the scaling behavior 
\begin{equation}
	\ln T(\tilde{\mathscr{F}},k) = \lambda k +o(k)\,,
\end{equation}
then we can distinguish subsystem symmetries by their growing parameter $\lambda$.
We call a symmetry irregular if $0<\lambda<\lambda_{\text{lat}}$, and regular if $\lambda=\lambda_{\text{lat}}$.
For the irregular patterns, the density of the support vanishes in the thermodynamic limit $k\rightarrow \infty$, and these patterns are comparable to fractal or chaotic symmetries in the Euclidean cases.
The hyperbolic geometry allows for a fractal pattern with treelike structure, such as a binary-tree [e.g., update rule Eq.~(\ref{eq_rule_tree}) with initial condition Eq.~(\ref{eq_rule_tree_ic1}) grows as $T(\tilde{\mathscr{F}},k)\sim 2^{k}$], which cannot be isometrically embedded into a Euclidean square lattice. 
Conversely, the symmetry support grows at the same rate as the lattice if $\lambda = \lambda_{\text{lat}}$, leading to regular symmetries.
For these regular symmetries, we find that the support occupies a finite density of the entire lattice, and thus they are analogous to membrane-like and checkerboard-like symmetries on a Euclidean lattice.
So far we have not discovered subsystem symmetry patterns exhibiting sub-exponential growth.

We show numerical fitting of the growth rate for regular and irregular patterns in Fig.~\ref{fig_sym_fit}, where we select as examples the patterns generated by (a,d) update rule Eq.~(\ref{eq_rule_reg_1}) with initial condition Eq.~(\ref{eq_reg1_ic1}), (b,e) update rule Eq.~(\ref{eq_rule_tree}) with initial condition Eq.~(\ref{eq_rule_tree_ic1}), and (c,f) update rule Eq.~(\ref{eq_54_cluster}) with initial condition Eq.~(\ref{eq_54_cluster_ic2}). 
The computation is performed on a lattice with $i_{\max}=16$ and $x^{16}$ contains approximately $5.7\times 10^6$ physical sites. 
For the SSPT model Eq.~(\ref{eq_54_cluster}), the computation is constrained to one sublattice, as a symmetry element acts only on one sublattice. 

\begin{figure*}[t]
	\includegraphics[width=0.9\textwidth]{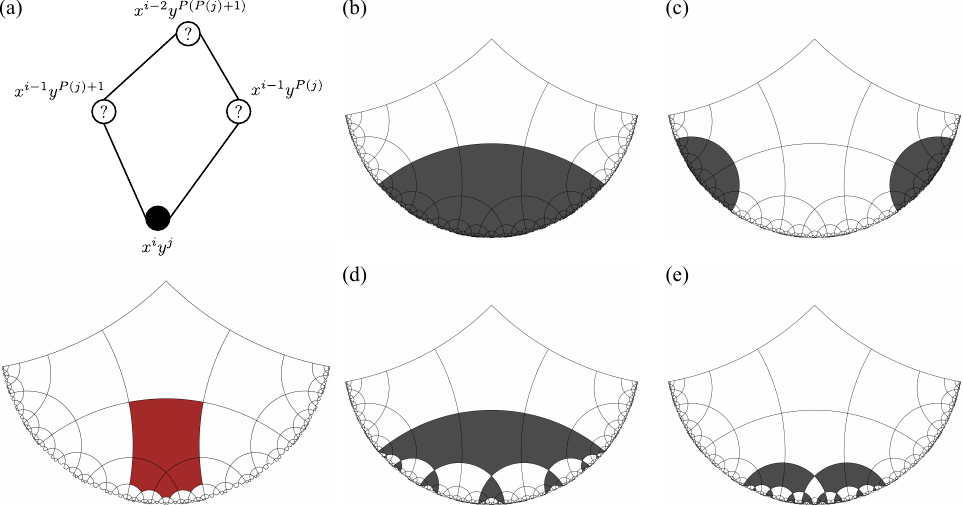}
	\caption{\label{fig_54_model_3} 
	Hamiltonian and symmetry patterns generated by the update rule Eq.~(\ref{eq_rule_reg_1}).
	(a) Coordinate of a general Hamiltonian term Eq.~(\ref{eq_Hamiltonian_reg_1}) on the deformed lattice. 
	The symbol ``?'' indicates that $J(\cdot)$ at this site can be $0$ or $1$. 
	The red region below illustrates the support of a Hamiltonian term on the hyperbolic lattice.
	Panels (b)--(e) show symmetry patterns generated by different initial conditions (b) $\mathbf{q}_1(x,y)$ in Eq.~(\ref{eq_reg1_ic1}), (c) $\mathbf{q}_2(x,y)$ in Eq.~(\ref{eq_reg1_ic2}), (d) $\mathbf{q}_3(x,y)$ in Eq.~(\ref{eq_reg1_ic3}), and (e) $\mathbf{q}_4(x,y)$ in Eq.~(\ref{eq_reg1_ic4}). Nontrivial Pauli $X$ actions of the symmetry are represented by black polygons. }
\end{figure*}

\subsection{NUCA-generated regular subsystem symmetries}
Now we present some examples of regular subsystem symmetries. 
To simplify the expressions, we only present the transposed update rules in the form of $\bar{\mathbf{f}}_j\cdot\bar{\mathbf{y}}$.
These transposed rules uniquely determine the corresponding update rules $\mathbf{f}_j(x)$ that are explicitly given in Appendix~\ref{app_NUCA54}.
A regular pattern has finite density in the thermodynamic limit as discussed above, typically showing a geodesic-wedge pattern discovered in the literature~\cite{yan2019hfm,yan2019hfm2}.
As an intuitive example, we consider the update rule written in the transposed form:
\begin{equation}
	\label{eq_rule_reg_1}
	\begin{aligned}
		\bar{\mathbf{f}}_j\cdot\bar{\mathbf{y}}& = x^{-1}y^{P(j)-j+1}+x^{-1}y^{P(j)-j}+x^{-2}y^{P(P(j)+1)-j}\,,\\
		&\qquad\qquad\qquad\qquad\qquad\qquad\qquad\qquad (i,j)\in \mathrm{S}_2\,,
	\end{aligned}
\end{equation}
where $\mathrm{S}_2 =\{(i,j)\mid i\ge2,2\le j\le A_{2i+1}-2,J(j)=1\}$, and $\bar{\mathbf{f}}_j(x)$ is trivial otherwise. 
In the formalism of Eq.~(\ref{eq_54_SSB}), the SSSB Hamiltonian of the transposed update rule Eq.~(\ref{eq_rule_reg_1}) is written as:
\begin{equation}
	\label{eq_Hamiltonian_reg_1}
	\begin{aligned}
		\mathscr{H} = -\sum_{(i,j)\in \mathrm{S}_2}&Z(x^iy^j(1 + x^{-1}(1+y)y^{P(j)-j}  \\
		&\qquad\qquad + x^{-2}y^{P(P(j)+1)-j}))\, .
	\end{aligned}
\end{equation}
A term in Eq.~(\ref{eq_Hamiltonian_reg_1}) is the interaction of four neighboring qubits around a vertex on the hyperbolic lattice, and we show the Hamiltonian term in Fig.~\ref{fig_54_model_3}(a).
This update rule also describes a hyperbolic fracton model studied in the literature~\cite{yan2019hfm,yan2019hfm2,yan2025hfm3}, which can be seen as a generalized plaquette Ising model with some holographic properties related to its subsystem symmetry.

We turn to specify the initial condition $\mathbf{q}$.
The transposed update rule $\bar{\mathbf{f}}_j(x)$ in Eq.~(\ref{eq_rule_reg_1}) can only be nontrivial for sites with $J(j)=1$.
Then $\mathbf{q}$ can be specified for sites with trivial $\bar{\mathbf{f}}_j(x)$, and the states of these sites are not determined by preceding configurations during the NUCA evolution.
For instance, we consider the particular $\mathbf{q}(x,y)$
\begin{subequations}
	\begin{align}
		\mathbf{q}_1(x,y) =& xy+\sum_{i=2}^{i_{\max}}  \sum_{j=L_{i-1}(1)}^{U_{i-1}(1)} (1-J(j))x^iy^j \label{eq_reg1_ic1}\\
		\mathbf{q}_2(x,y) =& x^2y+\sum_{i=3}^{i_{\max}}  \sum_{j=L_{i-2}(1)}^{U_{i-2}(1)} (1-J(j))x^iy^j\notag\\
		&+x^2y^6+\sum_{i=3}^{i_{\max}}  \sum_{j=L_{i-2}(6)}^{U_{i-2}(6)} (1-J(j))x^iy^j\,, \label{eq_reg1_ic2}
	\end{align}
\end{subequations}
where $L_1(j)=\lfloor \phi^2 j\rfloor,L_i(j)=\lfloor \phi^2 L_{i-1}(j)\rfloor, U_1(j)=\lfloor \phi^2 (j+1)\rfloor, U_i(j)=\lfloor \phi^2 (U_{i-1}(j)+1)\rfloor$.
These sites are mapped to pentagons on the hyperbolic $\{5,4\}$ lattice which are all on one side of a geodesic.
The symmetry patterns generated by NUCA$(5,4)$ are shown in Fig.~\ref{fig_54_model_3}(b)--(c), where the nontrivial support of the symmetry forms a geodesic-wedge pattern.

\begin{figure*}[t]
	\includegraphics[width=0.9\textwidth]{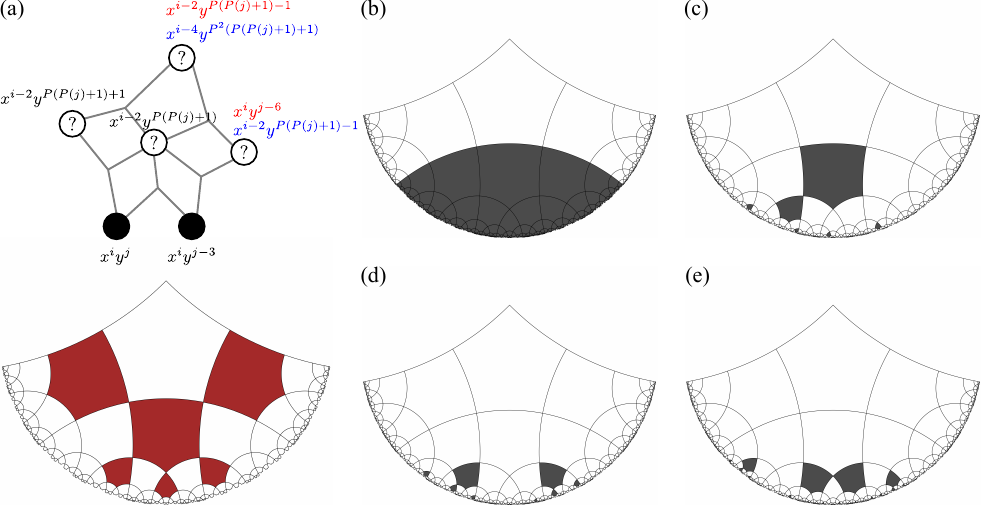}
	\caption{\label{fig_54_model_4} 
	Hamiltonian and symmetry patterns generated by the update rule Eq.~(\ref{eq_rule_reg_2}).
	(a) Coordinate of a general Hamiltonian term Eq.~(\ref{eq_Hamiltonian_reg_2}) on the deformed lattice. 
	The red and blue terms correspond to $J(P(P(j)+1))=0$ and $1$ respectively, and are physically equivalent. 
	The symbol ``?'' indicates that $J(\cdot)$ at this site can be $0$ or $1$. 
	The red region below illustrates the support of a Hamiltonian term on the hyperbolic lattice.
	Panels (b)--(e) show symmetry patterns generated by different initial conditions (b) $\mathbf{q}_1(x,y)$ in Eq.~(\ref{eq_reg2_ic1}), (c) $\mathbf{q}_2(x,y)$ in Eq.~(\ref{eq_reg2_ic2}), (d) $\mathbf{q}_3(x,y)$ in Eq.~(\ref{eq_reg2_ic3}), and (e) $\mathbf{q}_4(x,y)$ in Eq.~(\ref{eq_reg2_ic4}). Nontrivial Pauli $X$ actions of the symmetry are represented by black polygons.}
\end{figure*}

Another example of NUCA$(5,4)$-generated SSSB model that has a regular symmetry pattern is given by the following transposed update rule:
\begin{equation}
	\label{eq_rule_reg_2}
	\begin{aligned}
	\bar{\mathbf{f}}_j\cdot&\bar{\mathbf{y}} =  y^{-3} + (1-J(P(P(j)+1))) y^{-6} + x^{-2}y^{P(P(j)+1)+1-j} \\
	& + x^{-2}y^{P(P(j)+1)-j} + x^{-2}y^{P(P(j)+1)-1-j} \\
	&+ J(P(P(j)+1))x^{-4}y^{P(P(P(P(j)+1)+1))-j}\,,\quad (i,j)\in \mathrm{S}_3\,,
	\end{aligned}
\end{equation}
where $\mathrm{S}_3 =\{(i,j)\mid i\ge3,6\le j\le A_{2i+1}-8,J(j)=J(P(j)+1)=1\}$, and $\bar{\mathbf{f}}_j(x)$ is trivial otherwise. 
The Hamiltonian corresponding to this transposed update rule reads:
\begin{equation}
	\label{eq_Hamiltonian_reg_2}
	\begin{aligned}
		\mathscr{H} = -\sum_{(i,j)\in \mathrm{S}_3}&Z(x^iy^j(1 + y^{-3} + (1-J(P(P(j)+1))) y^{-6}  \\
		& + (y^{-1}+1+y)x^{-2}y^{P(P(j)+1)-j} \\
		& + J(P(P(j)+1))x^{-4}y^{P(P(P(P(j)+1)+1))-j}))\, .
	\end{aligned}
\end{equation}
Terms of this translationally invariant Hamiltonian are interactions of a qubit with its next-nearest-neighboring qubits on the physical $\{5,4\}$ lattice as shown in Fig.~\ref{fig_54_model_4}(a). 
The initial condition can be specified for all sites with trivial $\bar{\mathbf{f}}_j(x)$.
Then we choose the initial condition as
\begin{equation}
	\label{eq_reg2_ic1}
	\mathbf{q}_1(x,y) = xy+\sum_{i=2}^{i_{\max}}  \sum_{j=L_{i-1}(1)}^{U_{i-1}(1)} (1-J(j)J(P(j)+1)) x^iy^j\,, 
\end{equation}
and we obtain a regular symmetry pattern in Fig.~\ref{fig_54_model_4}(b). 
For examples considered here, the regular symmetries are found for SSSB models only due to the sublattices structure.

\subsection{NUCA-generated irregular subsystem symmetries}
Now we present the irregular symmetry patterns.
The subsystem symmetries of the SSPT examples discussed in Sec.~\ref{sec_NUCA54} are all irregular, as shown in Fig.~\ref{fig_54_model_1} and Fig.~\ref{fig_54_model_2}. 
Moreover, the update rules supporting regular symmetry patterns can also generate irregular patterns, similar to the behavior of rules supporting mixed symmetries on a Euclidean lattice.
The symmetry patterns generated by the update rule Eq.~(\ref{eq_rule_reg_1}) with initial condition
\begin{subequations}
	\begin{align}
		&\mathbf{q}_3(x,y) = x^1y^1 \label{eq_reg1_ic3}\\
		&\mathbf{q}_4(x,y) = x^2y^3+x^2y^4 \label{eq_reg1_ic4}
	\end{align}
\end{subequations} 
are shown in Fig.~\ref{fig_54_model_3}(d)--(e).
The symmetry patterns generated by the update rule Eq.~(\ref{eq_rule_reg_2}) with initial condition
\begin{subequations}
	\begin{align}
		&\mathbf{q}_2(x,y) = x^1y^1 \label{eq_reg2_ic2}\\
		&\mathbf{q}_3(x,y) = x^3y^7 \label{eq_reg2_ic3}\\
		&\mathbf{q}_4(x,y) = x^2y^3+x^2y^4 \label{eq_reg2_ic4}
	\end{align}
\end{subequations} 
are shown in Fig.~\ref{fig_54_model_4}(c)--(e).
Although both rules support regular geodesic-wedge symmetry patterns, their irregular symmetry patterns are different.

\begin{figure*}[t]
	\includegraphics[width=0.9\textwidth]{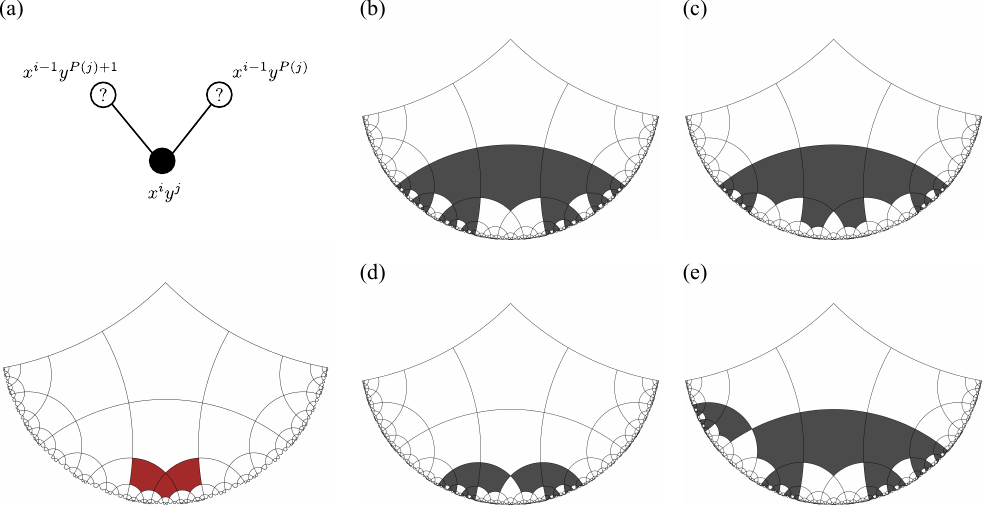}
	\caption{\label{fig_54_model_5} 
	Hamiltonian and symmetry patterns generated by the update rule Eq.~(\ref{eq_rule_tree}).
	(a) Coordinate of a general Hamiltonian term Eq.~(\ref{eq_Hamiltonian_rule_tree}) on the deformed lattice. 
	The symbol ``?'' indicates that $J(\cdot)$ at this site can be $0$ or $1$. 
	The red region below illustrates the support of a Hamiltonian term on the hyperbolic lattice.
	Panels (b)--(e) show symmetry patterns generated by different initial conditions  (b) $\mathbf{q}_1(x,y)$ in Eq.~(\ref{eq_rule_tree_ic1}), (c) $\mathbf{q}_2(x,y)$ in Eq.~(\ref{eq_rule_tree_ic2}), (d) $\mathbf{q}_3(x,y)$ in Eq.~(\ref{eq_rule_tree_ic3}), and (e) $\mathbf{q}_4(x,y)$ in Eq.~(\ref{eq_rule_tree_ic4}). Nontrivial Pauli $X$ actions of the symmetry are represented by black polygons.}
\end{figure*}

The irregular patterns above resemble mixed and chaotic patterns on Euclidean lattices. Hyperbolic lattices also allow a distinct treelike fractal subsystem symmetry, which is generated by the following rule.
We consider the transposed update rule 
\begin{equation}
	\label{eq_rule_tree}
	\bar{\mathbf{f}}_j\cdot\bar{\mathbf{y}} = x^{-1}y^{P(j)+1-j} + x^{-1}y^{P(j)-j}\,,\quad (i,j)\in \mathrm{S}_4\,,
\end{equation}
where $\mathrm{S}_4 =\{(i,j)\mid i\ge2,2\le j\le A_{2i+1}-2,J(j)=1\}$, and $\bar{\mathbf{f}}_j(x)$ is trivial otherwise. 
This rule is a generalization of the Sierpinski and Fibonacci rule discovered on the Euclidean lattice~\cite{Devakul2019fractal,Devakul2019fractalclass,zhang2024hoca}.
The Hamiltonian corresponding to this update rule is 
\begin{equation}
	\label{eq_Hamiltonian_rule_tree}
	\mathscr{H} = -\sum_{(i,j)\in\mathrm{S}_4}Z(x^iy^j(1 + x^{-1}(1+y)y^{P(j)-j}))\, ,
\end{equation}
which is not translationally invariant on the hyperbolic lattice and we visualize the Hamiltonian term in Fig.~\ref{fig_54_model_5}(a).
The initial conditions can be specified for all sites with $\bar{\mathbf{f}}_j(x)=\mathbf{0}$, and the following initial conditions
\begin{subequations}
	\begin{align}
		&\mathbf{q}_1(x,y) = x^1y^1  \,, \label{eq_rule_tree_ic1}\\
		&\mathbf{q}_2(x,y) = x^1y^1+x^2y^3+x^2y^4  \,, \label{eq_rule_tree_ic2}\\
		&\mathbf{q}_3(x,y) = x^2y^3+x^2y^4  \,, \label{eq_rule_tree_ic3}\\
		&\mathbf{q}_4(x,y) = x^1y^1+x^2y^6  \,, \label{eq_rule_tree_ic4}
	\end{align}
\end{subequations} 
generate symmetry patterns in Fig.~\ref{fig_54_model_5}(b)--(e).
Symmetries of this type have binary-treelike fractal structure, which cannot exist isometrically on a Euclidean square lattice.

\section{Non-uniform Clifford quantum cellular automata from classical NUCA}
\label{sec_QCA}

\subsection{Overview}
Quantum cellular automata (QCA) are quantum generalizations of classical CA and have been studied as models for universal quantum computation or proposals for quantum simulation~\cite{Raussendorf_2005_QCA,Osborne_2006_QCA,Arrighi2019qca,Farrelly2020reviewofquantum}.
An important class of QCA is Clifford QCA (CQCA), which map single Pauli operators to their tensor products under time evolution~\cite{Farrelly2020reviewofquantum}.
CQCA are useful tools in the study of many-body physics, in particular for investigating subsystem symmetries.
Translationally invariant one-dimensional CQCA are classified into periodic, glider and fractal classes~\cite{Gtschow2010,CQCA_class_2010} according to the trace of their transfer matrices.
These classes correspond to different classes of subsystem symmetries of Euclidean cluster states~\cite{Daniel2020computational,Stephen2019subsystem}.
However, this correspondence relies on the $\mathbb{Z}\times\mathbb{Z}$ translation symmetry of Euclidean lattices that leads to natural spatial-temporal structure.
For hyperbolic lattices, the exponential growth and non-Abelian translations make it unclear how to construct an underlying CQCA for hyperbolic cluster states.

In this section, we overcome this difficulty by developing a NUCA-guided partition method for the hyperbolic $\{5,4\}$ lattice.
Using the linear NUCA$(5,4)$ update rule of the hyperbolic cluster model in Eq.~(\ref{eq_54_cluster}), we partition the two-dimensional hyperbolic lattice into disjoint one-dimensional graphs whose endpoints coincide with sites having trivial transposed update rule $\bar{\mathbf{f}}_j$ and therefore match the NUCA initial conditions.
Based on this partition, we identify the tensor network description of the hyperbolic cluster state and construct its underlying non-uniform CQCA.
Through propagation of nontrivial Pauli operators, this non-uniform CQCA generates the corresponding subsystem symmetries of hyperbolic cluster state.
Although it is not translationally invariant in either spatial or temporal direction, physical locality is preserved on the target hyperbolic lattice.
Our construction provides an explicit hyperbolic extension of the correspondence between CQCA and subsystem symmetries, showing how the non-uniformity inherited from classical NUCA enables CQCA dynamics to incorporate hyperbolic geometry.

\subsection{Construction of non-uniform CQCA for the hyperbolic cluster state}
We first determine the tensor-network description of the cluster state on a two-dimensional lattice and then the underlying CQCA, following the procedure in Ref.~\cite{Stephen2019subsystem,Daniel2020computational}.
The cluster model Eq.~(\ref{eq_54clusterH}) can be mapped to a uniform form on both sublattices by applying Hadamard gates supported on one sublattice. 
After this basis change, the ground state $\left|\Psi \right>$ is obtained by applying controlled-Z (CZ) gates to neighboring qubits of a product state
\begin{equation}
	\left|\Psi \right> =\prod_{(a,b)\in E}{CZ_{a,b}\left| + \right>^{\otimes |Q|}} \,,
\end{equation}
where $\left| + \right>$ is the $+1$ eigenstate of $X$, $Q$ labels the qubit set and $E$ labels pairs of neighboring qubits.
We can partition the lattice into disjoint one-dimensional graphs, where connected polygons are mapped to vertices linked by edges of these graphs.
After this partition, each such graph defines a wire of a one-dimensional cluster state, while neighboring relations of qubits on different graphs are represented by CZ gates.
From the tensor network description of the one-dimensional cluster states, they are distinct wires where matrix-product state (MPS) tensors are defined on vertices with CZ gates connecting their physical bonds.

\begin{figure*}[t]
	\includegraphics[width=0.85\textwidth]{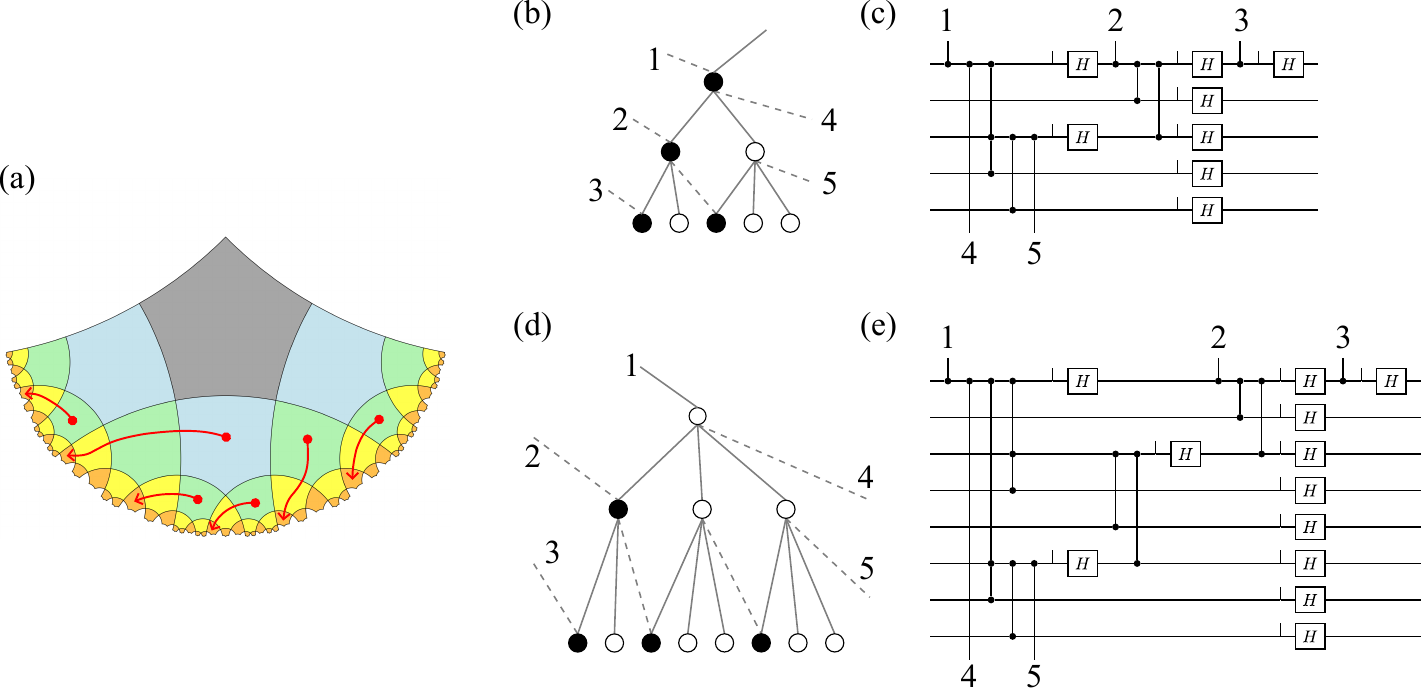}
	\caption{\label{fig_QCA_building_block} Construction of the CQCA for the cluster state on the $\{5,4\}$ lattice. 
	(a) Partition of the $\{5,4\}$ lattice into one-dimensional graphs, where the red lines represent a subset of those starting from the bulk. 
	(b,c) Mapping of a black node or white node with its offspring to Clifford circuits consisting of CZ gates, copy tensors and Hadamard gates. 
	Numbers indicate the correspondence between neighboring relations and dangling CZ gates.}
\end{figure*}

To obtain the tensor-network description of the hyperbolic cluster state, we use some properties of the MPS tensor of the one-dimensional cluster state. 
Firstly, such an MPS tensor is equivalent to a copy tensor with a Hadamard gate by symmetry, as
\begin{equation}
\includegraphics[height = 24pt]{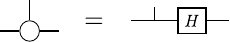}\,.
\end{equation}
Here, the horizontal bonds of the MPS tensor correspond to the virtual bond, and the vertical bond corresponds to the physical bond.
Then, the CZ gate acting on the physical bond can be pushed down to the virtual bond as
\begin{equation}
\includegraphics[height = 33pt]{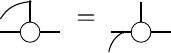}\,,
\end{equation}
where the dangling wire denotes half of a CZ gate.
Then we partition the network into different time slices that are vertical cuts across all wires.
Each time slice contains at most one MPS tensor and some CZ gates on each wire.
By replacing the MPS tensors by copy tensors and Hadamard gates, we can sort the components of the tensor network to obtain a proper temporal structure such that time direction is from left to right and different time slices are related by wires only.
Finally, the underlying Clifford circuit of the hyperbolic cluster state is obtained by moving the copy tensor to the front of the time slice.
The Clifford circuit may not be translationally invariant in spatial or temporal direction, as the underlying lattice may not possess $\mathbb{Z}\times \mathbb{Z}$ translation symmetry.

The partition of a hyperbolic lattice into one-dimensional graphs is not unique. 
However, not all possible partitioning methods yield Clifford circuits suitable for generating subsystem symmetries.
A suitable partition is one for which inputs of Pauli operators through virtual bonds propagate to physical Pauli supports that coincide with NUCA-generated symmetry patterns.
Here, we choose the partition associated with the NUCA$(5,4)$ update rule Eq.~(\ref{eq_54_cluster}), so that wire endpoints coincide with sites having trivial $\bar{\mathbf{f}}_j$ and hence are consistent with NUCA initial conditions.
Then the wires uniquely extend along paths consisting of black-node offspring successively, as shown in Fig.~\ref{fig_QCA_building_block}(a).
Consequently, the tensor network description of the cluster state on the $\{5,4\}$ lattice yields Clifford circuits with geometry as in Fig.~\ref{fig_QCA_building_block}(b)--(c).
Here we illustrate how the fundamental units of the lattice, i.e., black and white nodes with their offspring in the spanning tree, are mapped onto the corresponding elements of the circuit.
Due to the hyperbolic geometry, the resulting Clifford circuit is not translationally invariant in temporal or spatial direction.
The CZ gates can be nonlocal in the Clifford circuit, but their corresponding control and target qubits are neighboring on the hyperbolic lattice.
Therefore, the physical locality is preserved.

\subsection{Non-uniform CQCA generation of subsystem symmetries}
CQCA are defined on a lattice of qubits governed by Clifford operations.
By definition, CQCA map any tensor product of Pauli operators to another up to a phase factor~\cite{Farrelly2020reviewofquantum}.
In the Heisenberg picture, the dynamics of CQCA are characterized by the time evolution of initial Pauli operators.
At each time slice, the support of the propagated Pauli operator specifies the sites on which the corresponding physical Pauli operators act nontrivially.
The space-time configuration of the dynamics of CQCA has a one-to-one correspondence to the action of Pauli operators across the lattice.
The Clifford circuit we obtained serves as a non-uniform CQCA that can be utilized to generate subsystem symmetry of the hyperbolic cluster state, similar to translationally invariant CQCA for the cluster state on Euclidean lattices~\cite{Stephen2019subsystem,Daniel2020computational}.
The subsystem symmetries can be determined by the evolution of products of Pauli operators under CQCA.
When propagating a Pauli $X$ ($Z$) operator through the left virtual bond of a copy tensor, an $X$ ($I$) operator is left on the physical bond and an $X$ ($Z$) is left on the right virtual bond, i.e.,
\begin{equation}
\includegraphics[height = 20pt]{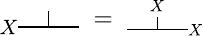}\,,
\end{equation}
and
\begin{equation}
\includegraphics[height = 15pt]{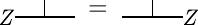}\,.
\end{equation}
By propagating a combination of $X$ operators as the initial condition through the virtual space of the tensor network, the left-behind $X$ operators in physical space constitute a subsystem symmetry element of the cluster state.

\begin{figure}[b]
	\includegraphics[width=0.8\columnwidth]{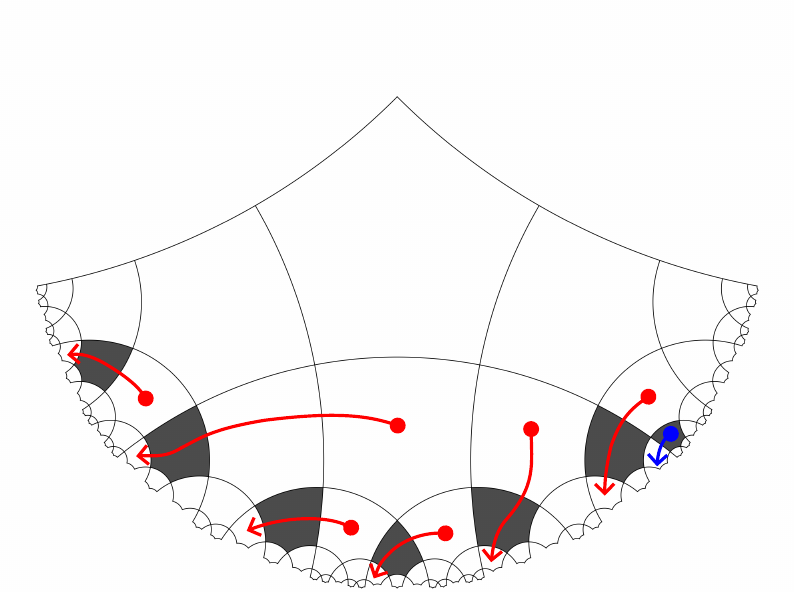}
	\caption{\label{fig_QCA_sym} 
A subsystem symmetry of the cluster state on the $\{5,4\}$ lattice generated by the underlying CQCA. Propagating an $X$ operator along the blue wire, with identity operators on all other wires, yields a subsystem symmetry whose nontrivial $X$ action is supported on the black pentagons.}
\end{figure}

Within the NUCA$(5,4)$ construction for the hyperbolic cluster model Eq.~(\ref{eq_54_cluster}), each product of the Pauli operators as input in the CQCA is uniquely mapped to a specific initial condition of the NUCA$(5,4)$.
Consequently, the full set of NUCA-generated subsystem symmetries is obtained by specifying different combinations of virtual $X$ operators as inputs and propagating them through the CQCA.
In Fig.~\ref{fig_QCA_sym}, we show a symmetry element generated by propagating an $X$ through the blue wire and $I$ through all the others.
This CQCA evolution is equivalent to NUCA$(5,4)$ evolution under the update rule Eq.~(\ref{eq_54_cluster}) with the initial condition $\mathbf{q}(x,y)=x^3y^4$.
Our construction here reveals that the CQCA preserves the non-uniformity of the classical NUCA, which is necessary for investigating physical systems on the hyperbolic lattice by cellular automata.

In the Euclidean case, the translationally invariant one-dimensional CQCA can be classified into periodic, glider and fractal classes~\cite{Gtschow2010,CQCA_class_2010} according to the trace of their transfer matrices. 
These CQCA are discovered to have a one-to-one correspondence with ribbon, cone and fractal subsystem symmetries of cluster states on Euclidean lattices~\cite{Daniel2020computational}.
The underlying non-uniform CQCA of the cluster state on the hyperbolic lattice does not preserve translation invariance in spatial and temporal directions, thus falling outside the scope of the above classification.
Nevertheless, it generates subsystem symmetries that cannot be obtained from translationally invariant CQCA within the above correspondence.
Therefore, this provides a hyperbolic analogue  of the established correspondence between translationally invariant CQCA and subsystem symmetries on Euclidean lattices, broadening our understanding of the interplay between CQCA, subsystem symmetries and lattice geometry.

\section{Probabilistic NUCA and directed percolation on the hyperbolic lattice}
\label{sec_DP}

\subsection{Overview}
\begin{figure*}[t]
	\centering
	\includegraphics[width=0.75\textwidth]{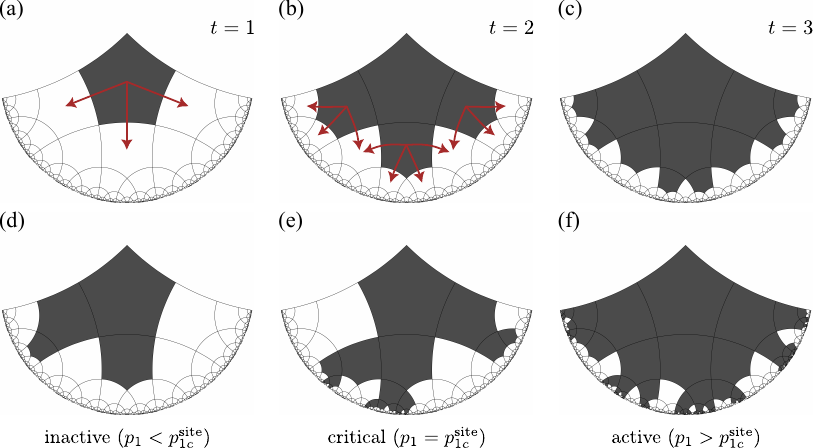}
	\caption{\label{fig_DP_steps} Visualization of directed percolation on the $\{5,4\}$ lattice. Occupied sites are represented by black polygons. 
	Panels (a)--(c) show three successive steps of the single seed evolution.
	Red arrows represent the direction of DP, corresponding to the direction from the center of the lattice to hyperbolic infinity.
	Panels (d)--(f) show the evolution patterns for site DP ($p_1=p_2$) in the inactive phase ($p_1=0.1<p_{1c}^{\text{site}}$), the critical point ($p_1=0.33= p_{1c}^{\text{site}}$), and the active phase ($p_1=0.6>p_{1c}^{\text{site}}$).
	}
\end{figure*}

Directed percolation (DP) is a fundamental model in non-equilibrium statistical physics particularly for the conjectured universality of DP~\cite{Hinrichsen2000,Janssen1981,Grassberger1982}, whose critical behavior has been experimentally verified~\cite{Takeuchi_2007_DP,Takeuchi_2009_DP,Lemoult2016,Lemoult2024}.
In a DP process, sites are either occupied or empty, and occupation spreads only along a specific direction through a stochastic process.
On hyperbolic lattices, numerical studies of undirected percolation suffer from intrinsic geometric and computational difficulties~\cite{Margenstern2018,hl_percolation}.
As emphasized in Ref.~\cite{hl_percolation}, the lack of a simple coordinate system makes it nontrivial to label sites and compute their neighborhood, while the exponential growth of lattice size makes the storage of the whole lattice computationally expensive.
For directed percolation, these difficulties are further coupled to the need to define direction of DP compatible with hyperbolic geometry.
On Euclidean lattices, DP can be simulated by uniform probabilistic CA, which are variants of CA with stochastic update rules~\cite{ROLLIER2025108362,PCA2018,DK1984,Pizzi2021,MIPT_DP_2021} as introduced in Sec.~\ref{sec_preliminaries}.
However, these one-dimensional uniform CA cannot be directly applied to hyperbolic lattice, as they cannot preserve physical locality or define dynamical direction on the hyperbolic lattice.

In this section, we overcome these difficulties by formulating a probabilistic NUCA$(5,4)$ to numerically simulate directed percolation on the $\{5,4\}$ lattice, a process that is visualized in Fig.~\ref{fig_DP_steps}.
By encoding dynamical direction into stochastic update rules Eq.~(\ref{eq_nuca54_percolation}), the DP process is represented by the space-time configuration of NUCA evolution.
Through the auxiliary function $J(j)$ and $P(j)$ in Eq.~(\ref{auxiliaryJ}) and Eq.~(\ref{auxiliaryP}), these non-uniform update rules intrinsically reflect the treelike structure of hyperbolic geometry while preserving physical locality.
Within the probabilistic NUCA framework for DP, our deformed-lattice coordinates and auxiliary functions provide a computational coordinate system for directly computing neighboring relation, avoiding the need to store the full lattice with neighborhood data.
Using this probabilistic NUCA$(5,4)$, we simulate several DP process including bond DP and site DP, and numerically estimate the DP thresholds from the scaling analysis of survival probability.
Through interpolation of the numerical results, we also obtain an approximate phase diagram.
Our results show that the probabilistic NUCA provides a local-to-global framework for studying non-equilibrium physics on hyperbolic lattices.

\subsection{Construction of probabilistic NUCA$(5,4)$ for DP on the $\{5,4\}$ lattice}
\begin{figure*}[t]
	\includegraphics[width=0.95\textwidth]{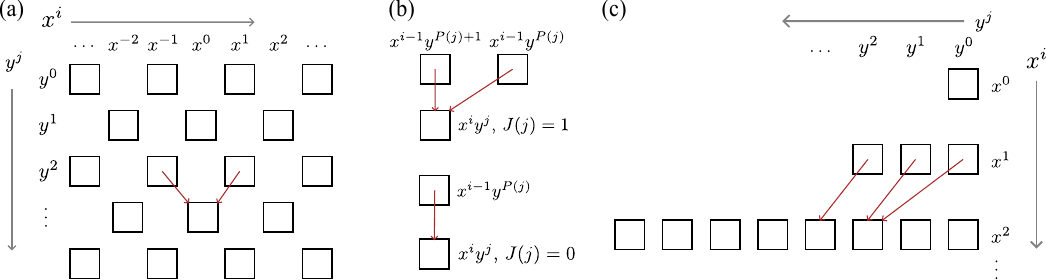}
	\caption{\label{fig_DP_process} 
	Simulation of directed percolation on the $\{4,4\}$ and $\{5,4\}$ lattices using probabilistic CA and NUCA. (a) Domany-Kinzel model for simulating DP on a diagonal Euclidean square lattice. The states of $x^{i-1}y^{j-1}$ and $x^{i+1}y^{j-1}$ determine the state of $x^{i}y^{j}$ through a stochastic process. (b) Update rule of the probabilistic NUCA$(5,4)$ for simulating DP on the $\{5,4\}$ lattice. Different sites have different update rules depending on their positions. (c) Visualization of the probabilistic NUCA$(5,4)$ on the deformed lattice. The NUCA evolution is along the $y^+$-direction, while the DP process is along the $x^+$-direction. The whole DP process is given by the space-time configuration of NUCA evolution.
	}
\end{figure*}

We start by briefly reviewing the application of a probabilistic CA, i.e., the Domany-Kinzel model for simulating DP on the Euclidean lattice~\cite{DK1984}.
We consider a diagonal square lattice with coordinate $x^iy^j,i+j\equiv1\pmod 2$ as shown in Fig.~\ref{fig_DP_process}(a), where $j=0,1,2,\cdots$ corresponds to discrete time steps of CA.
Then, the state $a_{ij}\in \mathbb{F}_2$ represents whether a site is occupied ($a_{ij}=1$) or empty ($a_{ij}=0$).
Initially, sites at $y^{0}$ are specified to be occupied seeds or empty, which is interpreted as the initial condition.
During the evolution of CA, the probability that site $x^iy^j$ is occupied $a_{ij}=1$ depends on the states of its neighboring sites $x^{i-1}y^{j-1}$ and $x^{i+1}y^{j-1}$.
Specifically, the state $a_{ij}$ is (\rmnum{1}) $0$ if both its neighboring sites are empty, (\rmnum{2}) $1$ with probability $p_1\in [0,1]$ if one and only one of its neighboring sites is occupied, and (\rmnum{3}) $1$ with probability $p_2\in [0,1]$ if both its neighboring sites are occupied.
Following the definition in Eq.~(\ref{eq_pnuca}), this stochastic, uniform and order-$1$ update rule can be written as:
\begin{equation}
	\begin{aligned}
		&P(a_{i,j}=1|a_{i-1,j-1}=1, a_{i+1,j-1}=0) = p_1\,, \\
		&P(a_{i,j}=1|a_{i-1,j-1}=0, a_{i+1,j-1}=1) = p_1\,, \\
		&P(a_{i,j}=1|a_{i-1,j-1}=1, a_{i+1,j-1}=1) = p_2\,,
	\end{aligned}
\end{equation}
for all sites and the rule is trivial otherwise. 
The Domany-Kinzel model on a diagonal square lattice is visualized in Fig.~\ref{fig_DP_process}(a), and we visualize the update rule of the DP process.

By performing the probabilistic CA under a certain initial condition, the time steps $j$ are equivalent to distinct DP steps and the process of DP is described by the evolution configuration $\mathscr{F}(x,y)$. 
Numerical simulations demonstrate that the DP process exhibits an absorbing-state phase transition across a critical line defined by $(p_{1c},p_{2c})$. 
In the inactive phase, the number of occupied sites decreases exponentially with time $j$ and the system reaches a unique absorbing state $r_{j}(x)=0$ if we specify a single seed as the initial condition.
Conversely, there is a finite probability that the resulting cluster is infinite in the active phase~\cite{Hinrichsen2000}.
When $p_1,p_2$ approach the thresholds $p_{1c},p_{2c}$ on the critical line, the system exhibits critical behavior characterized by the critical exponents of the DP universality class.
While $p_1$ and $p_2$ can be arbitrary values in $[0,1]$, there are some specific cases to note.
If $p_2=1$, the system is compact directed percolation, belonging to a distinct universality class.
The CA simulates site DP if $p_2=p_1$; while it simulates bond DP if $p_2=1-(1-p_1)^2=2p_1-p_1^2$.
The corresponding undirected site and bond percolation have been numerically studied on the hyperbolic lattices that are not in the same universality class as DP~\cite{hl_percolation,Hinrichsen2000}.

We construct a probabilistic NUCA$(5,4)$ to simulate the DP process on the $\{5,4\}$ lattice, which is performed on the deformed lattice introduced in Sec.~\ref{sec_NUCA54}.
The direction of DP flows from the center of the hyperbolic lattice (the bulk) to hyperbolic infinity (the boundary), as illustrated in Fig.~\ref{fig_DP_steps} (a)--(c).
On the deformed lattice, this DP direction is along the $x^+$-direction, and sites $x^iy^j$ sharing the same $i$ belong to the same DP step when a unique seed is specified as the initial condition.
The state $a_{i,j}=1$ for a site $x^{i}y^j$ is determined by the states of sites in the preceding DP step, while NUCA simulates this process through evolution along the $y^+$-direction.
Consequently, the entire history of the DP process on a finite hyperbolic lattice is represented by the whole NUCA evolution configuration $\tilde{\mathscr{F}}(x,y)$, rather than by a single intermediate stage of the evolution.

The number of physical sites $u_i$ grows exponentially with the power $i$ of $x^i$ on the deformed lattice, resulting in non-uniform update rules of DP on the hyperbolic lattice.
Geometrically, $J(j)$ determines the number of sites from the preceding DP step whose corresponding polygons are adjacent.
On the deformed lattice, for a site $x^iy^j$ with $J(j)=0$, there is a unique site $x^{i-1}y^{P(j)}$ from the preceding DP step whose corresponding pentagon is adjacent to that of $x^iy^j$.
Therefore, $a_{ij}$ can be $1$ with probability $p_1\in [0,1]$ if $a_{i-1,P(j)}=1$.
For a site $x^iy^j$ with $J(j)=1$, there are two sites $x^{i-1}y^{P(j)}$ and $x^{i-1}y^{P(j)+1}$ from the preceding DP step whose corresponding pentagons are adjacent to that of $x^iy^j$.
Therefore, $a_{ij}$ is $1$ with probability (\rmnum{1}) $p_1\in [0,1]$ if either $a_{i-1,P(j)}=1$ or $a_{i-1,P(j)+1}=1$, and (\rmnum{2}) $p_2\in [0,1]$ if $a_{i-1,P(j)}=a_{i-1,P(j)+1}=1$.
The probabilistic NUCA$(5,4)$ is defined as:
\begin{equation}
	\label{eq_nuca54_percolation}
	\begin{aligned}
		&P(a_{i,j}=1|a_{i-1,P(j)}=1,J(j)=0) =p_1\,,\\
		&P(a_{i,j}=1|a_{i-1,P(j)}=0,a_{i-1,P(j)+1}=1,J(j)=1) = p_1\,,\\
		&P(a_{i,j}=1|a_{i-1,P(j)}=1,a_{i-1,P(j)+1}=0,J(j)=1) = p_1\,,\\
		&P(a_{i,j}=1|a_{i-1,P(j)}=1,a_{i-1,P(j)+1}=1,J(j)=1) =p_2\,,
	\end{aligned}
\end{equation}
and all the other cases have probability $0$ for $a_{ij}=1$.
In Fig.~\ref{fig_DP_process}(b), we visualize the update rule of the DP process on the deformed lattice.

This DP process intrinsically reflects the treelike structure of the hyperbolic lattice, and the non-uniform update rule is designed to follow this structure.
Analogous to the Euclidean case, site DP is defined by $p_2=p_1$ and bond DP is defined by $p_2=2p_1-p_1^2$.
Both cases reduce to the same probability $p_1$ for sites with $J(j)=0$ and the distinction between site and bond DP appears only at sites with $J(j)=1$.
Although each time step $j$ of NUCA is not a discrete step of the DP process, the whole space-time configuration of evolution $\tilde{\mathscr{F}}(x,y)$ represents the DP process.
In Fig.~\ref{fig_DP_process}(c), we visualize the evolution of probabilistic NUCA$(5,4)$.

For the NUCA defined on the deformed lattice, the update rules can be determined by the auxiliary functions on a lattice of arbitrary size.
Moreover, it is unnecessary to store all lattice sites and their neighborhood which can be directly computed through the auxiliary functions in the numerical implementation.
Through the design of probabilistic NUCA$(5,4)$ with update rule Eq.~(\ref{eq_nuca54_percolation}), the DP on the $\{5,4\}$ lattice can be studied, overcoming the difficulties in the numerical study of percolation on hyperbolic lattices~\cite{hl_percolation}.

\subsection{Numerical results of directed percolation on the $\{5,4\}$ lattice}
\begin{figure*}[t]
	\includegraphics[width=0.95\textwidth]{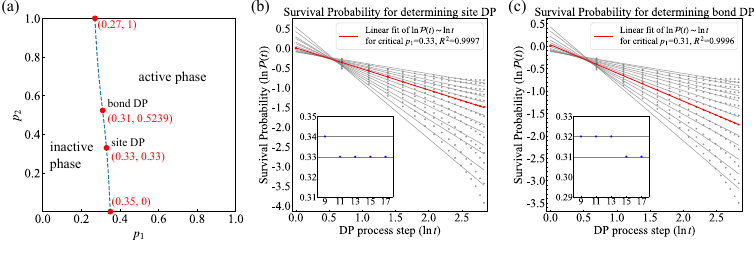}
	\caption{\label{fig_DP_diagram} 
	Numerical results for directed percolation on the $\{5,4\}$ lattice using probabilistic non-uniform CA. (a) Phase diagram of DP on the $\{5,4\}$ lattice. The red points are numerically computed, and the dashed line is an estimate obtained by interpolation. The thresholds for site DP (b) and bond DP (c) are obtained by comparing the coefficient of determination $R^2$ for fitting the data $\ln \mathcal{P}(t) \sim \ln t$ with a total of $t=17$ DP steps, giving the thresholds $p_{1c}=0.33$ in (b) and $p_{1c}=0.31$ in (c). Data corresponding to the critical probability are highlighted in red. Insets show thresholds individually computed with $9,11,13,15,17$ DP steps. Data in (b) and (c) are obtained from $10,000$ independent probabilistic NUCA$(5,4)$ evolutions.
	}
\end{figure*}

To numerically estimate the phase boundary $p_{1c}$ and $p_{2c}$, we apply finite-size scaling analysis~\cite{Hinrichsen2000}.
Because the coordinate on the deformed lattice does not directly measure geodesic distance, here we consider a dimensionless quantity to compute the probability threshold.
The survival probability $\mathcal{P}(t)$ is the probability that a cluster grown from a single seed still has some sites occupied after $t$ discrete DP steps along the $x^+$-direction.
We assign the single seed at $x^0y^0$ as $a_{00}=1$, such that $t=i+1$.
As the hyperbolic lattice with finite surface-to-volume ratio in the thermodynamic limit can be regarded as being infinite dimensional~\cite{hl_percolation,Heydenreich2017,Wang2013DP}, here we use the power-law form for the survival probability:
\begin{equation}
	\mathcal{P}(t)\simeq t^{-\delta}\,.
\end{equation}
This serves as a finite-size scaling ansatz for locating the transition.
The survival probability is expected to decay exponentially in the inactive phase and to saturate to a nonzero value in the active phase.
At the transition when $p_1,p_2$ approach $p_{1c},p_{2c}$ on the critical line, the survival probability will have power-law decay~\cite{Hinrichsen2000}.
For data corresponding to $p_1,p_2$ obtained by the NUCA simulation, we fit $\ln \mathcal{P}(t)$ linearly as a function of $\ln t$ and use the coefficient of determination $R^2$ to identify the data closest to the power-law decay, which estimates the threshold.

The numerical computation is performed on a lattice with $i_{\max}=16$ and $x^{16}$ contains approximately $5.7\times 10^6$ physical sites. 
For each combination $(p_{1},p_{2})$, we perform $10,000$ independent probabilistic NUCA$(5,4)$ evolutions to compute the survival probability $\mathcal{P}(t)$.
The threshold $(p_{1c},p_{2c})$ for a specific DP process is estimated by comparing the coefficient of determination from the linear fit of the data.
Our results estimate that the site DP threshold is $p_{1c}^{\text{site}}=0.33$, while the bond DP threshold is $p_{1c}^{\text{bond}}=0.31$.
Moreover, the threshold for the compact DP limit $p_{2}=1$ is estimated at $p_{1c}=0.27$, while that for the $p_{2}=0$ limit is estimated at $p_{1c}=0.35$. 
Through interpolation of these results, we obtain the approximate DP phase diagram on the hyperbolic $\{5,4\}$ lattice as shown in Fig.~\ref{fig_DP_diagram}(a). 
We explicitly visualize the scaling behavior of $\mathcal{P}(t)$ for $(p_{1},p_{2})$ near the site DP and bond DP thresholds in Fig.~\ref{fig_DP_diagram}(b)--(c).
In Fig.~\ref{fig_DP_steps}(d)--(f), we illustrate examples of evolution patterns of site DP process in the inactive phase, the critical point, and the active phase simulated by probabilistic NUCA$(5,4)$.

As expected for directed percolation, which is more constrained than undirected percolation, the probability threshold for DP is larger than in the corresponding undirected case~\cite{Hinrichsen2000}.
In Ref.~\cite{hl_percolation}, the most accurate numerical results to date for undirected site and bond percolation on hyperbolic lattices are reported.
Our investigation of site and bond DP defined for polygons on the $\{5,4\}$ lattice is compared to undirected site and bond percolation defined for vertices on the $\{4,5\}$ lattice.
This comparison is natural because polygon adjacency on the $\{5,4\}$ lattice is dual to vertex adjacency on the $\{4,5\}$ lattice.
The numerical results $p_{c,\text{undir}}^{\text{site}}= 0.29890539(6)$ for undirected site percolation and $p_{c,\text{undir}}^{\text{bond}}=0.2689195(3)$ for undirected bond percolation are smaller than the corresponding thresholds for DP.
We also note that $p_{1c}^{\text{site}}$ and $p_{1c}^{\text{bond}}$ are numerically close. 
According to the generation of the lattice in Sec.~\ref{sec_NUCA54}, the update rules on sites with $J(j)=0$ are independent of $p_{2}$ and are uniquely determined by $p_{1}$.
The distinction between site DP and bond DP is exclusively manifested at the sites corresponding to the black nodes in the spanning tree, reflecting the treelike structure of the hyperbolic lattice.

\section{Summary and Outlook}
\label{sec_conclusion}
In this paper, we develop a \textit{higher-order non-uniform cellular automata} (NUCA) framework for studying geometry-dependent physics on translationally invariant regular Euclidean and hyperbolic lattices, as illustrated in Fig.~\ref{fig_NUCA}. 
By embedding hyperbolic lattices into a Euclidean square lattice, the lattice-deforming procedure induces position-dependent update rules that encode hyperbolic geometry while preserving physical locality on the target lattice.
This construction generalizes conventional uniform CA by replacing geometry-independent update rules with higher-order and geometry-dependent update rules derived from the lattice-deforming procedure.

Using the linear NUCA$(5,4)$, we systematically design and analyze SSPT and SSSB models associated with NUCA-generated subsystem symmetries.
We classify these subsystem symmetries into regular [e.g., Eq.~(\ref{eq_rule_reg_2})] and irregular [e.g., Eq.~(\ref{eq_54_cluster})] classes, revealing the imprint of the exponential expansion of the hyperbolic lattice. 
To diagnose the resulting SSPT order, we construct NUCA-generated nonlocal MPSC whose support is dictated by subsystem symmetries on hyperbolic lattices, and derive a sufficient condition for NUCA-generated Hamiltonians to be invariant under non-Abelian translation symmetry.
As a quantum generalization, we construct a non-uniform CQCA through classical NUCA.
Based on the update rule Eq.~(\ref{eq_54_cluster}) for the cluster state on the $\{5,4\}$ lattice, we obtain a CQCA from the tensor network description of the hyperbolic cluster state to generate its subsystem symmetries.
This provides an explicit hyperbolic extension of the correspondence between translationally invariant CQCA and subsystem symmetries on Euclidean lattices~\cite{Gtschow2010,CQCA_class_2010,Daniel2020computational}.
Finally, we formulate a probabilistic NUCA$(5,4)$ for directed percolation on the hyperbolic lattice.
The stochastic update rules Eq.~(\ref{eq_nuca54_percolation}) incorporate the treelike structure induced by negative curvature, allowing the geometric growth of the hyperbolic lattice to enter directly into the DP dynamics.
Through NUCA evolutions, we numerically estimate several DP thresholds, including site DP and bond DP, and obtain the phase diagram of DP on the $\{5,4\}$ lattice through interpolation in Fig.~\ref{fig_DP_diagram}.
These results establish NUCA as a unified local-to-global framework for constructing, classifying, and diagnosing geometry-driven quantum phases and non-equilibrium dynamics beyond the Euclidean setting.

Several future directions naturally follow from this work.
Beyond Euclidean and hyperbolic lattices, one promising direction is to extend the NUCA construction to other complex geometries such as fractal lattices~\cite{Mandelbrot_fractal,Gefen1980fractal,Kempkes2018,zhou_2024_fractal}, where non-uniform update rules may encode self-similar geometric structures.
Another direction is to generalize CA-type constructions to three-dimensional lattices, where two-dimensional CA can be applied and can host diverse dynamics~\cite{biswas2022mca,sfairopoulos2025cellularautomataddimensions,Devakul2019fractal}.
Given that three-dimensional lattices support fracton topological order~\cite{fu2015fracton,fu2016fracton,chen2019foliated,Shirley2018fracton,song2022fractonsubsystem,Nandkishore2019fractons,wang2026fracton,lmy2020fracton,lmy2021fracton,lmy2022fracton,lmy2023erg,li2026SEE,Ma2017fracton,Canossa2024fracton}, it would be interesting to investigate whether CA can be systematically used to design and analyze fracton topological order with subsystem symmetries.
Subsystem symmetries are also intimately related to symmetry-enriched topological (SET) order~\cite{Barkeshli2019set,Stephen2020set,Stephen2022fractionalization,zhang2025set,Hsin2025fractionalization}.
Although recent studies have shown that SET states on Euclidean lattices can be generated by uniform higher-order CA with anyon mobility governed by uniform update rules~\cite{zhang2025set}, it remains an open question whether NUCA can reveal anyon properties and mobility structures within SET states on hyperbolic lattices.
It would also be useful to investigate nonlinear NUCA, where spontaneous breaking of translation symmetry can be considered~\cite{Sfairopoulos2023,sfairopoulos2025cellularautomataddimensions,sfairopoulos2026spinmodelsnonlinearcellular}.
Finally, the NUCA framework itself raises mathematical questions.
Determining whether all subsystem symmetries on hyperbolic lattices can be generated by linear NUCA is tied to mathematical properties of NUCA such as topological transitivity~\cite{zhang2024hoca,nuca2026,ROLLIER2025108362,DENNUNZIO2024119942}.
Alongside this, a comprehensive study of Clifford QCA on hyperbolic lattices merits further effort, and may provide further insight into the interplay between many-body physics, quantum information theory, and curved geometry.

\acknowledgements
This work was supported
by the National Natural Science Foundation of China (NSFC) under Grant
No.~12474149, the Research Center for Magnetoelectric Physics of Guangdong
Province under Grant No.~2024B0303390001, and the Guangdong Provincial Key
Laboratory of Magnetoelectric Physics and Devices under Grant
No.~2022B1212010008.

\appendix

\section{Brief review of hyperbolic geometry}
\label{app_hyperbolic}
In this appendix, we review some basic mathematics for the hyperbolic plane to complete our discussion, including hyperbolic geometry and our lattice setup~\cite{Ratcliffe2019hyperbolic,Crystallography2022}.

The hyperbolic plane $\mathbb{H}^2$ is a maximally symmetric two-dimensional Riemannian manifold with constant negative Riemann curvature. 
In particular, $\mathbb{H}^2$ is the Euclidean counterpart of two-dimensional anti-de Sitter space $\text{AdS}_2$, which is a maximally symmetric Lorentzian manifold. 
The Poincar\'e disk model is a conformal disk model of two-dimensional hyperbolic geometry defined in a unit disk in the complex plane, with a nontrivial metric
\begin{equation}
	\label{metric}
	ds^2=(2\kappa)^2\frac{(d\xi)^2 + (d\eta)^2}{(1-\lvert z\rvert^2)^2}\,,
\end{equation}
where $z=\xi+\mathrm{i}\eta$, $\kappa$ is the curvature radius whose corresponding constant Riemann curvature is $K = -\kappa^{-2}$.
In this model, all points are inside a unit disk and the boundary is at hyperbolic infinity $\lvert z \rvert^2 = 1$. 
The hyperbolic distance between two points $z,z'$ within the disk is given by:
\begin{equation}
	d(z,z')=\kappa \operatorname{arcosh} \left( 1+ \frac{2 \lvert z-z' \rvert^2}{(1-\lvert z \rvert^2)(1-\lvert z' \rvert^2)} \right)\,.
\end{equation}
Within the Poincar\'e disk model, the geodesics in $\mathbb{H}^2$ correspond to circular arcs that perpendicularly intersect the boundary.

The orientation-preserving isometries of the Poincar\'e disk are fractional linear transformations
\begin{equation}
	z\mapsto Mz\coloneqq \frac{az+b}{b^*z+a^*}\,,
\end{equation}
where the complex numbers $a$ and $b$ satisfy $\lvert a\rvert^2- \lvert b\rvert^2=1$ and $M$ is a $\text{SU}(1,1)$ matrix
\begin{equation}
	M=\begin{pmatrix}
	a&		b\\
	b^*&		a^*\\
	\end{pmatrix} \,.
\end{equation}
The isometric maps preserve the hyperbolic distance under the metric in Eq.~(\ref{metric}).
These maps form the orientation-preserving isometry group of the disk model, i.e. the projective special unitary group $\text{PSU}(1,1)=\text{SU}(1,1)/\{\pm I\}$, which is isomorphic to the orientation-preserving isometry group $\text{Iso}^+(\mathbb{H}^2)$ of $\mathbb{H}^2$. 
Notably, there are two other models for hyperbolic geometry: the upper-half plane model with orientation-preserving isometry group $\text{PSL}(2,\mathbb{R})$ and the hyperboloid model with $\text{SO}^+(1,2)$, and these symmetry groups are isomorphic.

Regular tessellations of maximally symmetric Riemannian manifolds are denoted by the Schl\"afli symbol, which leads to the study of discrete symmetries of Coxeter groups.
In the case of the Poincar\'e disk model for $\mathbb{H}^2$ and tessellation of regular polygons $\{p,q\}$, the full space group of a hyperbolic lattice $\{p,q\}$ is given by a triangle group $\Delta(2,q,p)$ which is an infinite Coxeter group:
\begin{equation}
	\Delta(2,q,p) = \left< a,b,c\, |\, a^2,b^2,c^2,(ab)^2,(bc)^q,(ca)^p \right>\,,
\end{equation}
where $a,b,c$ are some reflections and the listed products are imposed as identity relations. 
As we consider the orientation-preserving operations, the isometry group is a discrete subgroup of $\text{PSU}(1,1)$ that maps the tessellation to itself. 
This subgroup is called a Fuchsian group, whose elements can be classified into the elliptic, hyperbolic and parabolic elements.
The elliptic element is
\begin{equation}
	R(\theta)=\begin{pmatrix}
	\exp(\mathrm{i}\theta/2)&		0\\
	0&		\exp(-\mathrm{i}\theta/2)\\
	\end{pmatrix} 
\end{equation}
which represents a rotation by an angle $\theta$ with $\mathrm{Tr}(R)<2$. 
A hyperbolic element representing translation is defined as
\begin{equation}
	T(\tau)=\begin{pmatrix}
	\cosh(\tau/(2\kappa))&		\sinh(\tau/(2\kappa))\\
	\sinh(\tau/(2\kappa))&		\cosh(\tau/(2\kappa))\\
	\end{pmatrix} 
\end{equation}
with the parameter $\tau>0$.
The hyperbolic elements can be regarded as a generalization of Euclidean translations to the hyperbolic case. 
For a hyperbolic element, the trace of its matrix satisfies $\mathrm{Tr}(T)>2$. 
However, all the hyperbolic elements together with the identity do not form a translation subgroup of the space group, as the product of hyperbolic elements can be an elliptic element. 
The Fuchsian translation group of the hyperbolic lattice is torsion-free, meaning that it contains no nontrivial finite-order elements, and is associated with the construction of a Bravais lattice. 
This leads to a generalized notion of translation invariance and novel properties for physical systems, particularly for non-interacting fermions~\cite{Maciejko2020HBT,Maciejko2022AHBT,Lenggenhager2023supercell,Vidal2023dos}.

A hyperbolic lattice has non-Abelian translation symmetry that is not isomorphic to $\mathbb{Z} \times \mathbb{Z}$ with a well-defined thermodynamic limit.
Remarkably, the number of vertices of a hyperbolic lattice grows exponentially with the radial distance from a chosen origin.
In contrast, the lattice size in a two-dimensional Euclidean lattice grows only polynomially with its linear size.
The mismatch of translation symmetry makes it problematic to deform and embed a hyperbolic lattice into a square lattice.
Consequently, the usual setup of uniform CA for Euclidean lattices (with or without sublattice structure) is not applicable to the hyperbolic lattice~\cite{Devakul2019fractal,zhang2024hoca,Daniel2020computational}.
We therefore develop the NUCA framework to overcome this limitation.
The construction of the deformed lattice in Sec.~\ref{sec_NUCA54} for a hyperbolic lattice based on the splitting method avoids the need to compute complex coordinates explicitly. 
To visualize the hyperbolic lattice in the Poincar\'e disk model, there are many methods to obtain the complex coordinates of the vertices.
Here we adopt the vertex-inflation method which has been extensively detailed in the literature to generate lattice vertices layer by layer~\cite{huang2025hyperbolicEE,chen_2023_h_haldane,Jahn_2020_holography,Boyle2020conformal,Schrauth_2024_hypertiling}.

\section{Applicability and consistency of the NUCA framework}
\label{app_applicability}
In this appendix, we discuss the applicability of our NUCA construction for regular Euclidean and hyperbolic lattices in detail.
Additionally, we show that our NUCA$(4,4)$ for the Euclidean square [i.e., $\{4,4\}$] lattice with uniform update rules is consistent with uniform higher-order CA that are restricted to the Euclidean lattice~\cite{Devakul2019fractal,zhang2024hoca}.

\subsection{Applicability of the NUCA construction}
In Sec.~\ref{sec_NUCA54} and Appendix~\ref{app_NUCA66}, we develop a lattice-deforming procedure for the hyperbolic lattice based on the splitting method and the language of splitting, to design the deformed lattice. 
Therefore, we first review which $\{p,q\}$ lattices the splitting method and the language of the splitting can be applied to~\cite{Margenstern2018}.
These methods were originally developed by M. Margenstern in the study of two-dimensional hyperbolic cellular automata directly defined on hyperbolic space~\cite{margenstern_1999,Margenstern_2000,MARGENSTERN2001,Margenstern2007,Margenstern2018}.
We note that our work focuses on one-dimensional non-uniform CA whose space-time configuration is mapped to a two-dimensional lattice, which exhibits distinct dynamical behaviors.

Let $X$ be a metric space that can be a Euclidean plane $\mathbb{R}^2$ or a hyperbolic plane $\mathbb{H}^2$.
Such a space can be split into finite isometric images (copies) of $\mathcal{Q}$, where $\mathcal{Q}$ is named a quarter.
Then $\mathcal{Q}$ can be split into its leading polygon $\mathcal{P}_1$ (a $p$-gon), some copies of $\mathcal{Q}$, and a remainder strip denoted as $\mathcal{S}$. 
A similar statement is valid for $\mathcal{S}$ as in Sec.~\ref{sec_NUCA54}, and the polygons form a regular tessellation of the plane.
This iterative procedure is combinatorial if the tessellation can be generated from finite rules on the leading polygon.
Then the spanning tree generated by some finite and local rules on the leading polygon $\mathcal{P}_1$ is topologically equivalent to the tessellation.
The tessellations $\{p,q\}$ with $q\ge 4$, and $\{p,3\}$ with $p\ge 7$ are combinatorial~\cite{Margenstern2007}, including both Euclidean and hyperbolic cases.

After the construction of the spanning tree, each node is labeled by a unique natural number $v=1,2,3\cdots$ similar to the example of the $\{5,4\}$ lattice.
Based on the matrix of the splitting Eq.~(\ref{splittingmatrix}), these numbers can be represented in the basis of $\{u_k\}_{k\in \mathbb{N}}$ as:
\begin{equation}
	\label{eq_representation}
	\begin{aligned}
		v = \sum_{i=0}^{i_{\max}} \alpha_i u_i\,\rightarrow\mathcal{A}(v)= \alpha_{i_{\max}}\cdots \alpha_1\alpha_0 \,,
	\end{aligned}
\end{equation}
where $\alpha_i\in\{0,1,\cdots, (p-3)(\lfloor q/2 \rfloor-1+\epsilon)\},\,\epsilon = q \bmod 2$ are coefficients determined by the lattice $\{p,q\}$.
The representation $\mathcal{A}(v)$ with largest possible $i_{\max},\alpha_{i_{\max}}\ne 0$ labels the location of nodes. 
The representation $\mathcal{A}(v)$ is denoted as the standard language of the splitting. 
In general, the representation of a positive integer in the basis of $\{u_k\}$ may not be unique.
We choose $\mathcal{A}(v)$ with the longest length as the standard representation that is determined by the matrix of the splitting.

A language is regular if a legitimate coordinate can be accepted by a finite automaton, and thus the addition is efficient.
As proved by M. Margenstern, for $\{p,q\}$ with $p\ge4,q\ge 4$ the language of the splitting is regular, and there is an algorithm linear in the coordinate length to calculate the location of a node~\cite{Margenstern2007}.
We require that the neighboring relation of an arbitrary node can be computed with memory usage that does not scale too rapidly [e.g., logarithmic to the system size], such that we need not store the neighboring relation of nodes. 
It is helpful to introduce \textit{parent} and \textit{child} to denote the relationship of nodes in the offspring-based generation.
If the language of the splitting is regular, the neighboring relation can be obtained based on the parent and child relation.

From these results, the splitting method is combinatorial and the language is regular for a hyperbolic $\{p\ge4,q\ge4\}$ lattice.
Therefore, we can use the lattice-deforming procedure introduced in Sec.~\ref{sec_NUCA54} for these $\{p\ge4,q\ge4\}$ lattices and perform NUCA.
Based on the construction, a non-uniform NUCA update rule can be defined and calculated in the thermodynamic limit.
Both the $\{5,4\}$ and $\{6,6\}$ tessellations studied in detail in this paper fulfill these conditions, and thus NUCA construction can be implemented directly. 
However, the sublattice structure and the neighboring relation should be specified separately, while the translation symmetry should be imposed as a constraint on the update rule.
We also note that the applicability of our NUCA construction is not equivalent to that of the splitting method.
By employing alternative coordinate systems, this construction can be extended to other lattice geometries.

\subsection{A Euclidean example of the splitting method}
\begin{figure*}[t]
	\includegraphics[width=0.85\textwidth]{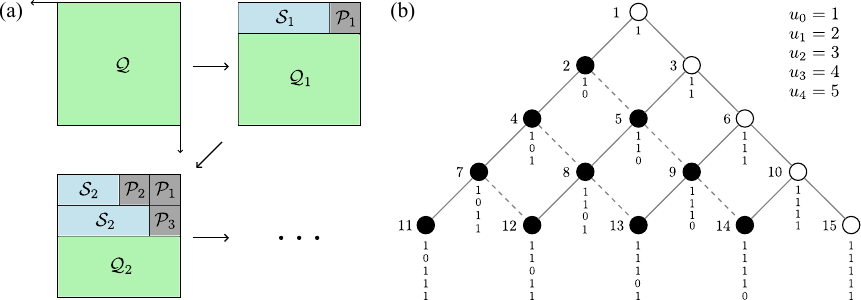}
	\caption{\label{fig_los44} 
	(a) The splitting method applied to a quarter of the $\{4,4\}$ lattice. Green regions denote copies of the quarter, blue regions denote strips, and gray regions denote the leading polygons. 
	Notations are defined in the main text.
	(b) The spanning tree of the $\{4,4\}$ lattice. Nodes are uniquely assigned discrete indices represented in the basis of $\{u_k\}$. 
	The neighboring relation is represented by lines (offspring) and dashed lines (non-offspring).}
\end{figure*}

As a simple example of how the splitting method works, we apply it to the Euclidean square lattice and construct the language of the splitting.
Consider a quarter $\mathcal{Q}$ of an infinite two-dimensional square lattice tessellated by regular rectangles, each of which is labeled as a node. 
Without loss of generality, we set the quarter in $x<0,y<0$ of $\mathbb{R}^2$, which is shown in Fig.~\ref{fig_los44}(a).
By choosing the leading rectangle related to the origin as $\mathcal{P}_1$, $\mathcal{Q}$ can be split into $\mathcal{P}_1$, a copy of $\mathcal{Q}$ as $\mathcal{Q}_1$, and a strip $\mathcal{S}_1$. 
For the strip $\mathcal{S}_1$, by choosing the leading rectangle as $\mathcal{P}_2$, it can be split into a leading rectangle $\mathcal{P}_2$ and a remaining strip $\mathcal{S}_2$. 
Thereafter, the strips and quarters can be split in the same way. 
We denote the leading rectangle of a quarter as a white node 
\begin{tikzpicture}[baseline=-0.5ex]
    \node[wcirc] (c1) {};
\end{tikzpicture} and the leading rectangle of a strip as a black node \begin{tikzpicture}[baseline=-0.5ex]
    \node[bcirc] (c1) {};
\end{tikzpicture} as in the main text.
Then the generating rule is \begin{tikzpicture}[baseline=-0.5ex, node distance=0.3ex]
    \node[wcirc] (c1) {};
    \node[right=of c1] (arrow) {$\rightarrow$};
    \node[bcirc, right=of arrow] (c2) {};
    \node[wcirc, right=of c2] (c3) {};
\end{tikzpicture}
and 
\begin{tikzpicture}[baseline=-0.5ex, node distance=0.3ex]
    \node[bcirc] (c1) {};
    \node[right=of c1] (arrow) {$\rightarrow$};
    \node[bcirc, right=of arrow] (c2) {};
\end{tikzpicture}. 
The splitting matrix Eq.~(\ref{splittingmatrix}) is given by:
\begin{equation}
	\label{splittingmatrix_44}
	S=\begin{pmatrix}
	1&		1\\
	0&		1\\
	\end{pmatrix} \,.
\end{equation}
The characteristic polynomial of Eq.~(\ref{splittingmatrix_44}) is $x^2 - 2x+1$ with the positive root $1$. This leads to the following relation for the number of nodes on each level $u_{k+2} = 2u_{k+1} -u_k$. Considering that $u_0=1$ and $u_1=2$, we have
\begin{equation}
	u_k = k+1\,,\quad k=0,1,2\cdots\,.
\end{equation}

Now we can assign a positive integer number $v=1,2,\cdots$ as indices to each rectangle $\mathcal{P}$.
The numbers can be represented in the basis of $\{u_k\}$.
The spanning tree is visualized in Fig.~\ref{fig_los44}(b), where we have represented the index $v = \sum_{i=0}^{i_{\max}} {\alpha_i u_i},\, \alpha_i \in \{0,1\}, u_i\in \{u_k\}$. 
The representation of nodes on the first two levels can be uniquely determined. 
By directly using mathematical induction, the following properties are easily derived: 
(\rmnum{1}) For a white node $v$, $\mathcal{A}(v)$ contains no $0$, whereas that for a black node $\mathcal{A}(v)$ contains exactly one $0$. The representation of the parent-node $p(v)$ of $v$ is $\mathcal{A}(p(v))=\mathcal{A}(v')$ where $\mathcal{A}(v)=\mathcal{A}(v')b,b\in\{0,1\}$.
(\rmnum{2}) For a white node $v$ represented as $\mathcal{A}(v)$, its black-node child is represented as $\mathcal{A}(v)0$, while that of its white-node child is $\mathcal{A}(v)1$. In the case of a black node, its only black-node child is $\mathcal{A}(v)1$. 
The child-node ending in $1$ with representation ending in $1$ is denoted as $c(v)$ for both white and black nodes $v$.
The neighboring relation of nodes is summarized in Table~\ref{tab_neighborhood44}.
It is worth noting that the nodes in odd/even levels have neighboring nodes in even/odd levels only, a sublattice structure similar to the case on the hyperbolic $\{5,4\}$ lattice.

\begin{table}[b]
\caption{\label{tab_neighborhood44} Neighboring nodes of a node $v$ in $\{4,4\}$ spanning tree. $p(v)$ labels the parent node of a node $v$, while $c(v)$ labels the child-node whose representation ends in $1$. The node $p(v)+1$ in the second line lies in the neighboring quarter. }
\begin{ruledtabular}
\begin{tabular}{ccccc}
	Node $v$ & \multicolumn{4}{c}{Neighboring nodes} \\
	\hline
	black node \begin{tikzpicture}[baseline=-0.5ex, node distance=0.3ex]
		\node[bcirc] (c1) {};
	\end{tikzpicture} & $p(v)$ & $p(v)-1$ & $c(v)$ & $c(v)+1$  \\
	white node \begin{tikzpicture}[baseline=-0.5ex, node distance=0.3ex]
		\node[wcirc] (c1) {};
	\end{tikzpicture} & $p(v)$ & $p(v)+1$ & $c(v)-1$ & $c(v)$  \\
\end{tabular}
\end{ruledtabular}
\end{table}

\subsection{Euclidean consistency of NUCA$(4,4)$}
Next, we elaborate details and properties related to NUCA$(4,4)$ in Sec.~\ref{ssec_nuca44}, where we defined NUCA$(4,4)$ directly on the square lattice. 
The unit translation of a site in terms of the polynomial representation $x^iy^j$ is realized by multiplication by monomials in $\{x,y,x^{-1},y^{-1}\}$. 
The neighboring sites of a site $x^iy^j$ are obtained by a one-unit translation, i.e., $\{x^iy^{j-1},x^iy^{j+1},x^{i-1}y^j,x^{i+1}y^j\}$, which are all independent of the position. 
Therefore, if the Hamiltonian is translationally invariant, the associated update rule is a function vector of $x$ as:
\begin{equation}
	\label{Euclidean_update_rule}
	\mathbf{f}(x) \coloneqq (\mathrm{f}_1(x),\mathrm{f}_2(x),\cdots ,\mathrm{f}_n(x))^T \,,
\end{equation}
which is irrelevant to $i,j$ and thus $n$ is a constant. 
Thus, we focus on the uniform update rule $\mathbf{f}(x)$, and the transition operator $F^{(k)}_j$ introduced in Eq.~(\ref{eq_transition_operator}) is irrelevant to $j$ as $F^{(k)}$. 
As the rule is uniform, the NUCA$(4,4)$ is consistent with uniform higher-order CA in the sense that their evolution configurations are in one-to-one correspondence~\cite{zhang2024hoca,zhang2025set}, while the latter introduce supercells for sublattice structure.

A Toeplitz matrix $\mathsf{T}$ is defined as $\mathsf{T}_{ab}=\mathsf{T}_{a-b}$, which is intimately related to the Euclidean translation symmetry~\cite{Boxi2024toeplitz,Boxi2025toeplitz,Lee_2014,lee_ye_2015_wannier}.
For a uniform update rule of NUCA$(4,4)$, the action of $F^{(k)}$ on $x^iy^j$ influences the states of sites in $r_{j+k}(x)$:
\begin{equation}
	F^{(k)}[x^i] = \sum_{m} c_{k;i,m} x^m = \sum_{m} c_{k;i-m} x^m = x^i \mathrm{f}_{k}(x)\,,\quad \forall i\,.
\end{equation}
The matrix element $(\mathbf{F}^{(k)})_{a+c,b+c} = c_{k;a+c,b+c}= c_{k;a-b}=(\mathbf{F}^{(k)})_{a,b}$.
From the perspective of linear NUCA, the matrix representation of transition operator $\mathbf{F}^{(k)}$ is a Toeplitz matrix for all $k$ provided that it is a nonzero matrix. 
Such a Toeplitz form of $\mathbf{F}^{(k)}$ facilitates an underlying relation between $\mathbf{f}(x)$ and $\bar{\mathbf{f}}(x)$. 
As $[(\mathbf{F}^{(k)})^T]_{ij}=(\mathbf{F}^{(k)})_{ji}$, this leads to the action of $\bar{F}^{(k)}$ on $x^iy^j$ as
\begin{equation}
	\bar{F}^{(k)} [x^i] = \sum_{m} \bar{c}_{k;i,m} x^m = \sum_{m} c_{k;m,i} x^m = x^i \bar{\mathrm{f}}_{k}(x)\,,\quad \forall i\,,
\end{equation}
and thus $x^i \bar{\mathrm{f}}_{k}(x) = \sum_{m} c_{k;m-i} x^m$.
If $c_{k;i,l}=1$ such that $\mathrm{f}_{k}(x)=\cdots + x^{l-i}+\cdots$, we also have $\bar{\mathrm{f}}_{k}(x)=\cdots + x^{i-l}+\cdots$ for any $l$. This is satisfied if and only if $\bar{\mathrm{f}}_{k}(x) = \mathrm{f}_{k}(x^{-1})$. 
Therefore, the (transposed) update rule automatically satisfies the constraint
\begin{equation}
	\label{eq_f_inverse}
	\bar{\mathbf{f}}(x) = \mathbf{f}(x^{-1})\,,
\end{equation}
such that the orders $n=\bar{n}$ are constant and treated as equivalent.
This is exactly the property of the update rule of uniform CA studied in the literature~\cite{zhang2024hoca,Devakul2019fractal}. 
Under the uniform update rule, the Hamiltonian terms are uniformly defined at different lattice sites and thus translationally invariant. 
From the definition in Eq.~(\ref{eq_initial_condition_NUCA}), sites with $\bar{\mathrm{f}}_k(x) = 0,\forall k$ can be specified as initial conditions.
Only boundary sites $x^iy^j$ with $j<n$ satisfy this condition, because a nontrivial $\bar{\mathbf{f}}(x)$ of order-$n$ cannot be applied to them. Therefore, these sites can be assigned initial conditions. 

We further discuss the proof of the commutation relation between the symmetry and the Hamiltonian. 
In the semi-infinite lattice where $j\ge0$, we consider a general Hamiltonian and the corresponding symmetry $X(\mathscr{F}(x,y))$. 
The commutation polynomial for a Hamiltonian term and a symmetry operator is
\begin{equation}
	\begin{aligned}
		&\mathrm{C}(\mathscr{F}(x,y), x^iy^j(1+\bar{\mathbf{f}}\cdot\bar{\mathbf{y}}_{1,\bar{n}})) \\
		& = x^{-i}y^{-j} (1+\mathbf{f}\cdot\mathbf{y}_{1,n}) \sum_{k=0}^{\infty} y^k r_k(x)  \\
		& =  x^{-i}y^{-j} \left[\sum_{k=0}^{\infty} y^k r_k(x) + \sum_{k=n}^{\infty} y^k r_k(x) + \sum_{k=0}^{n-1} y^k \tilde{r}_k(x)\right]  \\
		& =  x^{-i}y^{-j} \left[\sum_{k=0}^{n-1} y^k r_k(x) + \sum_{k=0}^{n-1} y^k \tilde{r}_k(x)\right]  \, ,
		\label{symmetry_commupoly}
	\end{aligned}
\end{equation}
where we consider the action $Z(x^iy^j(1+\bar{\mathbf{f}}\cdot\bar{\mathbf{y}}_{1,\bar{n}}))$ on the same sublattice with the symmetry.
In the last line $\tilde{r}_k(x)$ is defined as:
\begin{equation}
	\label{r_res}
	\sum_{k=0}^{n-1} y^{k} \tilde{r}_k(x) \equiv  \sum_{k=0}^{\infty} y^{k} r_{k}(x) \mathbf{f} \cdot\mathbf{y}_{1,n} - \sum_{k=n}^{\infty} y^{k} r_{k}(x)\, .
\end{equation}
The first term in the r.h.s. of Eq.~(\ref{r_res}) reproduces the second term for $k\ge n$, which is by definition the time evolution of CA. Thus Eq.~(\ref{r_res}) only contains terms with $y$-power lower than $n$. 
Therefore, $[\mathrm{C}(\mathscr{F}(x,y), x^iy^j(1+\bar{\mathbf{f}}\cdot\bar{\mathbf{y}}_{1,\bar{n}}))]_{x^0y^0}=0$ if we exclude all terms with support $x^iy^j(1+\bar{\mathbf{f}}\cdot\bar{\mathbf{y}}_{1,\bar{n}})$ extending outside the lattice, and all Hamiltonian terms in Eq.~(\ref{eq_44_SSPT}) commute with the symmetry.

Next we consider the construction of the formalism commonly adopted for uniform CA in the literature~\cite{Devakul2019fractal,zhang2024hoca}, where the sublattices $(a)$ and $(b)$ are represented by supercell structure.
To apply such evolution, each supercell $x^iy^j$ of the square lattice consists of two neighboring sites belonging to different sublattices $(a)$ and $(b)$, and the lattice topology is still $\mathbb{Z}\times \mathbb{Z}$.
Therefore, the Hamiltonian Eq.~(\ref{eq_44_SSPT}) is rewritten in the following form
\begin{equation}
	\begin{aligned}
		\mathscr{H}=&-\sum_{i,j}Z^{(a)}(x^{i}y^{j}(1+\bar{\mathbf{f}}\cdot\bar{\mathbf{y}}))Z^{(b)}(x^{i}y^{j})\\
		&-\sum_{i,j}X^{(b)}(x^{i}y^{j}(1+\mathbf{f}\cdot\mathbf{y}))X^{(a)}(x^{i}y^{j}) \, ,
	\end{aligned}
\end{equation}
where $i,j$ label supercells and $O^{(a)}$ labels a Pauli operator defined on one sublattice.
In our NUCA formalism, by mapping two sites $x^{2k}y^{2l}$ and $x^{2k}y^{2l-1}$ ($x^{2k+1}y^{2l+1}$ and $x^{2k+1}y^{2l}$) to a new supercell at $x^ky^{2l}$ ($x^ky^{2l+1}$), Eq.~(\ref{eq_44_SSPT}) reduces to the above Hamiltonian.
From these results, our NUCA construction can also be applied to the translationally invariant Euclidean lattice.
With uniform update rules, the NUCA$(4,4)$ reduces to uniform higher-order CA and agrees with the standard Euclidean formalism~\cite{Devakul2019fractal,zhang2024hoca}.

\section{Details of NUCA$(5,4)$ for hyperbolic $\{5,4\}$ lattice}
\label{app_NUCA54}
In Sec.~\ref{sec_NUCA54}, we designed a deformed lattice for constructing the linear and probabilistic NUCA$(5,4)$ for the hyperbolic $\{5,4\}$ lattice.
We define auxiliary functions that incorporate geometric data into the update rules. 
This incorporation leads to the non-uniformity of NUCA$(5,4)$ as the deformed lattice induces distortion. 
By using these auxiliary functions, we define update rules that preserve physical locality on the hyperbolic lattice.
For complex update rules we only present their transposed form in the main text, as the rule is completely fixed when the transposed rule is specified. 
In this appendix, we prove and derive the auxiliary functions, and provide these (transposed) update rules as well as the calculation for the commutation for completeness.

\subsection{Derivation of auxiliary functions}
First we provide the details of obtaining the neighboring relation of nodes.
For the $\{5,4\}$ lattice, we use an alternative Fibonacci representation rather than the standard language of the splitting to simplify calculation~\cite{Margenstern2007}.
We recall that the generating rule is 
\begin{tikzpicture}[baseline=-0.5ex, node distance=0.3ex]
    \node[wcirc] (c1) {};
    \node[right=of c1] (arrow) {$\rightarrow$};
    \node[bcirc, right=of arrow] (c2) {};
    \node[wcirc, right=of c2] (c3) {};
    \node[wcirc, right=of c3] (c4) {};
\end{tikzpicture} and 
\begin{tikzpicture}[baseline=-0.5ex, node distance=0.3ex]
    \node[bcirc] (c1) {};
    \node[right=of c1] (arrow) {$\rightarrow$};
    \node[bcirc, right=of arrow] (c2) {};
    \node[wcirc, right=of c2] (c3) {};
\end{tikzpicture} in Fig.~\ref{fig_los54}.
On the $\{5,4\}$ spanning tree, if a node $v$ is black, its preferred child-node $c(v)$ is the leftmost child.
If a node $v$ is white, its preferred child-node $c(v)$ is the middle child.
For the spanning tree of the $\{5,4\}$ lattice and the Fibonacci representation, we have the following preferred child property for nodes~\cite{margenstern_1999,Margenstern_2000,MARGENSTERN2001,Margenstern2007,Margenstern2018}:

\textit{Theorem 1.} In the Fibonacci representation, for a parent-node represented as $\alpha_{l}\alpha_{l-1}\cdots \alpha_2\alpha_1$, the node represented as $\alpha_{l}\alpha_{l-1}\cdots \alpha_2\alpha_100$ is its preferred child-node.

\textit{Corollary 1.} In the Fibonacci representation, if the representation of a node $v$ ends in $01$ ($10$), the node is white (black). Meanwhile, if a node's representation ends in $00$, then it is white (black) if $v-1$ is black (white).  

\textit{Corollary 2.} In the Fibonacci representation, the parent-node $p(v)$ of a node $v$ represented as $\mathcal{A}(v) = \mathcal{A}(u)\alpha_1\alpha_0$ is $p(v) = u+\alpha_1$, where $u$ is a node that can be calculated from the Fibonacci representation $\mathcal{A}(u)$.

In the $\{5,4\}$ spanning tree, the parent- and child-nodes are also neighboring nodes.
Following these results, the neighboring relation can be readily calculated and is summarized in Table~\ref{tab_neighborhood54}.

\begin{table}[b]
\caption{\label{tab_neighborhood54} Neighboring nodes of a node $v$ in $\{5,4\}$ spanning tree. $p(v)$ labels the parent node of a node $v$, while $c(v)$ labels the preferred-child node.}
\begin{ruledtabular}
\begin{tabular}{cccccc}
	Node $v$ & \multicolumn{5}{c}{Neighboring nodes} \\
	\hline
	black node \begin{tikzpicture}[baseline=-0.5ex, node distance=0.3ex]
		\node[bcirc] (c1) {};
	\end{tikzpicture} & $p(v)$ & $p(v)-1$ & $c(v)$ & $c(v)+1$ & $c(v)+2$ \\
	white node \begin{tikzpicture}[baseline=-0.5ex, node distance=0.3ex]
		\node[wcirc] (c1) {};
	\end{tikzpicture} & $p(v)$ & $c(v)-1$ & $c(v)$ & $c(v)+1$ & $c(v)+2$ \\
\end{tabular}
\end{ruledtabular}
\end{table}

Now we prove that there is a sequence to determine whiteness/blackness of physical sites for the $x$-coordinate on the deformed lattice. 
The physical sites are the first $u_k = A_{2k+1}$ sites in each $x^k$, which are mapped to polygons of the $\{5,4\}$ lattice.
The white/black sequence of physical sites in $x^k$ is identical to the white/black sequence of the first $A_{2k+1}$ sites in $x^{k+1}$ and thus is a prefix.
To prove this, we notice that the generating rule $\sigma$ can be written as 
\begin{tikzpicture}[baseline=-0.5ex, node distance=0.3ex]
    \node[wcirc] (c1) {};
    \node[right=of c1] (arrow) {$\rightarrow$};
    \node[wcirc, right=of arrow] (c2) {};
    \node[wcirc, right=of c2] (c3) {};
    \node[bcirc, right=of c3] (c4) {};
\end{tikzpicture}
and 
\begin{tikzpicture}[baseline=-0.5ex, node distance=0.3ex]
    \node[bcirc] (c1) {};
    \node[right=of c1] (arrow) {$\rightarrow$};
    \node[wcirc, right=of arrow] (c2) {};
    \node[bcirc, right=of c2] (c3) {};
\end{tikzpicture} for deformed lattice. 
We denote the white/black sequence of $x^k$ as $L_k$ and the generating process is denoted as $L_{k+1} = \sigma(L_k)$.

We start from the fact that, during the generation process according to the generating rules, the physical sites of $x^1$ are $L_1 = \sigma(L_0) = 
\begin{tikzpicture}[node distance=0.3ex]
    \node[wcirc] (c1) {}; 
    \node[wcirc, right=of c1] (c2) {}; 
    \node[bcirc, right=of c2] (c3) {}; 
\end{tikzpicture} = L_0\cdot \begin{tikzpicture}[node distance=0.3ex]
    \node[wcirc] (c1) {}; 
    \node[bcirc, right=of c1] (c2) {}; 
\end{tikzpicture}$ and thus the sequence of physical sites in $x^0$ is contained in that of $x^1$. 
Now, the physical sites in the $x^{k+1} (k\ge 1)$ are generated from physical sites in the $x^{k}$ according to the same rule, i.e., $L_{k+1} = \sigma(L_k)$.
We assume the white/black sequence of the first $A_{2k+1}$ sites in the $x^{k+1}$ is identical to the sequence of all physical sites in the $x^{k}$, i.e., $L_{k+1} = L_k\cdot S_{k+1}$, where $S_{k+1}$ is a unique sequence in $x^{k+1}$ that can be determined but is not relevant.
By definition, for $x^{k+2}$ we have $L_{k+2} = \sigma(L_{k+1}) = \sigma(L_k\cdot S_{k+1})$. 
As the generating rule can be applied under concatenation, we obtain $L_{k+2} =\sigma(L_k)\cdot \sigma(S_{k+1}) = L_{k+1}\cdot \sigma(S_{k+1})$. 
Consequently, the sequence of the first $A_{2k+3}$ sites in $x^{k+2}$ is identical to the entire sequence of physical sites in $x^{k+1}$.
Using mathematical induction, we establish that the sequence $L_k$ of $x^{k}$ is a prefix of the sequence of $L_{k+1}$ of $x^{k+1}$ for all $k\ge 1$. Therefore, the generated lattice as a whole possesses self-replication.

Then we prove that the whiteness/blackness sequence in the deformed lattice is identical to the sequence of nodes of the spanning tree, up to a constant shift. 
That is, we can assign a representation $\mathcal{A}^{\text{dfm}}(y^j) = \mathcal{A}(j)$ to sites with $j=1,2,3\cdots$ on the deformed lattice, and the neighboring relation of $x^iy^j$ can be derived from the spanning tree.
We still proceed with the proof of this result by mathematical induction. The physical sites of $x^0,x^1$ are $L_0 = \begin{tikzpicture}[]
    \node[wcirc] {}; 
\end{tikzpicture}$ and $L_1 = \sigma(L_0) = 
\begin{tikzpicture}[node distance=0.3ex]
    \node[wcirc] (c1) {}; 
    \node[wcirc, right=of c1] (c2) {}; 
    \node[bcirc, right=of c2] (c3) {}; 
\end{tikzpicture}$ which is generated according to $L_0$ under rule $\sigma$. 
Upon removal of the first site $x^1y^0$ from $L_1$, the remaining sequence $L_1^{\text{cut}} = \begin{tikzpicture}[node distance=0.3ex]
    \node[wcirc] (c1) {}; 
    \node[bcirc, right=of c1] (c2) {}; 
\end{tikzpicture}$ is identical to the sequence of nodes in the spanning tree.
The sequence in $L_{k+1}$ $(k\ge 1)$ is generated according to $L_k$ by $L_{k+1} = \sigma(L_k)$. 
Now we assume that upon removal of the first site from $L_{k+1}$, the remaining sequence $L_{k+1}^{\text{cut}}$ is identical to the sequence of nodes in the spanning tree, i.e., $L_{k+1} =\begin{tikzpicture}[]
    \node[wcirc] {}; 
\end{tikzpicture} \cdot L_{k+1}^{\text{cut}}$. 

We define $\sigma'$ as the generating rule of nodes in the spanning tree such that \begin{tikzpicture}[baseline=-0.5ex, node distance=0.3ex]
    \node[wcirc] (c1) {};
    \node[right=of c1] (arrow) {$\rightarrow$};
    \node[bcirc, right=of arrow] (c2) {};
    \node[wcirc, right=of c2] (c3) {};
    \node[wcirc, right=of c3] (c4) {};
\end{tikzpicture}
and 
\begin{tikzpicture}[baseline=-0.5ex, node distance=0.3ex]
    \node[bcirc] (c1) {};
    \node[right=of c1] (arrow) {$\rightarrow$};
    \node[bcirc, right=of arrow] (c2) {};
    \node[wcirc, right=of c2] (c3) {};
\end{tikzpicture}.
Moreover, by reordering the sequence it can be proved that $\begin{tikzpicture}[]
    \node[bcirc] {}; 
\end{tikzpicture} \cdot \sigma(L) \equiv \sigma'(L) \cdot \begin{tikzpicture}[]
    \node[bcirc] {}; 
\end{tikzpicture}$.
By definition, $L_{k+2} =\sigma(L_{k+1})= \sigma(\begin{tikzpicture}[]
    \node[wcirc] {}; 
\end{tikzpicture}) \cdot \sigma(L_{k+1}^{\text{cut}})$. 
For the first site \begin{tikzpicture}[]
    \node[wcirc] {}; 
\end{tikzpicture}, we have $\sigma(\begin{tikzpicture}[]
    \node[wcirc] {}; 
\end{tikzpicture}) = \begin{tikzpicture}[baseline=-0.5ex, node distance=0.3ex]
    \node[wcirc] (c1) {};
    \node[wcirc, right=of c1] (c2) {};
    \node[bcirc, right=of c2] (c3) {};
\end{tikzpicture}$.
Therefore, $L_{k+2} = \begin{tikzpicture}[baseline=-0.5ex, node distance=0.3ex]
    \node[wcirc] (c1) {};
    \node[wcirc, right=of c1] (c2) {};
\end{tikzpicture}\cdot \begin{tikzpicture}[]
    \node[bcirc] {}; 
\end{tikzpicture} \cdot \sigma(L_{k+1}^{\text{cut}}) = \begin{tikzpicture}[baseline=-0.5ex, node distance=0.3ex]
    \node[wcirc] (c1) {};
    \node[wcirc, right=of c1] (c2) {};
\end{tikzpicture} \cdot \sigma'(L_{k+1}^{\text{cut}})\cdot \begin{tikzpicture}[]
    \node[bcirc] {}; 
\end{tikzpicture}$. 
As $L_{k+1}^{\text{cut}}$ contains a sequence of nodes in the spanning tree starting from $1$ and $\sigma'$ is the generating rule for the spanning tree, $\sigma'(L_{k+1}^{\text{cut}})$ is exactly a sequence of nodes in the spanning tree starting from $2$. 
Together with a prefix \begin{tikzpicture}[]
    \node[wcirc] {}; 
\end{tikzpicture} representing node $v=1$, as well as a suffix \begin{tikzpicture}[]
    \node[bcirc] {}; 
\end{tikzpicture} as the black-child node of the sequential node of $L_{k+1}^{\text{cut}}$, the sequence $L_{k+2}$ with the removal of the first \begin{tikzpicture}[]
    \node[wcirc] {}; 
\end{tikzpicture} node is a sequence of the spanning tree starting from $1$. 
Using mathematical induction and the results we obtained above, we can assign a representation $\mathcal{A}^{\text{dfm}}(j) = \mathcal{A}(j),j=1,2,3,\cdots$ for every physical site $x^iy^j$ on the deformed lattice.

Now we show that the index $v_n$ of the $n$-th black node in the spanning tree can be directly determined by the sequence $v_n= \lfloor \phi^2n \rfloor,\, n=1,2,3,\cdots$, where $\phi=(1+\sqrt{5})/2$ is the golden ratio and $\lfloor \cdot \rfloor $ is the floor function.
In the Fibonacci representation Eq.~(\ref{eq_Fibonacci_Rep}), the preferred child of a node is equivalent to a basis shifted by two units. 
Terms in the Fibonacci sequence $A_n$ can be given by Binet's formula:
\begin{equation}
	\label{Binet}
	A_n = \frac{1}{\sqrt{5}}\left[\left(\frac{1+\sqrt{5}}{2}\right)^n - \left(\frac{1-\sqrt{5}}{2}\right)^n\right]\,.
\end{equation}
As the absolute value of the second term in the r.h.s. of Eq.~(\ref{Binet}) is less than $1$, for sufficiently large $n$, a two-unit shift of the representation of $n$ in the Fibonacci basis is approximately equivalent to multiplication by $\phi^2$, i.e., $A_{n+2} \approx \phi^2 A_n$ and thus the preferred child of a node $v$ is approximated by $c(v) \approx \phi^2 v,\, v\rightarrow \infty$. 
The final precise result will differ by a rounding operation, which is realized by either the floor $\lfloor \cdot \rfloor $ or ceiling function $\lceil \cdot \rceil$. 

We denote $b_v=1$ if a node $v$ is black while $b_v=0$ if it is white.
For a node $v>1$, the number of black nodes in the preceding $v-1$ nodes is given by $\sum^{v-1}_{u=1}b_u=\lfloor v / \phi^2 \rfloor$.
Here, $\lfloor v / \phi^2 \rfloor$ counts the number of nodes whose preferred child index is smaller than $v$, as a black node is either a preferred child or its index minus one.
Through the generating rule $\sigma'$ of the spanning tree, the first child-node $c_1(v)$ of $v$ is 
\begin{equation}
	c_1(v) = 2 + \sum^{v-1}_{u=1}(3-b_u)=2+3(v-1)-\lfloor v / \phi^2 \rfloor\,,
\end{equation}
where the constant $2$ comes from the generating rule $\sigma'$ rather than $\sigma$.
We notice that $ 1/ \phi^2 = 3-\phi^2$, thus $\lfloor \phi^2 v \rfloor = \lfloor 3v - v / \phi^2 \rfloor = 3v - \lfloor v / \phi^2 \rfloor - 1$.
Therefore, the first child-node $c_1(v)$ of $v$ is:
\begin{equation}
	c_1(v) = \lfloor \phi^2 v \rfloor\,.
\end{equation}
Considering that the first child-node of both a white and black node is a black node, we derive the result.
From these results, we find that for a black node $v$, its preferred child is given by $\lfloor \phi^2v \rfloor$. 
For a white node $v$, the preferred child is $\lceil \phi^2v \rceil$ and therefore, its black child is also $\lfloor \phi^2v \rfloor$. 
That is, all black nodes are obtained from both black and white nodes $v$ by $\lfloor \phi^2v \rfloor$. 
The sequence $v_n= \lfloor \phi^2n \rfloor,\, n=1,2,3,\cdots$ completely determines the sequence of black nodes in the spanning tree.

We use some auxiliary functions for NUCA$(5,4)$ in Sec.~\ref{sec_NUCA54}: $J(j)$ defined in Eq.~(\ref{auxiliaryJ}), $P(j)$ defined in Eq.~(\ref{auxiliaryP}), and $C(j)$ defined in Eq.~(\ref{auxiliaryC}).
These functions contain scaling of the coordinate, capturing the exponential growth of the lattice size of the hyperbolic lattice.
Based on the conclusions above, we can derive these functions directly.
As they only involve multiplication, floor and ceiling operations of integers, the implementation of these functions in numerical computation becomes straightforward. 

The sequence of sites corresponding to black nodes, which is equivalent to the function $J(j)$, can be derived from the sequence $j_n$ established above. 
The positions of physical sites $x^iy^{j_n}$ that correspond to black nodes are directly given by $j_n = \lfloor \phi^2 n\rfloor,n=1,\cdots,A_{2i-1}$.
Because each site $x^iy^{j_n}$ with $j_n= \lfloor \phi^2n \rfloor$ has a parent site $x^{i-1}y^{n-1}$, each site $x^{i-1}y^{j}$ has a black-child site $x^iy^{\lfloor \phi^2 (j+1)\rfloor}$.
Therefore, if a site $x^iy^j$ is black, it should satisfy the condition $j=\lfloor \phi^2 \lceil j/\phi^2 \rceil \rfloor$. Note that $j$ starts from $1$ and we define $J(0) = 0$.

The function $P(j)$ can be obtained by noticing the parent-child relation in the original spanning tree sequence.
The parent node of a node $v$ is $p(v) = \lceil v/\phi^2 \rceil$ if $v$ is black, while $p(v) = \lfloor v/\phi^2 \rfloor$ if $v$ is a sibling white node of a black node $v-1$. 
As $j$ starts from $0$ in the deformed lattice, for a black site $x^iy^j$ we have $P(j) = \lceil j/\phi^2 \rceil - 1=\lfloor j/\phi^2 \rfloor$. 
A white site $x^iy^j$ as a sibling child of a black site $x^iy^{j+1}$ has the same parent site. 
Since all the white sites $x^iy^j$ are bounded by some black site and the transformation $j\rightarrow \phi^2 j$ is monotonic, $P(j)$ for a white site $x^iy^j$ which is a sibling site of a white site $x^iy^{j+1}$ is straightforward.
Therefore, we conclude that $P(j) = \lfloor j/\phi^2 \rfloor$ if $J(j)=0$ as in Eq.~(\ref{auxiliaryP}). 

The preferred-child function $C(j)$ is straightforward as the black sequence is known. For any site $x^iy^j$, its black-child is given by $x^{i+1}y^{\lfloor \phi^2 (j+1)\rfloor}$. For a black site $x^iy^j$ the black-child itself is preferred, while for a white site $x^iy^j$ the middle white child $x^{i+1}y^{\lfloor \phi^2 (j+1)\rfloor -1}$ is preferred.
Therefore, we obtain $C(j)=\lfloor \phi^2 (j+1)\rfloor$ if $J(j)=1$ and $C(j)=\lfloor \phi^2 (j+1)\rfloor -1$ if $J(j)=0$ in Eq.~(\ref{auxiliaryC}).

\subsection{Details of NUCA$(5,4)$-generated Hamiltonians}
\label{app_update_rule_forms}
We first verify that the NUCA$(5,4)$-generated symmetry elements Eq.~(\ref{symmetry}) commute with all the Hamiltonian terms Eq.~(\ref{eq_54_SSPT}). 
We consider the nontrivial action of a Hamiltonian term $Z(x^iy^j(1+\bar{\mathbf{f}}_j\cdot\bar{\mathbf{y}}_{1,\bar{n}}))$ on the same sublattice with the symmetry $X(\mathscr{F}(x,y)) = X(\sum_k  r_k(x)y^k)$. 
We note that the update rule and its transposed form uniquely determine each other as discussed in Sec.~\ref{sec_preliminaries}.
The state $a_{ij}$ of $x^iy^j$ in $\mathscr{F}(x,y)$ is given by Eq.~(\ref{eq_state_evolution_54}). 
For all sites $x^iy^j$ with nontrivial $\bar{\mathbf{f}}_j$ that define a Hamiltonian term, $a_{ij}$ is given by the evolution of NUCA:
\begin{equation}
	a_{ij} = \left( \sum_{k=1}^{\bar{n}(j)}{\sum_{m:x^m\in \left\{ x^i\bar{\mathrm{f}}_{j,k} \right\}}{a_{m,j-k}}} \right)\, \bmod 2 \,.
\end{equation} 
We then examine the commutation relation through the commutation polynomial:
\begin{equation}
	\label{commutation}
	\begin{aligned}
		[\mathrm{C}&(x^iy^j(1+\bar{\mathbf{f}}_j\cdot\bar{\mathbf{y}}), \mathscr{F}(x,y))]_{x^0y^0} \\
		&=\left[x^iy^j (1+\bar{\mathbf{f}}_j\cdot\bar{\mathbf{y}}) \sum_{k=0}^{\infty} y^{-k} r_k(x^{-1}) \right]_{x^0y^0} \\
		& = a_{ij} + \sum_{k=1}^{\bar{n}(j)}{\sum_{m:x^m\in \left\{ x^i\bar{\mathrm{f}}_{j,k} \right\}}{a_{m,j-k}}} \\
		& = 2 a_{ij} \equiv 0 \pmod 2 \,,
	\end{aligned}
\end{equation}
which is satisfied by all $j\ge \bar{n}(j)$. 
Therefore, we exclude all terms with support extending outside the boundary, and the generated symmetry commutes with all Hamiltonian terms.
The calculation for NUCA$(5,4)$-generated SSSB models and the corresponding symmetries is analogous.

Now we supplement the update rules designed in Sec.~\ref{sec_NUCA54} and Sec.~\ref{sec_diversity}.
Because the forms of the (transposed) update rules are complex, we write them in the form of $\mathbf{f}_j \cdot \mathbf{y}$ or $\bar{\mathbf{f}}_j \cdot \bar{\mathbf{y}}$.
Because the lattice is finite, some terms are excluded for sites near the boundary.
The update rule of the cluster model on the $\{5,4\}$ lattice Eq.~(\ref{eq_54_cluster}) is
\begin{equation}
	\begin{aligned}
	\mathbf{f}_j &\cdot \mathbf{y} = J\left( j+1 \right) y + J\left( j+2 \right)y^2+ J\left( j \right) J\left( j+3 \right) y^3  \\
	& + x^2 y^{C\left( C\left( j \right) \right) - j - 5} + x^2 y^{C\left( C\left( j \right) \right) - j - 2}+ J\left( j \right)x^2 y^{C\left( C\left( j \right) \right)-j} \\
	& + (1-J\left( j \right))x^2y^{C\left( C\left( j \right) \right)-j+1} + (1-J\left( j \right))x^2y^{C\left( C\left( j \right) \right)-j+3} 
	\end{aligned}
\end{equation}
if $0<j<A_{2i+1}-3$, and
\begin{equation}
	\mathbf{f}_j \cdot \mathbf{y} = x^2y^{C\left( C\left( j \right) \right)-j+1} + x^2y^{C\left( C\left( j \right) \right)-j+3} 
\end{equation}
if $j=0,i\ge 1$, and 
\begin{equation}
	\begin{aligned}
	\mathbf{f}_j &\cdot \mathbf{y} = x^2 y^{C\left( C\left( j \right) \right) - j - 5} + x^2 y^{C\left( C\left( j \right) \right) - j - 2}\\
	& + (1-J\left( j \right))x^2y^{C\left( C\left( j \right) \right)-j+1} + (1-J\left( j \right))x^2y^{C\left( C\left( j \right) \right)-j+3}
	\end{aligned}
\end{equation}
if $0<A_{2i+1}-3 \le j \le A_{2i+1}-1$.
We specify $\mathbf{f}_j \cdot \mathbf{y}=x^2y^{C\left( C\left( j \right) \right)-j+1}$ for $x^0y^0$.
The Hamiltonian and symmetries are shown in Fig.~\ref{fig_54_model_1}.
We also check the commutation relation between Hamiltonian terms. 
Considering that $k-i=1\bmod2$, the constraint Eq.~(\ref{constraint}) for the cluster model is equal to 
\begin{equation}
	\begin{aligned}
		[&y^{j-P(l)}(1+y^{-1}+y^{-2}+(1-J(P(j)))y^{-3}) \\
		&+ y^{l-P(j)}(J(P(l))y^{P(P(l))-l+1}+y^{P(P(l))-l})]_{y^0}=0
	\end{aligned}
\end{equation}
for any $J(j)=1, J(l)=1$. 
This can be directly verified by substituting the explicit form of $P(j)$ in Eq.~(\ref{auxiliaryP}).

The transposed update rule associated with the Hamiltonian and subsystem symmetry in Fig.~\ref{fig_54_model_2} is 
\begin{widetext}
	\begin{equation}
		\label{eq_54_sspt_2}
		\begin{aligned}
			\bar{\mathbf{f}}_j \cdot \bar{\mathbf{y}}=&y^{-2}+y^{-8}+y^{-10}+(1-\chi_1)\left( y^{-16}+y^{-18} \right) \\
			&+x^{-2}y^{P^{2}\left( j \right) -j}\left( \chi_1y^{-4}+\left( 1-\chi_1 \right) y^{-3}+y^{-2}+y^{-1}+1+y+y^2+\chi_1y^4 \right)  \\
			&+\left( 1-\chi_1 \right) \left( 1-\chi_3 \right) \left( \chi_4\left( x^{-4}y^{P^{4}\left( j \right) -j-1}+x^{-2}y^{P^{2}\left( j \right) -j+4} \right) +\left( 1-\chi_4 \right) \left( x^{-4}y^{P^{4}\left( j \right) -j+1}+x^{-2}y^{P^{2}\left( j \right) -j-5} \right) \right)  \\
			&+x^{-4}y^{P^{4}\left( j \right) -j}\left( 1+\chi_1y \right) +\left( 1-\chi_1 \right) \chi_3x^{-4}y^{P^{4}\left( j \right) -j}\left( y+y^{-1} \right) \\
			&+\chi_1\left( \chi_2+\chi_3 \right) x^{-6}y^{P\left( P^{5}\left( j \right) +1 \right) -j}+\chi_1\left( 1-\chi_2 \right) x^{-4}y^{P^{4}\left( j \right) -j+2}+\chi_1\left( 1-\chi_3 \right) x^{-4}y^{P^{4}\left( j \right) -j-1} \,,
		\end{aligned}
	\end{equation}
\end{widetext}
for all sites $x^iy^j$ satisfying $P^3(j)\in [3,A_{2i-5}-4]$, and the rule is trivial otherwise. 
$P^{n}(j) =(P\circ\cdots\circ P)(j)$ is $n$ iterations of $P(j)$.
To simplify the expression, we have written the update rule in $\bar{\mathbf{f}}_j \cdot \bar{\mathbf{y}}$ and abbreviated the following functions $\chi_1=J\left( P^{3}\left( j \right) \right)$, $\chi_2=J\left( P^{4}\left( j \right) +1 \right)$, $\chi_3=J\left( P^{4}\left( j \right) \right)$, $\chi_4= (1-J\left( P^{3}\left( j \right) \right))(C\left( P^{4}\left( j \right) \right) -P^{3}\left( j \right)) $, whose values are all in $\{0,1\}$.
In the formalism of Hamiltonians of SSPT models Eq.~(\ref{eq_54_SSPT}), the only support of a general Hamiltonian term in another sublattice is given by $u(j)=3$, $m(j)=P^{3}\left( j \right)$, i.e., $x^{i-3}y^{P^3(j)}$. 
Now we check the commutation relation of Hamiltonian terms. 
We suppose two general $Z$ and $X$ Hamiltonian terms at $x^iy^j$ and $x^ky^l$ generated by $\bar{\mathbf{f}}_j$ and $\bar{\mathbf{f}}_l$. 
Without loss of generality, we set $i>k$ and $i-k=1\bmod 2$.
If $i-3 = k$ and $C(x^{i-3}y^{P^3(j)}, x^ky^l\bar{\mathbf{f}}_{l}\cdot \bar{\mathbf{y}})=1$, then $C(x^{k-3}y^{P^3(l)}, x^{i-6}y^{P\left( P^{5}\left( j \right) +1 \right)})=1$.
If $i-1 = k$ and $C(x^{i-3}y^{P^3(j)}, x^ky^l\bar{\mathbf{f}}_{l}\cdot \bar{\mathbf{y}})=1$, then $C(x^{k-3}y^{P^3(l)}, x^{i-4}y^{P^{4}\left( j \right)}+ (1-\chi_1(j))x^{i-4}y^{P^{4}\left( j \right)+1})=1$.
Therefore, all the Hamiltonian terms commute with each other.

The update rule $\mathbf{f}_j(x)$ of Eq.~(\ref{eq_rule_reg_1}) is written as
\begin{equation}
	\begin{aligned}
	\mathbf{f}_j& \cdot \mathbf{y} = x y^{C(j)-2-j} + J(j)xy^{C(j)-j}  \\
	& + (1-J(j))xy^{C(j)+1-j} + x^2 y^{C(C(j))-5-j} \\
	& + x^2 y^{C(C(j))-2-j} + (1-J(j))x^2 y^{C(C(j))+1-j}
	\end{aligned}
\end{equation}
if $0<j<A_{2i+1}-1$, and
\begin{equation}
	\mathbf{f}_j \cdot \mathbf{y} =xy^{C(j)+1-j} + x^2 y^{C(C(j))-2-j} + x^2 y^{C(C(j))+1-j}
\end{equation}
if $j=0,i\ge 1$, and
\begin{equation}
	\mathbf{f}_j \cdot \mathbf{y} = x y^{C(j)-2-j} + x^2 y^{C(C(j))-5-j} + x^2 y^{C(C(j))-2-j}
\end{equation}
if $j=A_{2i+1}-1,i\ge 1$.
We specify $\mathbf{f}_j \cdot \mathbf{y}=x^2 y^{C(C(j))+1-j} + x^2 y^{C(C(j))-2-j}$ for $x^0y^0$.
The Hamiltonian and symmetries are shown in Fig.~\ref{fig_54_model_3}.

The update rule corresponding to Eq.~(\ref{eq_rule_reg_2}) is 
\begin{equation}
	\begin{aligned}
	\mathbf{f}_j& \cdot \mathbf{y} = (1-J(j+1)) x^2 y^{C^{2}(j)+9-j} \\
	&  + J(j+1) x^2 y^{C^{2}(j)+6-j} + x^2 y^{C^{2}(j)+1-j}  +  x^2 y^{C^{2}(j)-7-j}\\
	& + x^4 y^{C^{4}(j)+6-j} + x^4 y^{C^{4}(j)-15-j} + x^4 y^{C^{4}(j)-36-j}
	\end{aligned}
\end{equation}
if $J(j)=0$, and
\begin{equation}
	\begin{aligned}
	\mathbf{f}_j& \cdot \mathbf{y} = (1-J(P(j)))(1-J(P(j)+1)) y^{3}\\
	&+J(P(j))(1-J(P(j)+1))J(P(P(j)+1)) y^{3}\\
	&+J(P(j))(1-J(P(j)+1))(1-J(P(P(j)+1))) y^{6}\\
	& + x^2 y^{C^{2}(j)+6-j} + x^2 y^{C^{2}(j)-2-j} + x^2 y^{C^{2}(j)-7-j}\\
	&+ x^4 y^{C^{4}(j)-15-j} + x^4 y^{C^{4}(j)-36-j} 
	\end{aligned}
\end{equation}
if $J(j)=1$, for a physical site $x^iy^j$ satisfying $0\le j\le A_{2i+1}-1$.
Here, $C^{n}(j) =(C\circ\cdots\circ C)(j)$ is $n$ iterations of $C(j)$.
We note that, if a term $x^{k}y^{l}$ in $\mathbf{f}_j \cdot \mathbf{y}$ for a site $x^iy^j$ satisfies $l+j<6$ or $l+j > A_{2(k+i)+1}-8$, it is excluded.
The Hamiltonian and symmetries are shown in Fig.~\ref{fig_54_model_4}.

The update rule $\mathbf{f}_j(x)$ of Eq.~(\ref{eq_rule_tree}) is written as follows:
\begin{equation}
	\begin{aligned}
	\mathbf{f}_j& \cdot \mathbf{y} = xy^{C(j)-j-2} + J(j)xy^{C(j)-j} + (1-J(j))xy^{C(j)-j+1}
	\end{aligned}
\end{equation}
if $0<j<A_{2i+1}-1$, and
\begin{equation}
	\mathbf{f}_j \cdot \mathbf{y} = xy^{C(j)-j+1}
\end{equation}
if $j=0,i\ge1$, and
\begin{equation}
	\mathbf{f}_j \cdot \mathbf{y} = xy^{C(j)-j-2}
\end{equation}
if $j=A_{2i+1}-1,i\ge1$.
The Hamiltonian and symmetries are shown in Fig.~\ref{fig_54_model_5}.

\subsection{Non-Abelian translation invariance of NUCA-generated Hamiltonians}
We derive a sufficient condition for a NUCA-generated Hamiltonian to be translationally invariant in Sec.~\ref{sec_NUCA54}, while we do not discuss in detail why a uniform update rule $\bar{\mathbf{f}}_{j}(x)\equiv \bar{\mathbf{f}}(x)$ does not guarantee translation invariance. 
In this appendix, we discuss the translation invariance condition on a hyperbolic lattice with periodic boundary conditions (PBC). 
We refer readers to Ref.~\cite{Maciejko2020HBT, Maciejko2022AHBT,Crystallography2022,Lux_2023_PBC,Lenggenhager2023supercell,Tummuru_2024_hyperbolicsemimetal,A_Chen_2024_localization} for more details of PBCs for the hyperbolic lattice.

As reviewed in Appendix~\ref{app_hyperbolic}, we denote the space group of a hyperbolic lattice $\{p,q\}$ as $\Delta(2,q,p)$.
To define a PBC lattice for a $\{p,q\}$ lattice, one should find the Fuchsian translation group $\Gamma$ as the largest normal torsion-free subgroup of orientation-preserving elements of $\Delta(2,q,p)$.
Then one can define the unit cell $\mathcal{U}$ (with $N_\mathcal{U}$ vertices in each $\mathcal{U}$) of a Bravais lattice on the hyperbolic plane by the quotient $\Delta(2,q,p)/\Gamma$. 
We further need to determine a normal subgroup $\Gamma_{\text{PBC}}$ of $\Gamma$ of index $N$, as the factor group $\Gamma /\Gamma_{\text{PBC}}$ is a finite group of order $N$. 
By the coset decomposition $\Gamma =\Gamma_{\mathrm{PBC}}\sqcup g_2\Gamma_{\mathrm{PBC}}\sqcup \cdots \sqcup g_N\Gamma_{\mathrm{PBC}}$, where $G=\left\{ g_1=I\, ,g_2,\cdots ,\, g_N \right\} \subset \Gamma $ is a right transversal, the elements of the order-$N$ factor group $\Gamma /\Gamma_{\text{PBC}}$ are $[g_i]\equiv g_i \Gamma_{\text{PBC}}$ for $g_i \in G$. 

A size-$N$ PBC cluster $\mathcal{C}_N$ is constructed by $N$ copies of the unit cell, and $\mathbb{H}^2$ is tiled by the PBC clusters as $\mathbb{H}^2=\bigsqcup_{\gamma_{\mathrm{PBC}}\in \Gamma_{\mathrm{PBC}}}{\gamma_{\mathrm{PBC}}\mathcal{C}_N}$ where $\mathcal{C}_N=\bigsqcup_{i=1}^N{g_i\mathcal{U}}$.
Hence, an $N$-unit-cell PBC lattice $\mathcal{C}_N$ is constructed by wiring the boundary sites of the PBC cluster together, on which $\Gamma_{\text{PBC}}$ has trivial action. 
The action of $[g_i]$ can be considered as a permutation of unit cells~\cite{Maciejko2022AHBT}. 
Now each of these $N$ unit cells is labeled by a unique $[g_i]$ as $\mathcal{U}_{g_i}$, such that two unit cells $\mathcal{U}_{g_j}$ and $\mathcal{U}_{g_i}$ are related by $[g_k]$ if $[g_j] = [g_k g_i]$. 
As the choice of $\Gamma_{\text{PBC}}$ is not unique even for a given $N$, the PBC lattice is also not unique. 

Suppose there are $N_\mathcal{U}$ vertices in each $\mathcal{U}$ labeled as $\alpha_1,\alpha_2,\cdots,\alpha_{N_\mathcal{U}}$, on which the qubits are located.
We denote the translation operator corresponding to $[g_k]$ as $\mathcal{T}_{[g_k]}$, whose action on a Pauli operator $O_{[g_i],\alpha}$ defined at the $\alpha$ vertex in unit cell $\mathcal{U}_{g_i}$ is $\mathcal{T}_{[g_k]} O_{[g_i],\alpha} \mathcal{T}_{[g_k]}^{-1} = O_{[g_kg_i],\alpha}$. 
A translationally invariant Hamiltonian is defined as 
\begin{equation}
	\mathcal{T}_{[g_k]} H \mathcal{T}_{[g_k]}^{-1} = H\,,\, \forall [g_k] \in \Gamma /\Gamma_{\text{PBC}}\,.
\end{equation}

Now we consider the general form of a many-body Hamiltonian:
\begin{equation}
	\label{H_general_form}
	\begin{aligned}
		H=&\sum_{n=1}^{N\times N_{\mathcal{U}}} \sum_{[g_1],\cdots,[g_n]\in\Gamma/\Gamma_{\text{PBC}}} \sum_{\alpha_1,\cdots,\alpha_n=1}^{N_{\mathcal{U}}} \\
		&J^{\left( n \right)}([g_1],\cdots ,[g_n];\alpha_1,\cdots ,\alpha_n)O_{[g_1],\alpha_1}\cdots O_{[g_n],\alpha_n}\,,
	\end{aligned}
\end{equation}
where $n$ labels the $n$-body interaction, $[g_1],[g_2],\cdots$ label the indices of unit cells and $\alpha_1,\alpha_2,\cdots$ label the indices of vertices on a unit cell. 
$J^{\left( n \right)}([g_1],\cdots ,[g_n];\alpha_1,\cdots ,\alpha_n)$ is the coupling constant that depends on $[g_1],[g_2],\cdots$ and $\alpha_1,\alpha_2,\cdots$ in general. 
Under translation by $[g_k]$, the Hamiltonian is transformed as:
\begin{widetext}
\begin{equation}
	\begin{aligned}
		\mathcal{T}_{[g_k]}H\mathcal{T}_{[g_k]}^{-1} &= \sum_{n=1}^{N\times N_{\mathcal{U}}} \sum_{[g_1],\cdots,[g_n]\in\Gamma/\Gamma_{\text{PBC}}} \sum_{\alpha_1,\cdots,\alpha_n=1}^{N_{\mathcal{U}}} J^{\left( n \right)}([g_1],\cdots,[g_n];\alpha_1,\cdots ,\alpha_n) \mathcal{T}_{[g_k]}O_{[g_1],\alpha_1}\mathcal{T}_{[g_k]}^{-1}\mathcal{T}_{[g_k]}O_{[g_2],\alpha_2}\mathcal{T}_{[g_k]}^{-1}\cdots  \\
		&=\sum_{n=1}^{N\times N_{\mathcal{U}}} \sum_{[g_1],\cdots,[g_n]\in\Gamma/\Gamma_{\text{PBC}}} \sum_{\alpha_1,\cdots,\alpha_n=1}^{N_{\mathcal{U}}} J^{\left( n \right)}([g_1],\cdots ,[g_n];\alpha_1,\cdots ,\alpha_n) O_{[g_kg_1],\alpha_1}O_{[g_kg_2],\alpha_2}\cdots \\
		&= \sum_{n=1}^{N\times N_{\mathcal{U}}} \sum_{[g_1],\cdots,[g_n]\in\Gamma/\Gamma_{\text{PBC}}} \sum_{\alpha_1,\cdots,\alpha_n=1}^{N_{\mathcal{U}}} J^{\left( n \right)}([g_{k}^{-1}g_1'],\cdots ,[g_{k}^{-1}g_n'];\alpha_1,\cdots ,\alpha_n) O_{[g_1'],\alpha_1} O_{[g_2'],\alpha_2}\cdots \,.
	\end{aligned}
\end{equation}
\end{widetext}
In the third line, we have denoted $[g_i'] = [g_kg_i]$ and so on. 
As translation does not change interactions $n$, compared to Eq.~(\ref{H_general_form}), we have the translation-invariance condition for the coupling constant $J^{\left( n \right)}$:
\begin{equation}
	\label{eq:coupling_constant}
	\begin{aligned}
		&J^{\left( n \right)}([g_1],\cdots ,[g_n];\alpha_1,\cdots ,\alpha_n) \\
		&= J^{\left( n \right)}([g_{k}^{-1}g_1],\cdots ,[g_{k}^{-1}g_n];\alpha_1,\cdots ,\alpha_n),\,\forall k\,.
	\end{aligned}
\end{equation}
That is, $J^{\left( n \right)}$ is a function of relative relation among $[g]$ only.

From Eq.~(\ref{eq:coupling_constant}), the coupling constant $J^{\left( n \right)}([g_1],\cdots,[g_n];\alpha_1,\cdots,\alpha_n)$ is invariant under all choices of $[g_k]$ which are determined by $\Gamma/\Gamma_{\text{PBC}}$.
As the quotient group $\Gamma/\Gamma_{\text{PBC}}$ is not unique, there are no intrinsic orthogonal axes such as horizontal and vertical directions on the two-dimensional Euclidean lattice as translation directions.
However, when generating Hamiltonians by the NUCA, the determination of the coupling constant $J^{\left( n \right)}$ is irrelevant to the construction of the non-Abelian translations and how we wire together the boundary. 
The $x$- and $y$-directions of the deformed lattice do not correspond to the geodesic directions on the hyperbolic lattice.
Although a uniform update rule is invariant under translation in $x$- and $y$-directions, it is not equivalent to translation invariance of Hamiltonians under the hyperbolic translation symmetry. 

A sufficient condition for the Hamiltonian to be translationally invariant regardless of the choice of $\Gamma/\Gamma_{\text{PBC}}$ is that $J^{\left( n \right)}$ should not have a specific orientation, implying that the interaction of qubits on the polygons is isotropic.
This can be satisfied if $J^{\left( n \right)}$ is a function of orbits of a rotation at either a vertex or face.
These orbits can be characterized by the graph distance, or equivalently by the geodesic distance among vertices.
For the NUCA$(5,4)$-generated models, the coupling constant is given by the transposed update rule Eq.~(\ref{update_rule}) depending on the positions of sites $x^iy^j$ in general. 
The parent-child relation of two nodes corresponding to sites $x^{i_1}y^{j_1},x^{i_2}y^{j_2}$ is not preserved when they are translated by some $[g_k]$.
Consequently, this sufficient condition leads to the constraints derived for $\mathbf{f}_j$ in Sec.~\ref{sec_NUCA54}, and the NUCA-generated Hamiltonians on the hyperbolic $\{5,4\}$ lattice satisfying this sufficient condition are translationally invariant.
This is directly demonstrated in the Hamiltonian terms visualized in the corresponding figures, and can be checked by the neighboring relation of polygons summarized in Table~\ref{tab_neighborhood54} as the lattice is regular.

Additionally, uniformity implies translation symmetry on the Euclidean lattice.
We discuss the application of NUCA$(4,4)$ on the Euclidean square lattice in Sec.~\ref{sec_preliminaries} and Appendix~\ref{app_applicability} with uniform update rules. 
On the Euclidean plane $\mathbb{E}^2$ the discrete normal translation subgroup corresponding to $\text{Iso}(\mathbb{E}^2) =  \mathbb{R}^2\rtimes O(2)$ is given by $\mathbb{Z}\times \mathbb{Z}$, with normal subgroup $N_x\mathbb{Z} \times N_y \mathbb{Z}$. 
Both of these groups are Abelian, so the factor group is also Abelian and isomorphic to $\mathbb{Z}_{N_x} \times \mathbb{Z}_{N_y}$. 
This defines an $N_x \times N_y$-unit-cell PBC lattice with 2-torus topology. 
Therefore, a sufficient condition for a translationally invariant Hamiltonian is that the term defined by the update rule Eq.~(\ref{Euclidean_update_rule}) is invariant under translations in $\mathbb{Z}_{N_x} \times \mathbb{Z}_{N_y}$.
Therefore, the term need not be isotropic. 
This is intrinsically guaranteed by a uniform update rule $\bar{\mathbf{f}}(x)$. Thus the condition for translation invariance in the Euclidean case is trivial.

\section{Construction of linear NUCA$(6,6)$ on the $\{6,6\}$ lattice}
\label{app_NUCA66}
In this Appendix, we show the construction of the NUCA to a more general hyperbolic lattice, i.e., linear NUCA$(6,6)$ for the bipartite $\{6,6\}$ lattice to investigate SSPT states.
The NUCA$(5,4)$ in Sec.~\ref{sec_NUCA54}, whose update rules vary temporally, captures some key non-uniformity of hyperbolic geometry.
However, embedding a hyperbolic lattice into a Euclidean lattice will induce a more complex distortion than that on the $\{5,4\}$ lattice, leading to a spatially and temporally non-uniform update rule.
Therefore, the realization of our NUCA for the $\{6,6\}$ lattice includes the general non-uniform properties of NUCA$(p,q)$.

\begin{figure*}[t]
	\includegraphics[width=0.8\textwidth]{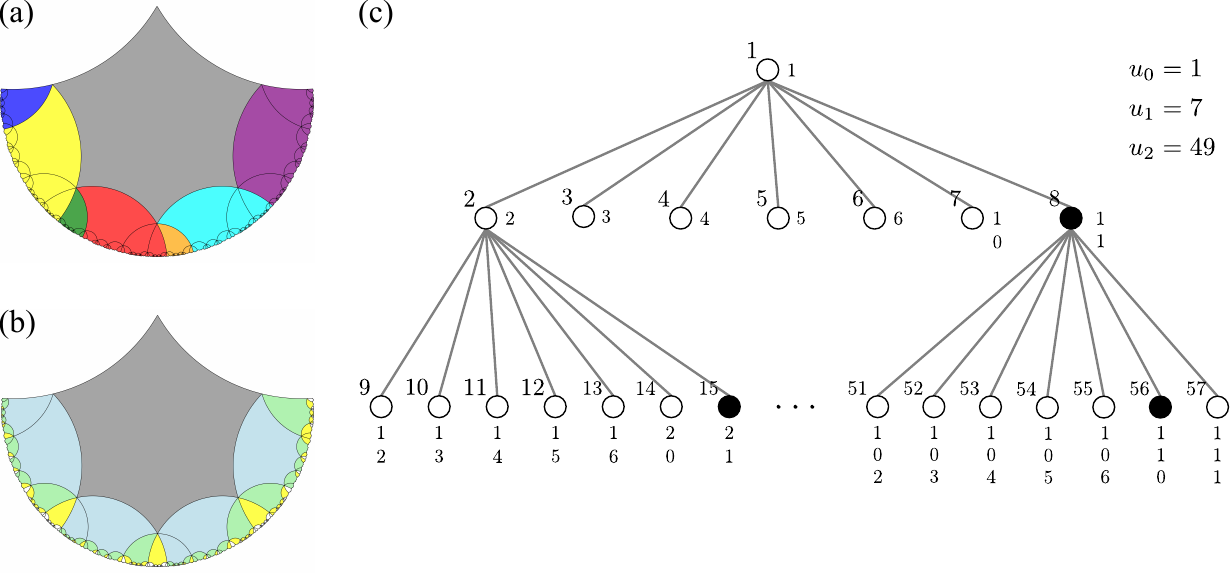}
	\caption{\label{fig_los66} 
	(a) The splitting method applied to a quarter of the $\{6,6\}$ lattice. The gray region denotes the leading hexagon, the purple region denotes a strip which corresponds to a black node, while the other colored regions denote copies of the quarter which correspond to white nodes.
	(b) Lattice generated by the splitting method. 
	Hexagons with the same color correspond to the same level. Only the first four levels are colored.
	(c) The spanning tree of the $\{6,6\}$ lattice. The nodes (hexagons) are uniquely assigned discrete indices represented in the basis of $\{u_k\}_{k\in \mathbb{N}}$, where $\{u_k\}$ is the number of nodes on each level. The lines represent offspring relation of nodes. Neighboring relation can be calculated based on the representation. }
\end{figure*}

We visualize the $\{6,6\}$ lattice in Fig.~\ref{fig_lattices} and apply the splitting method to it.
The splitting matrix Eq.~(\ref{splittingmatrix}) of the $\{6,6\}$ lattice is:
\begin{equation}
	S_{6,6}=\begin{pmatrix}
	6&		1\\
	6&		1\\
	\end{pmatrix} \,,
\end{equation}
such that both a quarter and a strip are split into six quarters and a strip.
The characteristic polynomial gives the number of nodes $\{u_k\}_{k\in \mathbb{N}}$ on each level $k$:
\begin{equation}
	u_{k} = 7^k\,,\quad k=0,1,2,\cdots \,,
\end{equation}
and the generating rule is given by
\[\begin{tikzpicture}[baseline=-0.5ex, node distance=0.3ex]
    \node[wcirc] (c1) {};
    \node[right=of c1] (arrow) {$\rightarrow$};
    \node[wcirc, right=of arrow] (c2) {};
    \node[wcirc, right=of c2] (c3) {};
    \node[wcirc, right=of c3] (c4) {};
    \node[wcirc, right=of c4] (c5) {};
    \node[wcirc, right=of c5] (c6) {};
    \node[wcirc, right=of c6] (c7) {};
    \node[bcirc, right=of c7] (c8) {};
\end{tikzpicture}\,,\qquad
\begin{tikzpicture}[baseline=-0.5ex, node distance=0.3ex]
    \node[bcirc] (c1) {};
    \node[right=of c1] (arrow) {$\rightarrow$};
    \node[wcirc, right=of arrow] (c2) {};
    \node[wcirc, right=of c2] (c3) {};
    \node[wcirc, right=of c3] (c4) {};
    \node[wcirc, right=of c4] (c5) {};
    \node[wcirc, right=of c5] (c6) {};
    \node[bcirc, right=of c6] (c7) {};
    \node[wcirc, right=of c7] (c8) {};
\end{tikzpicture}. \]
The generating procedure is visualized in Fig.~\ref{fig_los66}. 
After assigning indices $v$ to the nodes, the standard representation $\mathcal{A}(v)$ of the indices is obtained in the basis of $\{u_k\}_{k\in \mathbb{N}}$ as:
\begin{equation}
	v= \sum_{i=0}^{i_{\max}} \alpha_i u_i,\,\alpha_i\in \{0,1,\cdots,6\},\,u_i=7^i \,,
\end{equation}
which can be uniquely determined by a greedy algorithm. 
Suppose that a node $v\equiv\mathcal{A}(v)=\alpha_{i_{\max}}\cdots\alpha_1\alpha_0$ and $v_-=v-1\equiv\mathcal{A}(v_-)$.
The children of $v$ are $\{\mathcal{A}(v_-)\alpha|_{\alpha=2,3,4,5,6},\mathcal{A}(v)0,\mathcal{A}(v)1\}$, which are identical for both black and white parent nodes.

\begin{table}[b]
\caption{\label{tab_neighborhood66} Neighboring nodes of a node $v$ in the $\{6,6\}$ spanning tree. The labels leftmost  and rightmost indicate the location of a node $v$ within the children of its parent node. $c_R(v)$ and $c_L(v)$ label the rightmost and leftmost child of $v$ respectively, while $p(v)$ labels the parent node of $v$. $\mu_-=\mu-1,\mu_+=\mu+1,v_-=v-1$ are simplified notations. 
$\mathcal{N}_w(v)=\{\mathcal{A}(v_-)3,\mathcal{A}(v_-)5,\mathcal{A}(v)0,\mathcal{A}(v)1\}$ and $\mathcal{N}_b(v)=\{\mathcal{A}(v_-)2,\mathcal{A}(v_-)4,\mathcal{A}(v_-)6,\mathcal{A}(v)0\}$ are sets of nodes. Other notations are defined in the main text.
}
\begin{ruledtabular}
\begin{tabular}{ccccc}
	Node $v$ & Condition & \multicolumn{3}{c}{Neighboring nodes} \\
	\hline
	black & -- & $\mathcal{N}_b(v)$ & $\mathcal{A}(v)1$ & $p(v)$\\
	leftmost & $\mu_-$ white, $\mu$ marked & $\mathcal{N}_w(v)$ & $c_R^{2l}(\mu_-)$ & $v+1$ \\
	leftmost & $\mu_-$ white, $\mu$ endnode & $\mathcal{N}_w(v)$ & $c_R^{2l+2}(\mu_-)$ & $v+1$ \\
	leftmost & $\mu_-$ white, $\mu$ black, $l=1$ & $\mathcal{N}_w(v)$ & $c_R^{2l}(\mu_-)$ & $p(v)$  \\
	leftmost & $\mu_-$ white, $\mu$ black, $l>1$ & $\mathcal{N}_w(v)$ & $c_R^{2l}(\mu_-)$ & $v+1$  \\
	leftmost & $\mu_-$ black, $\mu$ white & $\mathcal{N}_w(v)$ & $c_R^{2l-1}(\mu_-)$ & $v+1$ \\
	rightmost & $\mu$ endnode, $\mu_+$ marked & $\mathcal{N}_w(v)$ & $c_L^{\frac{l}{2}}(\mu+1)$ & $p(v)$ \\
	rightmost & $\mu$ marked, $\mu_+$ endnode & $\mathcal{N}_w(v)$ & $c_L^{\frac{l}{2}-1}(\mu+1)$ & $p(v)$ \\
	rightmost & $\mu$ white, $\mu_+$ black & $\mathcal{N}_w(v)$ & $c_L^{\frac{l}{2}}(\mu+1)$ & $p(v)$ \\
	rightmost & $\mu$ black, $\mu_+$ white & $\mathcal{N}_w(v)$ & $c_L^{\frac{l+1}{2}}(\mu+1)$ & $p(v)$ \\
	marked & -- & $\mathcal{N}_w(v)$ & $p(v)$ & $v_-$ \\
	endnode & -- & $\mathcal{N}_w(v)$ & $v+1$ & $c_R^2(v_-)$ \\
\end{tabular}
\end{ruledtabular}
\end{table}

Then we label the polygons by indices in the representation of the language of the splitting.
The neighboring relation can be calculated by an algorithm linear in the length of the representation $\mathcal{A}(v)$.
We calculate and summarize the results for the neighborhood of an arbitrary node $v$ in Table~\ref{tab_neighborhood66}, and we refer readers to Ref.~\cite{Margenstern2007} for more details. 
In Table~\ref{tab_neighborhood66}, a white node $v$ is a marked-child if it is a neighbor of its parent node, and is an end-child otherwise.
For a leftmost node $v$ with $\mathcal{A}(v)$ ending in $2$, we follow the parent-search path from $v$ until we reach the first node whose representation does not end in $2$ and we denote this node as $\mu$.
For a rightmost node $v$ with $\mathcal{A}(v)$ ending in $1$, $\mu$ is defined analogously as the first node on the parent-search path whose representation does not end in $1$.
In both cases, $l>0$ denotes the length of this path.
Within this offspring-based generation, the children of a node may not be neighbors of it due to the distortion, leading to a complex lattice geometry. 
Consequently, to prevent the finite lattice constructed via the spanning tree from being dominated by outer-boundary hexagons, the boundary conditions must be defined otherwise.
Here, we adopt the smooth boundary geometry generated by the vertex-inflation method~\cite{Jahn_2020_holography,Boyle2020conformal,chen_2023_h_haldane,huang2025hyperbolicEE}.
Within this procedure, the lattice is generated layer by layer and all vertices of inner layers have exactly $q=6$ neighboring hexagons as visualized in Fig.~\ref{fig_los66}(b). 
Therefore, specific outer-level nodes of the spanning tree that are not encompassed by the lattice constructed via the vertex-inflation method are excluded.

\begin{figure}[b]
	\centering
	\includegraphics[width=0.9\columnwidth]{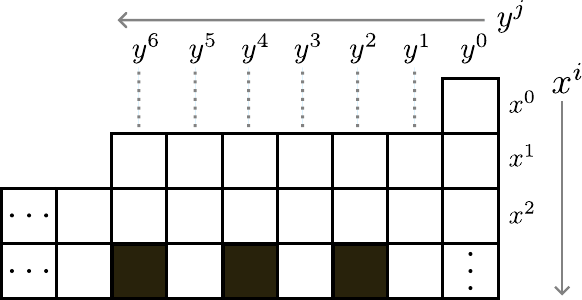}
	\caption{\label{fig_66coor} Visualization of the deformed lattice on which NUCA$(6,6)$ is performed. Sites $x^iy^j$ with $j<u_{i}$ are mapped to hexagons on the $\{6,6\}$ lattice. Nearest-neighbor relations on the deformed lattice do not represent physical locality of the corresponding hexagons. To obtain a smooth boundary geometry, we adopt the vertex-inflation method to define the boundary. The black sites are excluded by specifying trivial update rules $\mathbf{f}_{(i,j)}(x)=\mathbf{0}$ and $\bar{\mathbf{f}}_{(i,j)}(x)=\mathbf{0}$.}
\end{figure}

Now we introduce the deformed lattice where we perform NUCA$(6,6)$ as visualized in Fig.~\ref{fig_66coor}.
Similar to the $\{5,4\}$ case, the deformed lattice is a square lattice with coordinate $i=0,1,2,\cdots$ and $j=0,1,2,\cdots$. 
Each node $v$ of the spanning tree is uniquely mapped to a physical site on a two-dimensional square lattice $x^iy^j,j\le u_i-1$ by $v = \sum_{l=0}^{i} u_{l}-j$.
Sites sharing the same $y^j$ belong to the same time step $j$.

We define some auxiliary functions on the deformed lattice. 
The auxiliary functions for NUCA$(6,6)$ can be obtained by a similar procedure in Appendix~\ref{app_NUCA54}. 
The indicator function is based on both $i$ and $j$ as:
\begin{equation}
	J\left( i,j \right) =\begin{cases}
	1, \quad \text{if } x^iy^j \text{ is mapped to a black node}\\
	0, \quad \text{if } x^iy^j \text{ is mapped to a white node}\\
	\end{cases}\,.
\end{equation}
The parent function is:
\begin{equation}
	P(j) = \lfloor j /7\rfloor\,.
\end{equation}
The rightmost-child function $c_R$ and the leftmost-child function $c_L$ are:
\begin{equation}
	c_R(j) = 7j\,,\quad c_L(j) = 7j+6\,.
\end{equation}
Notably, the parent and child functions do not imply physical locality on the $(6,6)$ lattice as in NUCA$(5,4)$, capturing the general non-uniformity of NUCA.

Following the standard setup in Sec.~\ref{sec_preliminaries}, we construct NUCA$(6,6)$. 
The state of a site $x^iy^j$ is represented as
$a_{ij}x^iy^j,\, a_{ij}\in \mathbb{F}_2$.
The configuration of the states of physical sites is
\begin{equation}
	\mathscr{F}(x,y) = \sum_{i=0}^{\infty}\sum_{j=0}^{\infty} a_{ij}x^{i}y^{j}\, .
\end{equation}
As the boundary condition is given by the vertex-inflation method, only the generated sites are mapped to hexagons of the $\{6,6\}$ lattice.
The configuration of the states at a time step $j$ is represented by
\begin{equation}
	r_{j}(x)\equiv\sum_{i=0}^{\infty} a_{ij}x^{i}\, ,
\end{equation}
where sites with $j<u_i$ are physical.

On the deformed lattice, the physical neighboring relation is a function of both $i$ and $j$ as summarized in Table~\ref{tab_neighborhood66}.
Therefore, the update rule $\mathbf{f}_{(i,j)}(x)$ and the transposed update rule $\bar{\mathbf{f}}_{(i,j)}(x)$ depend on $x^iy^j$ and thereby distinguish the sublattice structure.
The explicit form of the update rule is:
\begin{equation}
	\mathbf{f}_{(i,j)}(x) \equiv (\mathrm{f}_{(i,j),1}(x),\mathrm{f}_{(i,j),2}(x),\cdots ,\mathrm{f}_{(i,j),n}(x))^T\, ,
\end{equation}
where $n\equiv n(i,j)$ is the order of the update rule. 
The transposed update rule is similarly defined as:
\begin{equation}
	\bar{\mathbf{f}}_{(i,j)}(x) \equiv (\bar{\mathrm{f}}_{(i,j),1}(x),\bar{\mathrm{f}}_{(i,j),2}(x),\cdots ,\bar{\mathrm{f}}_{(i,j),\bar{n}}(x))^T\,,
\end{equation}
where $\bar{n}\equiv \bar{n}(i,j)$ is the order of the transposed update rule. 
By using the update rule, the time evolution of NUCA$(6,6)$ is:
\begin{equation}
	\begin{aligned}
	r_j(x) = \sum_{k=1}^{n_{\max}} \sum_{i} a_{i,j-k}x^i \mathrm{f}_{(i,j-k),k}(x) +q_j(x)\, ,
	\end{aligned}
\end{equation}
where $n_{\max}$ is defined in Eq.~(\ref{state_evolution}). The initial condition $\mathbf{q}(x)\equiv (q_0(x),q_1(x),\cdots)^T$ is defined in Eq.~(\ref{eq_initial_condition_NUCA}).

The general form of a NUCA$(6,6)$-generated SSPT Hamiltonian is
\begin{equation}
	\label{eq_hoca66sspt}
	\begin{aligned}
		&\mathscr{H}=\\
		&-\sum_{\substack{(i,j) \in \Lambda^{(a)}\\ \bar{\mathbf{f}}_{(i,j)}\neq \mathbf{0}}}Z(x^iy^j(1+\bar{\mathbf{f}}_{(i,j)}\cdot\bar{\mathbf{y}}_{1,\bar{n}(i,j)}))Z\left( x^{u(i,j)}y^{m\left( i,j \right)} \right)\\
		&-\sum_{\substack{(k,l) \in \Lambda^{(b)}\\ \bar{\mathbf{f}}_{(k,l)}\neq \mathbf{0}}}X(x^ky^l(1+\bar{\mathbf{f}}_{(k,l)}\cdot\bar{\mathbf{y}}_{1,\bar{n}(k,l)}))X\left( x^{u(k,l)}y^{m\left( k,l \right)} \right) \, ,
	\end{aligned}
\end{equation}
where $x^{u(i,j)}y^{m(i,j)}$ and $x^iy^j(1+\bar{\mathbf{f}}_{(i,j)}\cdot\bar{\mathbf{y}})$ are defined on distinct sublattices $\Lambda^{(a)}$ and $\Lambda^{(b)}$ respectively.
A term $h_{ij}$ that cannot be fully supported in a finite lattice is excluded, which is equivalent to assigning a trivial $\bar{\mathbf{f}}_{(i,j)}$ to $x^iy^j$.
Similar to the case for NUCA$(5,4)$, the commutation relation of Hamiltonian terms $C(h^Z_{ij},h^X_{kl})=0$ should be preserved, leading to constraints for the update rules and the functions $u(i,j),m(i,j)$. 

The Hamiltonian commutes with the following symmetries:
\begin{equation}
	S^{(a)}(\mathbf{q}^{(a)}) = X(\tilde{\mathscr{F}}^{(a)}(x,y))\, ,\quad  S^{(b)}(\mathbf{q}^{(b)}) = Z(\tilde{\mathscr{F}}^{(b)}(x,y))\, ,
\end{equation}
where $\tilde{\mathscr{F}}^{(a)}(x,y)$ is the truncated NUCA$(6,6)$ evolution configuration.
This commutation can be checked by an analogous calculation of the commutation polynomial in Eq.~(\ref{commutation}). 
The general form of NUCA$(6,6)$-generated SSSB Hamiltonian is defined as:
\begin{equation}
	\mathscr{H}=-\sum_{i,j:\bar{\mathbf{f}}_{(i,j)}\neq \mathbf{0}}Z(x^iy^j(1+\bar{\mathbf{f}}_{(i,j)}\cdot\bar{\mathbf{y}}_{1,\bar{n}}))\,,
\end{equation}
which commutes with the following symmetries:
\begin{equation}
	S(\mathbf{q}) = X(\tilde{\mathscr{F}}(x,y))\, .
\end{equation}

\begin{figure*}[t]
	\includegraphics[width=0.9\textwidth]{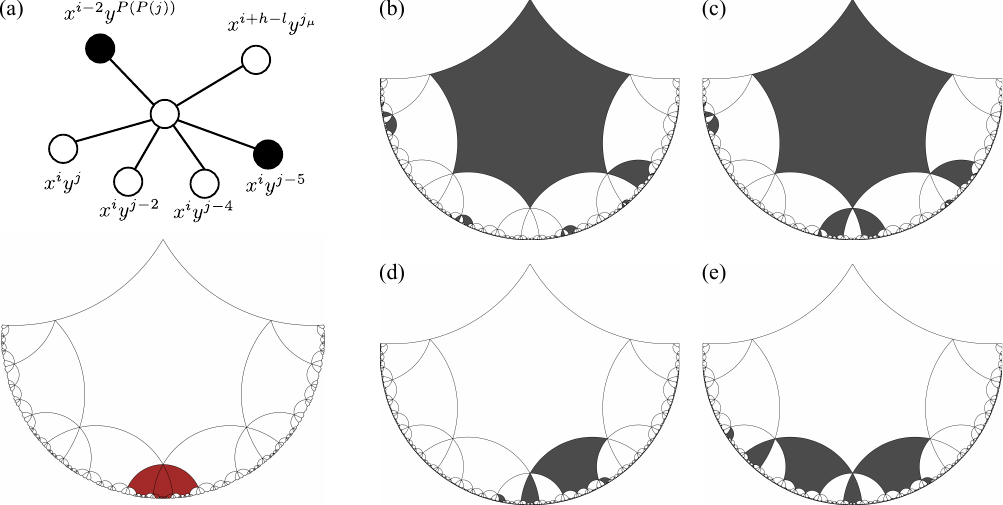}
	\caption{\label{fig_66_model_1} Hamiltonian and symmetry patterns of the cluster model on the $\{6,6\}$ lattice generated by the update rule given in Table~\ref{tab_66hcmrule}. (a) Coordinate of a specific Hamiltonian term on the deformed lattice, where $l$ and $h$ are determined in Table~\ref{tab_66hcmrule}. The red region below illustrates the support of this Hamiltonian term on the hyperbolic lattice. Panels (b)--(e) show symmetry patterns generated by initial conditions (b) $\mathbf{q}_1(x,y)$ in Eq.~(\ref{eq_66_cluster_ic1}), (c) $\mathbf{q}_2(x,y)$ in Eq.~(\ref{eq_66_cluster_ic2}), (d) $\mathbf{q}_3(x,y)$ in Eq.~(\ref{eq_66_cluster_ic3}), and (e) $\mathbf{q}_4(x,y)$ in Eq.~(\ref{eq_66_cluster_ic4}). Nontrivial Pauli $X$ actions of the symmetry are represented by black polygons.}
\end{figure*}

We explicitly show an example of NUCA$(6,6)$-generated SSPT model. 
Similar to the case on the $\{5,4\}$ lattice, the cluster state can be defined on other $\{p,q\}$ lattices. 
Due to the distortion of the deformed lattice, the non-uniform update rule is complex.
To explicitly formulate the update rule, we represent it in the form of $\bar{\mathbf{f}}_{(i,j)}\cdot \bar{\mathbf{y}}$, and classify it into several distinct cases in Table~\ref{tab_66hcmrule}, where all nontrivial $\bar{\mathbf{f}}_{(i,j)}$ have five nonzero terms.
In this table, for a leftmost node $v$ with $\mathcal{A}(v)$ ending in $2$, we follow the parent-search path from $v$ until we reach the first node whose representation does not end in $2$; this node is denoted by $\mu$.
For a rightmost node $v$ with $\mathcal{A}(v)$ ending in $1$, $\mu$ is defined analogously as the first node on the parent-search path whose representation does not end in $1$.
In both cases, $l>0$ denotes the length of this path.
The node $v$ is chosen as $v\equiv x^{i-1}y^{P(j)}$ in case 2 and $v\equiv x^{i}y^{j}$ in case 4 and case 5.
$x^{i_\mu}y^{j_\mu}\equiv c_R^h(\mu+1)$, where $h$ is determined from Table~\ref{tab_neighborhood66}.
The Hamiltonian of the cluster model in the standard form is written as:
\begin{equation}
	\label{eq_66clusterH1}
	\begin{aligned}
		\mathscr{H}=&-\sum_{ij}Z(x^iy^j(1+\bar{\mathbf{f}}_{(i,j)}\cdot\bar{\mathbf{y}}_{1,\bar{n}}))Z\left( x^{i-1}y^{P\left( j \right)} \right)\\
		&-\sum_{kl}X(x^ky^l(1+\bar{\mathbf{f}}_{(k,l)}\cdot\bar{\mathbf{y}}_{1,\bar{n}}))X\left( x^{k-1}y^{P\left( l \right)} \right) 
	\end{aligned}
\end{equation}
for the first three cases in Table~\ref{tab_66hcmrule}, and
\begin{equation}
	\label{eq_66clusterH2}
	\begin{aligned}
		\mathscr{H}=&-\sum_{ij}Z(x^iy^j(1+\bar{\mathbf{f}}_{(i,j)}\cdot\bar{\mathbf{y}}_{1,\bar{n}}))Z\left( x^{i_\mu}y^{j_\mu} \right)\\
		&-\sum_{kl}X(x^ky^l(1+\bar{\mathbf{f}}_{(k,l)}\cdot\bar{\mathbf{y}}_{1,\bar{n}}))X\left( x^{k_\mu}y^{l_\mu} \right) 
	\end{aligned}
\end{equation}
for the other cases because the functions $u(i,j)$ and $m(i,j)$ in Eq.~(\ref{eq_hoca66sspt}) differ. 
Here, $x^{i_\mu}y^{j_\mu}$ is a neighbor of $x^iy^j$ and is explicitly determined in Table~\ref{tab_66hcmrule}. 
The summation indices $i,j,k,l$ should be chosen according to the sublattices to which the sites $x^iy^j$ and $x^ky^l$ belong, and the summation contains only sites with nontrivial transposed update rule.
In Fig.~\ref{fig_66_model_1}(a), we show a Hamiltonian term of case 2 and the coordinate of its nontrivial support.

\begin{table*}[t]
\caption{\label{tab_66hcmrule} Nontrivial update rule $\bar{\mathbf{f}}_{(i,j)}\cdot\bar{\mathbf{y}}$ for the cluster model on the $\{6,6\}$ lattice. $\bar{\mathbf{f}}_{(i,j)}\cdot\bar{\mathbf{y}}=\mathbf{0}$ otherwise. 
$\beta_0=y^{-2}+y^{-4}+y^{-5}$, $\beta_1(j_\mu)=x^{h-l+1}y^{c_R(j_\mu)-j}+x^{h-l+1}y^{c_R(j_\mu)-j+1}+x^{h-l+1}y^{c_R(j_\mu)-j+3}+x^{h-l+1}y^{c_R(j_\mu)-j+5}$.
Notations are defined in the main text.
}
\begin{ruledtabular}
\begin{tabular}{ccccc}
	Case & Condition & $\bar{\mathbf{f}}_{(i,j)}\cdot \bar{\mathbf{y}}$ \\
	\hline
	1 & $j\equiv 6\pmod7,J(i-1,P(j))=1$ &
	$\beta_0+y^{-6}+x^{-2}y^{P^2(j)-j}$ \\
	2 & $j\equiv 5\pmod7,P(j)\equiv 0\pmod7$, and $x^{i-1}y^{P(j)}$ is marked & $\beta_0+x^{-2}y^{P(P(j))-j}+x^{h-l}y^{j_\mu-j}$ \\
	3 & $j\equiv5\pmod7,P(j)\not\equiv 0,1\pmod7$, and $x^{i-1}y^{P(j)}$ is marked & $\beta_0+x^{-1}y^{P(j)-j+1}+x^{-2}y^{P^2(j)-j}$ \\
	4 & $j\equiv0\pmod7,j_\mu \equiv 6\pmod7$, and $x^{i+h-l}y^{j_\mu}$ is marked & $\beta_1(j_\mu)+x^{h-l-1}y^{P(j_\mu)-j}$ \\
	5 & $j\equiv0\pmod7$ and $x^{i+h-l}y^{j_\mu}$ is endnode & $\beta_1(j_\mu)+x^{h-l}y^{j_\mu-j-1}$ \\
\end{tabular}
\end{ruledtabular}
\end{table*}

Terms $x^iy^j(1+\bar{\mathbf{f}}_{(i,j)}\cdot\bar{\mathbf{y}}_{1,\bar{n}})$ that cannot be fully supported in a finite lattice are excluded by specifying trivial $\bar{\mathbf{f}}_{(i,j)}$ for the site $x^iy^j$.
Then we can assign initial conditions for sites $x^iy^j$ with trivial $\bar{\mathbf{f}}_{(i,j)}$ and we restrict the initial conditions to be defined on one sublattice.
The relevant sublattice can be obtained through the neighboring relation summarized in Table~\ref{tab_neighborhood66}.
In Fig.~\ref{fig_66_model_1}(b)--(e), we visualize the symmetries generated by the following $\mathbf{q}$:
\begin{subequations}
	\begin{align}
		&\mathbf{q}_1(x,y) = x^0y^0  \,, \label{eq_66_cluster_ic1}\\
		&\mathbf{q}_2(x,y) = x^0y^0 + x^1y^2+ x^2y^{21}  \,,\label{eq_66_cluster_ic2}\\
		&\mathbf{q}_3(x,y) = x^1y^1 \,, \label{eq_66_cluster_ic3}\\
		&\mathbf{q}_4(x,y) = x^1y^1 + x^1y^3\,,\label{eq_66_cluster_ic4}
	\end{align}
\end{subequations} 
where $\mathbf{q}_{1,2}$ and $\mathbf{q}_{3,4}$ are defined on different sublattices respectively.

Through the construction of the linear NUCA$(6,6)$ as an example of NUCA$(p,q)$ with the most general non-uniformity, we demonstrate the applicability of our NUCA construction to hyperbolic lattices.

\section{Properties of the hyperbolic cluster state}
\label{app_hcm}
\begin{figure}
	\centering
	\includegraphics[width=0.6\columnwidth]{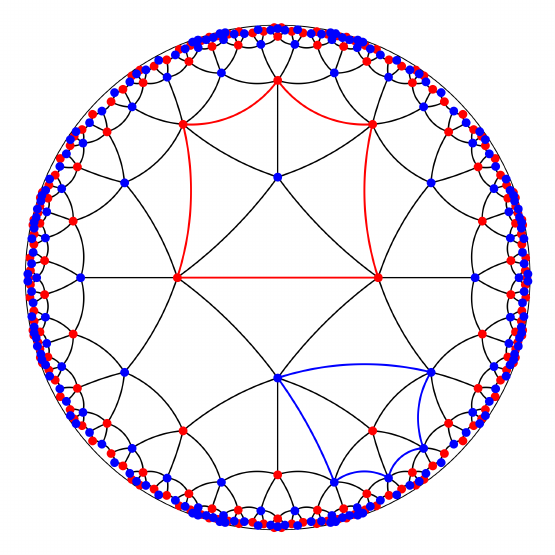}
	\caption{\label{fig_cluster}The bipartite $\{4,5\}$ lattice where qubits are located on vertices. The red and blue vertices denote the two sublattices. The model is defined on vertices of the $\{4,5\}$ lattice, dual to the polygon model on $\{5,4\}$ in Sec.~\ref{sec_NUCA54}. The five red (blue) vertices in the red (blue) plaquette with the central blue (red) vertex form the support of a Hamiltonian term. }
\end{figure}

We have constructed the hyperbolic cluster model with subsystem symmetries by utilizing a linear NUCA, and now we discuss some physical properties of this SSPT model.
The cluster model defined on polygons of the $\{5,4\}$ lattice is dual to that defined on vertices of the $\{4,5\}$ lattice. 
The two sublattices of vertices of the $\{4,5\}$ lattice are represented by $\Lambda^{(a)}$ and $\Lambda^{(b)}$. 
Under a change of basis $Z\leftrightarrow X$ realized by Hadamard gates, Eq.~(\ref{eq_54clusterH}) is equivalently defined by the following Hamiltonian: 
\begin{equation}
	\label{cluster_Hamiltonian}
	H=-\sum_{i\in\Lambda^{(a)}}{X_{i}^{\left( a \right)}\prod_{r\in\mathcal{N}(i)} Z_r^{(b)}} -\sum_{i\in\Lambda^{(b)}}{X_{i}^{\left( b \right)}\prod_{r\in\mathcal{N}(i)} Z_r^{(a)}}
\end{equation}
which is visualized in Fig.~\ref{fig_cluster}.
Here, $X$ and $Z$ are Pauli operators, and $\mathcal{N}(i)$ denotes neighboring vertices of $i$ that are connected by edges.
To determine the ground state of the cluster model, the decorated domain wall method can be applied similarly to the Euclidean case~\cite{you2018sspt}. 
The ground state $\left|\Psi \right>$ can be constructed by applying a local unitary circuit consisting of CZ gates to all pairs of nearest-neighboring qubits.

\textit{Degenerate edge modes:}
\label{subsec_degES}
Under open boundary conditions, there are degenerate edge modes which are robust against local perturbations respecting the subsystem symmetry. 
We consider the lattice boundary geometry shown in Fig.~\ref{fig_cluster}. 
There exist free spin-1/2 degrees of freedom that cannot be gapped by local perturbations preserving symmetries. 
We assume the boundary qubits $i$ are in the $(a)$ sublattice, and the corresponding edge operators can be written as
\begin{equation}
	\begin{aligned}
		&\mathscr{X}_i^{\left( a \right)}=X_{i}^{(a)}\prod_{r\in\mathcal{N}_{\partial}(i)}Z_{r}^{(b)}\, ,\\
		&\mathscr{Y}_i^{\left( a \right)}=Y_{i}^{(a)}\prod_{r\in\mathcal{N}_{\partial}(i)}Z_{r}^{(b)}\, ,\\
		&\mathscr{Z}_i^{\left( a \right)}=Z_{i}^{(a)}\,,
	\end{aligned}
\end{equation}
where $\mathcal{N}_{\partial}(i)$ denotes the neighboring vertices of the boundary vertex $i$ that remain inside the open lattice.
These are exactly the Hadamard-transformed edge operators defined in Eq.~(\ref{eq_edge_operators}) for the cluster model.
The scenario of the other sublattice is similar under the change of $(a) \leftrightarrow (b)$.
The edge operators for qubits satisfy the Pauli algebra and commute with the bulk Hamiltonian Eq.~(\ref{cluster_Hamiltonian}). 
Therefore, there are $2^{N_L}$-fold degenerate edge modes, where $N_L$ denotes the number of vertices on the boundary.

\textit{Transverse field phase transition:}
Now we append an external uniform transverse magnetic field $X$ to the hyperbolic cluster model. 
The Hamiltonian reads:
\begin{equation}
	H=-J\sum_{i}{X_{i}\prod_{r\in\mathcal{N}(i)} Z_r}-h\sum_{i}{X_{i}}\,,
\end{equation}
where $i$ runs over all vertices. 
By performing controlled-Z gates on all pairs of vertices we have:
\begin{equation}
	H\rightarrow H_{\text{dual}}=-h\sum_{i}{X_{i}\prod_{r\in\mathcal{N}(i)} Z_r}-J\sum_i{X_{i}}\,,
\end{equation}
where $h$ and $J$ are effectively exchanged.
When the transverse field reaches $J=h$, the transverse field cluster model is self-dual and is expected to have a first-order phase transition by analogy with the cluster model on the Euclidean lattice~\cite{you2018sspt}. 

\textit{Gauged theory:}
(Subsystem) symmetries of SPT models can be gauged and the (subsystem) symmetries become local gauge symmetries. Different phases under global (subsystem) symmetries can be mapped into different phases of the gauge theory. 
Here, we adopt the gauging procedure as described in Ref.~\cite{chen2019foliated} to gauge the Hamiltonian Eq.~(\ref{cluster_Hamiltonian}).

For clarity, we denote Pauli operators of the matter field as $\sigma^{z}$ and $\sigma^{x}$. The minimal coupling term that satisfies the subsystem symmetry is the product of the $\sigma^z$ operators of the five qubits neighboring a single vertex, i.e., $\sigma_{j}^{z}\sigma_{k}^{z}\sigma_{l}^{z}\sigma_{m}^{z}\sigma_{n}^{z}$ around a vertex $i$. To gauge the theory, we put a gauge field $\tau_{i}$ on each vertex $i$ of the $\{4,5\}$ lattice.
Then we specify the gauge-symmetry terms as: $A_i^{(a)}=\sigma_{i}^{x,\left( a \right)} \prod_{r\in\mathcal{N}(i)}  \tau_{r}^{x,\left( b \right)}$ 
and 
$A_i^{(b)}=\sigma_{i}^{x,\left( b \right)}\prod_{r\in\mathcal{N}(i)}  \tau_{r}^{x,\left( a \right)}$.
To make the model commute with the gauge symmetry, the Hamiltonian terms are modified to: 
\begin{equation}
	\sigma_{i}^{x,\left( a \right)}\prod_{r\in\mathcal{N}(i)} \sigma_{r}^{z,\left( b \right)} \rightarrow \sigma_{i}^{x,\left( a \right)} \tau_{i}^{z,\left( a \right)} \prod_{r\in\mathcal{N}(i)} \sigma_{r}^{z,\left( b \right)} 
\end{equation}
and 
\begin{equation}
	\sigma_{i}^{x,\left( b \right)}\prod_{r\in\mathcal{N}(i)} \sigma_{r}^{z,\left( a \right)} \rightarrow \sigma_{i}^{x,\left( b \right)} \tau_{i}^{z,\left( b \right)} \prod_{r\in\mathcal{N}(i)} \sigma_{r}^{z,\left( a \right)} \,.
\end{equation}
Due to the absence of loops in the subsystem symmetries, there are no local flux terms commuting with all the gauge-symmetry terms. 
Following the above procedures, the gauged Hamiltonian reads:
\begin{equation}
	\label{gaugedH}
	\begin{aligned}
		H_g=&-\sum_{i\in\Lambda^{(a)}}\left[ \sigma_i^{x,(a)}\tau_i^{z,(a)} \prod_{r\in\mathcal{N}(i)}\sigma_r^{z,(b)}+A_i^{(a)} \right] \\
		&-\sum_{i\in\Lambda^{(b)}}\left[ \sigma_i^{x,(b)}\tau_i^{z,(b)} \prod_{r\in\mathcal{N}(i)}\sigma_r^{z,(a)}+A_i^{(b)} \right]\,,
	\end{aligned}
\end{equation}
where the first term in each line is the gauged cluster term, and the second term is the gauge symmetry term.

The gauged model is a dual SSPT model of the original cluster model. By applying the symmetric unitary operator consisting of CNOT gates:
\begin{equation}
	V=\prod_{i}\prod_{r\in\mathcal{N}(i)}
	C_{\sigma_i}X_{\tau_r} \,,
\end{equation}
where the control and target qubits are on different sublattices, the Hamiltonian is transformed to:
\begin{equation}
	\begin{aligned}
		VH_{g}V^{\dagger} \cong &-\sum_{i\in\Lambda^{(a)}}{\tau_i^{z,(a)}\prod_{r\in\mathcal{N}(i)}\tau_r^{x,(b)}} \\
		& -\sum_{i\in\Lambda^{(b)}}{\tau_i^{z,(b)}\prod_{r\in\mathcal{N}(i)}\tau_r^{x,(a)}} \,.
	\end{aligned}
\end{equation}
This is a cluster model defined by the gauge field as we decouple the $\sigma$ qubits. The relation $\cong$ indicates that the original cluster model Eq.~(\ref{cluster_Hamiltonian}) and the guaged model $VH_{g}V^{\dagger}$ represent the same gapped phase. 



\providecommand{\noopsort}[1]{}\providecommand{\singleletter}[1]{#1}%

\end{document}